\documentclass[12pt,preprint]{aastex}
\usepackage{lscape,graphicx}

\def\deg      {{\ifmmode^\circ\else$^\circ$\fi}} 

 \shorttitle{Classification of COSMOS galaxies with ZEST \& morphologically-split LFs}
 \shortauthors{Scarlata et al.}
\begin{document}
\title{COSMOS$^{\star}$ morphological classification with ZEST \\
  (the Zurich Estimator of Structural Types)\\
  and the evolution since $z=1$ of the Luminosity Function of early-,
  disk-, and irregular galaxies}
\author{C. Scarlata\altaffilmark{1}, C. M. Carollo\altaffilmark{1}, S.J.
  Lilly\altaffilmark{1}, M. T. Sargent\altaffilmark{1}, R.
  Feldmann\altaffilmark{1}, P. Kampczyk\altaffilmark{1}, C.
  Porciani\altaffilmark{1}, A. Koekemoer\altaffilmark{2}, N.
  Scoville\altaffilmark{3}, J-P.  Kneib\altaffilmark{4}, A.
  Leauthaud\altaffilmark{4}, R.  Massey\altaffilmark{3}, J.
  Rhodes\altaffilmark{3,5}, L. Tasca\altaffilmark{4}, P.
  Capak\altaffilmark{3}, C.  Maier\altaffilmark{1}, H. J.
  McCracken\altaffilmark{6}, B.  Mobasher\altaffilmark{2}, A.
  Renzini\altaffilmark{7}, Y.  Taniguchi\altaffilmark{8}, D.
  Thompson\altaffilmark{3,9}, K. Sheth\altaffilmark{3,10}, M.
  Ajiki\altaffilmark{11}, H.  Aussel\altaffilmark{12,13}, T.
  Murayama\altaffilmark{11}, D. B.  Sanders\altaffilmark{12}, S.
  Sasaki\altaffilmark{11,14}, Y.  Shioya\altaffilmark{14}, M.
  Takahashi\altaffilmark{11,14}}
\altaffiltext{1}{Department of Physics, Swiss Federal Institute of
  Technology (ETH-Zurich), CH-8093 Zurich, Switzerland}
\altaffiltext{2}{Space Telescope Science Institute, 3700 SanMartin
  Drive, Baltimore, MD 21218} 
\altaffiltext{3}{California Institute of Technology, MC 105-24, 1200
  East California Boulevard, Pasadena, CA 91125}
\altaffiltext{4}{Laboratoire d'Astrophysique de Marseille, BP 8,
  Traverse du Siphon, 13376 Marseille Cedex 12, France}
\altaffiltext{5}{Jet Propulsion Laboratory, Pasadena, CA 91109}
\altaffiltext{6}{Institut d'Astrophysique de Paris, UMR 7095, 98 bis
  Boulevard Arago, 75014 Paris, France} 
\altaffiltext{7}{Dipartimento di Astronomia, Universit\'a di Padova,
  vicolo dell'Osservatorio 2, I-35122 Padua, Italy}
\altaffiltext{8}{Subaru Telescope, National Astronomical Observatory
  of Japan, 650 North Aohoku Place, Hilo, HI 96720}
\altaffiltext{9}{Large Binocular
  Telescope Observatory,University of Arizona, 933 N. Cherry Ave.,
  Tucson, AZ 85721}
\altaffiltext{10}{Spitzer Science Center, California Institute of
  Technology, Pasadena, CA 91125, USA} 
\altaffiltext{11}{Astronomical Institute, Graduate School of Science,
  Tohoku University, Aramaki, Aoba, Sendai 980-8578, Japan}
\altaffiltext{12}{Institute for Astronomy, 2680 Woodlawn Dr.,
  University of Hawaii, Honolulu, Hawaii, 96822}
\altaffiltext{13}{Service d'Astrophysique, CEA/Saclay, 91191
  Gif-sur-Yvette, France}
\altaffiltext{14}{Physics Department, Graduate School of Science and
  Engineering, Ehime University, Japan}

\altaffiltext{$\star$}{ Based on observations with the NASA/ESA {\em
    Hubble Space Telescope}, obtained at the Space Telescope Science
  Institute, which is operated by AURA Inc, under NASA contract NAS
  5-26555; also based on data collected at : the Subaru Telescope,
  which is operated by the National Astronomical Observatory of Japan;
  Kitt Peak National Observatory, Cerro Tololo Inter-American
  Observatory, and the National Optical Astronomy Observatory, which
  are operated by the Association of Universities for Research in
  Astronomy, Inc.  (AURA) under cooperative agreement with the
  National Science Foundation; and the Canada-France-Hawaii Telescope
  operated by the National Research Council of Canada, the Centre
  National de la Recherche Scientifique de France and the University
  of Hawaii.}

 \begin{abstract} 
   Motivated by the desire to reliably and automatically classify the
   structural properties of hundred-thousands of COSMOS galaxies,
   we present ZEST, the Zurich Estimator of Structural Types.\\
   To classify galaxy types and structure, ZEST uses: $(i)$ Five
   non-parametric diagnostics, i.e., asymmetry $A$, concentration $C$,
   Gini coefficient $G$, 2nd-order moment of the brightest 20\% of
   galaxy pixels M$_{20}$, and ellipticity $\epsilon$; and $(ii)$ The
   exponent $n$ of single--Sersic fits to the two-dimensional surface
   brightness distributions.  To fully exploit the wealth of
   information while reducing the redundancy present in these
   diagnostics, ZEST performs a principal component ($PC$) Analysis.
   We use a sample of $\sim 56,000$ $I_{AB}\le 24$ COSMOS galaxies to
   show that the main three $PC$s fully describe, without significant
   loss of information, the key aspects of the galaxy structure, i.e.,
   to calibrate a three-dimensional {\it ZEST classification grid} of
   axis $PC_1$, $PC_2$, and $PC_3$.  We demonstrate the robustness of
   the ZEST grid on the $z=0$ sample of \citet{frei1996}. The ZEST
   classification breaks most of the degeneracy between different
   galaxy populations that affects morphological classifications based
   on only some of the basic
   diagnostics included in the ZEST structural analysis.\\
   As a first application of the ZEST structural classification
   scheme, we present the evolution in the $0.2<z\le1$ redshift regime
   of the Luminosity Functions of COSMOS galaxies of different
   structural types (i.e., early, disk and irregular galaxies) and,
   for disk galaxies, of different bulge-to-disk ratios.  Overall, we
   find that the bright-end of the luminosity function up to a
   redshift $z=1$ is consistent with a pure-luminosity evolution (of
   about 0.95 magnitudes at $z \sim 0.7$).  We highlight however two
   trends, relative to the local Universe, that are in general
   agreement with a {\it down-sizing} scenario for galaxy formation,
   i.e.: $(1.)$ A deficit of a factor of about two at $z\sim0.7$ of
   $M_B>-20.5$ structurally--classified early--type galaxies; and
   $(2.)$ An excess of a factor of about three, at a similar redshift,
   of irregular galaxies.

\end{abstract}
 \section{Introduction}

 Galaxy morphologies are strongly correlated with the star formation
 histories of galaxies, are a proxy for their dynamical structure, and
 are thus a key diagnostic of galaxy evolution.  As state-of-the-art
 surveys of faint galaxies at (low and) high redshifts such as COSMOS
 now return millions of galaxies, it becomes essential to have robust
 tools for automatic morphological and structural classifications.

 Motivated by the need to automatically derive a quantitative
 description for the morphology of the large number of COSMOS
 galaxies, we present the Zurich Estimator of Structural Type (ZEST),
 a powerful classification scheme that combines:\\
 $-$ A Principal Component Analysis (PCA) of five non-parametric
 diagnostics of galaxy structure, i.e., asymmetry $A$, concentration
 $C$, Gini coefficient $G$, 2nd-order moment of the brightest 20\% of
 galaxy pixels M$_{20}$ \citep[e.g.,][]{abraham2003,lotz2004}, and the
 ellipticity $\epsilon$.  The PCA reduces the redundancy of
 information that is present in these five diagnostics, without
 substantial loss of information on the galaxy structure. \\ $-$ A
 parametric description of the galaxy light, i.e., the exponent $n$ of
 a single--Sersic fit to its surface
 brightness distribution \citep[][this volume]{sargent2006}.  \\
 Specifically, ZEST classifies galaxies on the basis of their location
 in the three-dimensional space defined by the main three eigenvectors
 $PC_1$, $PC_2$ and $PC_3$ that contain most of the variance of the
 original non-parametric quantities, while the $n$- Sersic indices are
 used to refine the classification of disk galaxies by splitting these
 in four separate bins of a bulgeness parameter (that is correlated
 with the bulge-to-disk ratio).

 We calibrate the ZEST grid with a sample of $\sim 56000$ $I_{AB} \le
 24$ COSMOS galaxies selected from a catalogue covering an area of
 about 40\% of the entire 2deg$^2$ COSMOS field, produced from the
 Cycle~12 HST ACS F814W images \citep{scoville2006a,leauthaud2006}.
 The ZEST calibration is also tested on the $z=0$ sample of
 \citet{frei1996}.  The ZEST grid assigns, to each galaxy: \\
 (1) A morphological type (=1, 2 or 3 for early-type, disk and
 irregular galaxies, respectively); \\
 (2) A bulgeness parameter for the type 2 disk galaxies, which
 coarsely correlates with the bulge-to-disk ratio (B/D); the disk
 galaxies are split in four bins, from pure disk galaxies ($T=2.3$)
 to bulge dominated disks ($T=2.0$);\\
 (3) An elongation parameter (four bins, respectively from $=0$ for
 face--on to $=3$ for edge--on galaxies); \\
 (4) An irregularity and a clumpiness parameter for types 1 and 2
 galaxies, which indicate whether the galaxy light distribution is
 regular/smooth or distorted/clumpy, respectively.

 Although the size of the galaxy does not enter directly in the ZEST
 classification, a measurement of the Petrosian radius
 \citep{petrosian1976} is produced as a by--product of the ZEST
 classification procedure, and is thus also
 available for all the ZEST--classified galaxies.\\

 The calibrated ZEST grid in the $PC_1$-$PC_2$-$PC_3$ space can be
 used to automatically classify galaxies according to their structural
 properties.

 While different combinations of some of the basic diagnostics that we
 use to construct the ZEST grid have been extensively used in the
 literature to classify galaxy types, these combinations are
 effectively ``projections'' on lower-dimensionality planes of the
 ZEST $PC_1$-$PC_2$-$PC_3$ space.  These projections mix very
 different galaxy populations (e.g., elliptical galaxies with regular
 edge-on disks in the $M_{20}$-$G$ and $C$-$M_{20}$ planes, etc.), and
 thus lead to heavy contamination of galaxy samples and erroneous
 statistical assessments concerning the evolution of specific galaxy
 populations.  The three-dimensional $PC_1$-$PC_2$-$PC_3$ space breaks
 the degeneracy between these different galaxy populations.

 As a first application of the ZEST structural classification of
 COSMOS galaxies, we present the redshift evolution since $z=1$ of the
 Luminosity Functions (LFs) of galaxies with different structural
 types (including different ``bulgeness'' for the disk galaxies).  In
 order to anchor the COSMOS measurements to the local Universe, we use
 the sample of 1813 Sloan Digital Sky Survey \citep[SDSS;
 ][]{york2000} galaxies constructed and discussed by Kampczyk et al.\
 (2006).  A detailed discussion of the LFs of more elaborated
 selections of galaxy samples that are structurally classified with
 ZEST will be presented elsewhere \citep[see, e.g.,][for an extensive
 discussion of the evolution of the LF of elliptical galaxy
 progenitors]{scarlata2006b}.

 This paper is structured as follows. In Section~2 we describe the
 methodology at the basis of the ZEST classification: in particular,
 Section~2.1 describes the non-parametric quantification of galaxy
 structure, Section~2.2 the principal component analysis in the
 $A,C,G,M_{20}$, and $\epsilon$ space, Section~2.3 presents the
 calibration of the ZEST morphological grid, and the use of the
 $n$-Sersic index values to establish the B/D ratio of disk galaxies
 (Section~2.3.2). The performance and reliability of ZEST are
 discussed in Section~3, where we present the test performed on a
 $z=0$ galaxy sample (Section~3.1), and a discussion of the error bars
 in the ZEST structural classification (Section~3.2). In Section~4 we
 show the advantage of the ZEST classification scheme over other
 approaches. These comparisons show that the ZEST classification is
 substantially more powerful, relative to other schemes that are based
 on only a few of the basic diagnostics that are contained in the ZEST
 grid, to separate galaxies with different structural properties.  In
 Section 5 we present the Luminosity Functions of the ZEST-classified
 COSMOS galaxies, the corrections applied to the LFs, and the sources
 of error.  The paper is briefly summarized in Section~6.

 Throughout the paper we use the cosmology that has been adopted
 throughout this volume, i.e., $\Omega_m=0.25$, $\Omega_m +
 \Omega_{\Lambda}=1$, and $H_{0}=70$ km s$^{-1}$ Mpc$^{-1}$.  All
 magnitudes are expressed in the AB system \citep{oke1974}.

\section{Methodology}

ZEST is based on a combination of non--parametric and parametric
quantification of galaxy structure, and a PCA to reduce the number of
variables. We describe below the steps that we followed to construct
the ZEST classification grid.

\subsection{Non-parametric quantification of structure}
\label{sec:parameters}

The ZEST PCA uses, as basic quantities, four widely-adopted
non-parametric measurements of the light distribution in galaxies
\citep[e.g., ][]{abraham2003,lotz2004}, plus a measurement of the
ellipticity of the galaxy light distribution. Specifically, ZEST uses:

\noindent {\it 1.) } The concentration $C$ ($=5
\log{\frac{r_{80}}{r_{20}}}$; with $r_{80}$ and $r_{20}$ the radii
including 80\% and 20\% of the total galaxy light, respectively),
which quantifies the central density of the galaxy light distribution;

\noindent {\it 2.)}  The asymmetry $A$, which quantifies the degree of
rotational symmetry of the light distribution. $A$ is measured by
calculating the normalized difference between the galaxy image and the
image rotated by $180^{\circ}$.  A correction for background noise is
also applied \citep[as in ][]{conselice2000,lotz2004}, i.e.:

\begin{equation}
A=\frac{\sum_{x,y}{ |I_{(x,y)}-I_{180(x,y)}| }} {2\,\sum{|I_{(x,y)}|}} -B_{180};
\end{equation}
\noindent
where $I$ is the galaxy flux in pixel $(x,y)$, $I_{180}$ is the image
rotated by 180$^o$ about the galaxy's central pixel, and $B_{180}$ is
the average asymmetry of the background;

\noindent {\it 3.)} The Gini coefficient $G$ which describes how
uniformly the flux is distributed among galaxy pixels. The Gini
statistic assumes values from 0 (if the galaxy light is homogeneously
distributed among galaxy pixels), up to 1 (if all the light is
concentrated in one pixel, regardless of its position in the galaxy).
Specifically, after ordering the pixels by increasing flux value, $G$
is given by:

\begin{equation}
G = \frac{1}{\bar{X} n (n-1)} \sum^n_i (2i - n -1) X_i,
\end{equation}

\noindent
where $n$ is the number of pixels assigned to a galaxy, and $\bar{X}$
is the mean pixel value \citep{glasser1962};

\noindent {\it 4.)} $M_{20}$ ($=\log{\sum{M_i}/M_{tot}}$, with
$\sum{f_{i}}<20$\% and $M_{tot}$ the total second order moment), i.e.,
the second order moment of the brightest 20\% of the galaxy flux.  For
centrally concentrated objects, $M_{20}$ correlates with the
concentration $C$; however, $M_{20}$ is also sensitive to bright
off--centered knots of light;

\noindent {\it 5.)} The ellipticity $\epsilon=1-b/a$ of the light
distribution, as measured by SExtractor \citep[version
2.4.3,][]{sextractor}.  SExtractor computes the semi-major axis $a$
and semi-minor axis $b$ from the second order moments of the galaxy
light; specifically, $a$ and $b$ are the maximum and minimum spatial
variance ({\it rms}) of the object, along the direction $\theta$ where
the variance is maximized:
\begin{equation}
a^2=\frac{\bar{x^2}+\bar{y^2}}{2}+\sqrt{\frac{(\bar{x^2}-\bar{y^2})^2}{4}+\bar{xy}^2},
\end{equation}
\begin{equation}
b^2=\frac{\bar{x^2}+\bar{y^2}}{2}-\sqrt{\frac{(\bar{x^2}-\bar{y^2})^2}{4}+\bar{xy}^2},
\end{equation}

\noindent
where the second order moments $\bar{x2}$, $\bar{y2}$, and
$\bar{xy}$ are given by:

\begin{equation}
\bar{x^2}=\frac{\sum_i I_ix^2_i}{\sum I_i} - \bar{x}^2,
\end{equation}

\begin{equation}
\bar{y^2}=\frac{\sum_i I_iy^2_i}{\sum I_i} - \bar{y}^2,
\end{equation}

\begin{equation}
\bar{xy}=\frac{\sum_i I_ix_iy_i}{\sum I_i} - \bar{x}\bar{y}.
\end{equation} 

The values of $a$ and $b$ are thus representative of the galaxy
ellipticity at large radii, and are rather insensitive to details in
the internal structure (e.g., bars in disk galaxies, knots of star
formation, etc).

\subsection{Principal Component Analysis}
\label{sec:pcaglobal}

The quantities above provide complementary, but also redundant
information on galaxy structure. We therefore performed a PCA using
the measurements of $A$, $C$, $M_{20}$, $G$ and ellipticity $\epsilon$
as basic variables.

The PCA is a classical statistical method for multivariate analysis,
which reduces the dimensionality of a dataset without a significant
loss of information. This is done by transforming the observed
variables into a new set of orthogonal variables, the {\it principal
  components} ``$PC_i$'', (with $i=1,..,n$, and $n$ the number of
basic parameters, i.e., variables). The $PC_i$ are ordered so that the
first few of them retain most of the variance present in the original
dataset. The principal components are a linear combination of the
original variables, and define a new coordinate system obtained by
rigid rotation of the original space. In the new system, the axes
represent the directions of maximum variability in the original
$n-$dimensional distribution of points.

In detail, the dataset is described by an $n \times m$ data matrix,
($n=5$ in the current version of ZEST), and $m$ is the number of
galaxies with measured basic parameters.  All variables are
standardized before performing the analysis by subtracting their
median value (indicated with the subscript $M$ in the following
expressions) and normalizing them with their standard deviation.
Therefore the five variables considered in the ZEST PCA are defined as
$x_1=(C-C_M)/\sigma_C$, $x_2=(M_{20}-M_{20,M})/\sigma_{M_{20}}$,
$x_3=(G-G_M)/\sigma_G$,
$x_{4}=(\epsilon-\epsilon_M)/\sigma_{\epsilon}$, and
$x_{5}=(A-A_M)/\sigma_A$.

The directions of the principal components are derived by calculating
the eigenvectors of the $n \times n$ covariance matrix of the $x_j$
variables ($S_{ij}=\langle(x_i-\langle x_i \rangle)(x_j-\langle
x_j\rangle)\rangle$). The matrix $S \ge 0$ is real and symmetric.
Thus, it admits real, positive eigenvalues $\lambda_i$.  By sorting
the eigenvectors in order of decreasing values of the eigenvalues, an
ordered orthogonal basis is obtained, with eigenvectors aligned along
directions of decreasing variance ($\lambda_i/ \sum_j \lambda_j$) in
the data. The first few principal components that account for most of
the power, i.e., most of the total variance -$\sum_j \lambda_j$- in
the dataset, are then used to replace the original $n$ variables
without any significant loss of information.

\subsection{ZEST calibration with 56000 $I_{AB}\le 24$ COSMOS
  galaxies}
 \label{sec:appliccosmos}

 We calibrate the ZEST classification grid on a sample of $\sim
 56000$, $I_{AB}\le 24$, COSMOS galaxies detected in the 260 ACS F814W
 images acquired during the HST Cycle~12 observing period
 \citep{scoville2006a}.  The total area covered by this fraction of
 COSMOS is 0.74 deg$^2$.  Details on the COSMOS sample are given in
 Appendix~\ref{app:sample}.

 For each COSMOS galaxy, we measured the basic non--parametric
 quantities described in Section~\ref{sec:parameters} by computing
 them on the galaxy pixels (defined using Petrosian apertures, see
 Appendix~\ref{app:dataanal}).  Figure~\ref{fig:allpar} shows the
 behavior of each basic non-parametric diagnostic as a function of the
 others.  The contours in each panel enclose respectively 30\%, 80\%
 and 98\% of the COSMOS galaxies in our sample.  Global correlations
 are known to exist between various non-parametric coefficients.  For
 example, relatively tight correlations exist between $G$, $M_{20}$,
 and $C$, with objects with high $C$ tending to have low $M_{20}$ and
 high $G$.  Any value of $C$ is observed for small values of $A$,
 while high values of $A$ are preferentially observed in low-$C$
 galaxies.  These trends have already been noted in the literature,
 and indeed highlight the redundancy of information present in these
 diagnostics.  As expected, the ellipticity $\epsilon$ does not
 correlate with any of the other parameters, except for a mild
 positive correlation with the concentration for $\epsilon> 0.6$.
 This is a geometric effect, since edge-on galaxies preferentially
 have high $C$ values.

 The results of the PCA on the normalized COSMOS dataset are presented
 in Table~\ref{tbl:1}.  In particular, Columns 2--6 refer to the five
 principal components derived in the analysis. The first row gives the
 eigenvalue (i.e., variance) of the data along the direction of the
 corresponding PC. The second row shows the fraction of the variance
 that is explained by each of the PCs, i.e., the fraction of the
 ``power" that is contained in each PC; the third row lists the
 cumulative fraction of the variance.  In the bottom part of the
 Table, each column lists the weights assigned to each input variable
 (listed in Column~1), in the linear combination that gives the
 direction of the specific principal component (e.g.,
 $PC_1=-0.54\times x_1 +0.60\times x_2 - 0.56\times x_3 + 0.20\times
 x_4+0.02\times x_5$).

 In Figure~\ref{fig:pca} we show the fraction of variance as a
 function of the corresponding principal component.  Solid circles
 refer to the PCA applied to the COSMOS galaxies in the considered
 $I_{AB} \le 24$ sample. Open squares refer to the same analysis,
 performed however on only the COSMOS galaxies brighter than
 $I_{AB}=22.5$.  The horizontal line represents the value of the
 eigenvalues that would be expected if the five variables were
 uncorrelated. The comparison of the results for the $I_{AB}\le 22.5$
 and the $I_{AB}\le 24$ samples demonstrates the stability of the
 analysis down to the faintest magnitudes in the sample.  

 As discussed in Appendix~\ref{app:simulations}, the scatter in the
 measured variables that define the PCs increases with increasing
 magnitude (the scatter at $I_{AB}=24$ is twice the scatter measured
 at $I_{AB}=22$). While the larger scatter in the data could
 potentially wash out correlations present in the original parameters,
 this effect is negligible down to the considered $I_{AB}=24$
 magnitude limit: e.g., the fraction of variance in $PC_1$ is only
 $\sim 3$\% larger in the $I_{AB}\le 22.5$ sample relative to the
 $I_{AB}\le 24$ sample.

 Several methods are proposed in the literature to establish the
 number of PCs which are sufficient to fully describe the properties
 of the sample; all methods require some degree of judgment. For
 example \cite{kaiser1960} proposes the rule--of--thumb of rejecting
 all components that contain less power than the variance expected for
 uncorrelated variables (in our case of 5 variables, less than 20\%);
 \cite{jolliffe1972} adopted instead a lower threshold value.  To
 classify galaxy structure, we will use the first three PCs; these
 explain 92\% of the total variance. The ZEST classification grid is
 therefore constructed in a three-dimensional space; specifically,
 galaxies are ranked, according to their structural properties, in
 unit cubes of the $PC_1$, $PC_2$, $PC_3$ ZEST space.

\subsubsection{Morphological classification of the $PC_1$-$PC_2$-$PC_3$ unit cubes}

(COSMOS) Galaxies with different structural properties occupy
different regions of the $PC_1$-$PC_2$-$PC_3$ space.  For example,
$PC_1$ is dominated by $C$, $M_{20}$, and $G$. Highly negative values
of $PC_1$\footnote{$M_{20}$ is always $<0$} are populated by highly
centrally-concentrated galaxies. $PC_2$ is, on the other hand,
dominated by ellipticity and asymmetry: round asymmetric objects are
found at negative values of $PC_2$, and symmetric flattened systems
are preferentially located at high positive values of $PC_2$. $PC_3$
is also mostly a combination of asymmetry and ellipticity, but in
$PC_3$ these parameters contribute both with positive weights to the
absolute value of $PC_3$.  Highly asymmetric and elongated objects are
thus located at high values of $PC_3$.

To associate a (dominant) morphological class to different regions of
the $PC_1$-$PC_2$-$PC_3$ space, the latter was partitioned into a
regular 3-D grid with unit steps in each of the coordinates, and {\it
  all galaxies} in our COSMOS sample within each unit
$PC_1$-$PC_2$-$PC_3$ cube were visually inspected.  Each unit cube was
then assigned a {\it morphological type} (T=1, 2 or 3 for early-type,
disk and irregular galaxies, respectively; face-on S0 galaxies would
of course be found in cubes classified as T=1, while more inclined S0
galaxies would be in cubes classified as T=2), and a {\it clumpiness
  parameter} (in unit steps, ranging from $=0$ for smooth surface
density distributions, to 2 for very clumpy morphologies).  For
galaxies of T=1 or =2, we furthermore assigned an {\it elongation
  parameter} (in unit steps, from 0 for face-on galaxies, to 3 for
edge-on galaxies), and an {\it irregularity parameter} (in unit steps,
ranging from $0$ for regular surface density distributions, to $2$ for
disturbed T=1 or =2 morphologies). A measure of the galaxy sizes
(i.e., their Petrosian radii, see Section~\ref{app:dataanal}) is also
available for all ZEST--classified galaxies, as a byproduct of our
analysis; see Appendix~\ref{app:dataanal}).

\subsubsection{Bulge--to--disk  ratios: Parametric surface brightness fits}

To refine the ZEST structural classification of $T=2$ disk galaxies,
we use additional information that we have available for a subsample
of the COSMOS galaxies considered in this paper, i.e., the GIM2D
single--Sersic fits to the $I_{AB} \le 22.5$ COSMOS galaxies of
\citet[][this volume]{sargent2006}. In particular, we use the
statistical distribution of Sersic index $n$ within each unit cube of
$PC_1$-$PC_2$-$PC_3$ with a T=2 classification, to assign a {\it
  bulgeness parameter} to each T=2 cube. Specifically, the T=2 unit
cubes are split in four bins, i.e, T= 2.3, 2.2, 2.1 and 2.0, depending
on the value of the median Sersic index $n$ of the galaxies in that
cube (T=2.3, 2.2, 2.1, 2.0 for $0<n_{median}< 0.75$, $0.75\le
n_{median} < 1.25$, $1.25\le n_{median}< 2.5$, and $n_{median} \ge
2.5$, respectively).  This refinement of the Type--2 ZEST
classification grid can be interpreted to correspond to a four--bins
classification of disk galaxies in terms of their bulge--to--disk
ratios, with the T=2.0, 2.1, 2.2 and 2.3 cubes hosting galaxies with
decreasing bulge--to--disk ratio (Type 2.0 = bulge--dominated
galaxies, including relatively inclined S0 galaxies, and Type 2.3 =
bulgeless disks).

\subsection{Summary: The ZEST $PC_1$-$PC_2$-$PC_3$ classification scheme and grid}

We summarize the final ZEST classification scheme in
Table~\ref{tbl:zest}, and the COSMOS-calibrated ZEST grid in
Table~\ref{tbl:class}.

To show the performance of ZEST in disentangling galaxies with
different structural properties, we plot in Figure~\ref{fig:examp} a
representative selection of the I$_{AB}\le 24$ COSMOS galaxies that
occupy four different unit cubes of $PC_1$-$PC_2$-$PC_3$.
Furthermore, in Figure~\ref{fig:examples6a} we show, in sequential
planes of constant $PC_3=-2,-1,0,+1,+2,+3$, a representative galaxy in
each of the ($PC_3$-)$PC_1$-$PC_2$ unit bins.  Arrows in the
bottom--left corners of each panel (i.e., $PC_3=$constant plane)
indicate the directions of the steepest (positive) variation for the
quantities specified as labels of the arrows; e.g., arrows labeled as
``bulgeness" or ``irregularity" respectively show the direction,
across the given $PC_3$=constant plane, of the maximum increase of the
``degree of bulgeness'' and of the irregularity of the galaxies
populating that plane.

In Figure~\ref{fig:class} we summarize the COSMOS-calibrated ZEST
classification grid in a schematic way.  In each $PC_3=$constant
plane, different symbols represent the different morphological types,
elongation and bulgeness parameter.  As indicated in the Figure,
ellipses represent the $T=1$ early-type galaxies; concentrical double
circles indicate $T=2$ disk galaxies, and stars represent $T=3$
irregular galaxies. The size of the internal ellipse of T=2 galaxies
is proportional to their bulgeness parameter. The continuity of
properties in the $PC_1$-$PC_2$-$PC_3$ space is immediately evident
from Figure~\ref{fig:class}. For example, the transition from
early-type morphologies ($T=1$), to bulge-dominated disks ($T=2.0$),
to pure-disks galaxies ($T=2.3$) is smooth both across the
$PC_1$-$PC_2$ planes, and along the $PC_3$ direction.  The
bulge-dominated galaxies are found preferentially at intermediate
values of $PC_3$, and, as expected, at low values of $PC_1$.

\section{The performance and reliability of ZEST}

Before demonstrating the robustness of the ZEST classification it is
important to stress that generally, galaxy appearance depends on the
rest-frame wavelength at which it is observed. Since only F814W ACS
images are available for the COSMOS galaxies it is important that any
comparison of morphology for galaxies at different redshifts is
treated with care.  In the low-to-intermediate redshift regime
($0.2<z\le1.0$) that is the focus of this paper, the central
wavelength of the F814W filter covers the rest-frame 4000-6700\AA\
window, where morphological $K-$corrections are negligible for most
galaxies \citep[see, e.g.,][]{lotz2004,cassata2005}.

\subsection{Testing ZEST on $z=0$ galaxies}
 
We assess the performance and reliability of the ZEST classification
grid by applying it to the \citet{frei1996} sample of 80 $z=0$
representative galaxies that (a) have Hubble types available from the
RC3 catalog \citep{devaucouleurs1991}, (b) have been observed (at the
1.1 meter telescope of the Lowell Observatory) in the B$_J$ band
($\lambda_{eff}= 4500$\AA) with a pixel scale of 1\farcs35 per pixel ,
and (c) have PSF-FWHM smaller than 5\farcs0. We excluded from the
analysis three galaxies for which the available Frei's images were too
small to reliably get an estimate of the background.  Frei's galaxies
span Hubble types from Ellipticals ($T=-5$) to Sd ($T=10$); they have
been used as a $z=0$ benchmark for assessing galaxy morphologies at
higher redshifts in a number of other works
\citep[e.g.][]{bershady2000,simard2002,lotz2004}.

In Figure~\ref{fig:visautonew} we show the fraction of objects with a
given RC3 classification (E, S0-Sab, Sb-Scd and Sd and later) that
have respectively Type=1, 2.0, 2.1, 2.2, 2.3, and T=3 ZEST
classification.  The comparison between the two classifications is
excellent, and highlights the power of the automatic ZEST
classification scheme to recover the physically-motivated Hubble types
of galaxies.
 
There are only a few objects which have significantly different
classifications between RC3 and ZEST.  In Figure~\ref{fig:visautonew2}
we show the postage stamps of these most discrepant galaxies.
A detailed analysis of these galaxies shows that:
\begin{itemize}
\item[--] NGC~4621 and NGC~4564, classified as $-5$ by the RC3 but as
  Type~2.0 by ZEST, have a disk component
  \citep{michard1994,scorza1995,mizuno1996,emsellem2004}.

\item[--] NGC~4710 has a ZEST Type$=2.2$ and a Hubble Type S0. As the
  image shows \citep[see also,][]{michard1994}, this edge-on galaxy
  displays a very bright ring and an important equatorial dust lane,
  which causes its visual classification to be highly uncertain:
  indeed, in the UGC catalog this galaxy is classified as an S0a.

\item[--] The two galaxies classified as Type~1 by ZEST and with an
  RC3 classification of S0-Sa are relatively face-on galaxies. Their
  surface brightness is rather smooth with no visible spiral arms, or
  star formation, with the exeption of a smooth ring around NGC~4340.

\item[--] NGC~4088, classified by ZEST as Type$=3$ and by the RC3 as
  Sb-Scd, is described as an irregular and distorted spiral by Dahari
  (1985).

\item[--] Finally, for the few galaxies which are classified by ZEST
  as bulge dominated (Type~2.0) galaxies and have RC3 classification
  of Sb-Scd, we performed a single Sersic fit to their surface
  brightness profile. They have concentrated light distributions, with
  Sersic's index $n\ge 3$, which confirms our ZEST classification as
  Type~2.0 galaxies.
\end{itemize}

\subsection{Error bars in the ZEST morphological classification of COSMOS galaxies}
\label{sec:ZESTerr}
In Appendix~\ref{app:simulations} we discuss in detail the
uncertainties and the systematic errors in the measured structural
parameters as a function of signal-to-noise ratio (S/N). To do so we
use a sample of bright COSMOS galaxies, which are progressively dimmed
to fainter magnitudes (lower S/N).  Here we summarize to what extent
the COSMOS-calibrated ZEST morphological classification grid is
affected by the S/N of the individual galaxies. We use the same sample
of bright and progressively S/N-degraded test galaxies described in
Appendix~\ref{app:simulations}, and compute for each of them all the
parameters involved in the ZEST classification, both on the original
and on the progressively S/N-degraded images. The $PC_i$ ($i=1,2,3$)
values are then computed for each galaxy at each S/N level, and the
ZEST morphological classification corresponding to the relevant unit
cube of $PC_1$-$PC_2$-$PC_3$ is assigned to each of the original and
artificially fainted galaxies.  The change in galaxy type ($\Delta T$)
that occurs due to degraded S/N is finally computed as a function of
magnitude.

The ZEST classification is robust down to I$_{AB}=24$.  For magnitudes
$I_{AB}\le 22.5$, more than 90\% of galaxies do not change
morphological class. The remaining few percent of galaxies changes
morphological type by smoothly moving through the $PC_1$-$PC_2$-$PC_3$
space; indeed, the change of ZEST morphological Type with varying S/N
happens typically for galaxies which are originally classified in
$PC_1$-$PC_2$-$PC_3$ cubes that are close to a ``type transition-wall"
in the PC's space. The fraction of galaxies with $\Delta T =0$ remains
larger than $\sim 75$\% down to magnitudes $I_{AB}=23.0$, and even in
the highest magnitude bin ($23.5<I<24$), the fraction of galaxies with
$\Delta T=0$ remains of order 65\%. This is illustrated in
Figure~\ref{fig:simclass}, where we show the distribution of the
average absolute variation in $PC_i$, namely $\langle
\Delta(PC)\rangle=(\sum_{i} |PC_{i,f}-PC_{i,0}|)/3$ (with $i=1,2,3$,
$PC_{i,0}$ the initial values of the $PC_i$'s, and $PC_{i,f}$ the
measured $PC_i$'s after S/N degradation).  Split in four magnitude
bins (from $I_{AB}=22$, top-left panel, down to $I_{AB}=24$,
bottom-right panel), the {\it solid } histograms show the $\langle
\Delta(PC) \rangle$ distribution for all galaxies in the considered
magnitude bin, and the {\it hatched} histograms show the $\langle
\Delta(PC) \rangle$ distribution for galaxies which change
morphological class due to S/N degradation.

Figure~\ref{fig:tracks} shows a few examples of how galaxies move
through PC space as the S/N of the galaxy images decreases.  Initial
and final values for the $PC_i$ coordinates of the example test
galaxies are indicated at the beginning and end of the track that
describes the movement, in the PC space of the specific test-galaxy.
Note that galaxies can have a $\langle \Delta(PC)\rangle$ as high as
$\langle \Delta(PC)\rangle=2$, without changing their morphological
class (as their path in $PC_i$ space occurs within a region uniformly
classified with a specific morphological type).  As stressed above,
virtually all galaxies that change morphological class are located, to
start with, in bins of $PC_1$-$PC_2$-$PC_3$ that border with bins with
a different morphological classification.

The most noticeable effect of S/N degradation is a contamination, at
the faintest magnitudes, of at most $\sim 30$\% from low S/N early
type galaxies to the Type~2 or Type~3 galaxy populations \citep[see
also, ][]{abraham1996,lotz2004}.  This suggests that at most 30\% of
early-type galaxies at the faintest magnitudes could drop from the
early-type sample and be misclassified as disk or irregular galaxies
due to their lower S/N values. We discuss in
Section~\ref{sec:LFsdsscomp} the implications of this effect on the
redshift evolution up to $z\sim1$ of the LFs of COSMOS galaxies
structurally classified with ZEST.

\section{The advantage of the ZEST classification scheme over other approaches}

Popular classifications of (nearby and high$-z$) galaxies in the past
few years have been typically based on combinations of two or three of
the non-parametric diagnostics that are used in ZEST \citep[e.g.,
][and references therein]{lotz2004,ferreras2005} or on a threshold in
Sersic index $n$ (most noticeably the SDSS defined early-type galaxies
the objects with $C \ge 2.87$ or $n \ge 2.5$, and late-type galaxies
all other galaxies).  Such approaches can be seen as
``lower-dimensionality" projections of the ZEST grid, and lead to
galaxy samples that are affected by large contamination of systems
with rather different structural properties.

As an example, in Figure~\ref{fig:projection} we show the
two-dimensional planes defined by $G$ and $M_{20}$ (top row), $C$ and
$M_{20}$ (middle row), and $C$ and $A$ (bottom row).  All plots show
the total density of galaxies in grey scale.  In the different panel
we indicate the region of space that is populated by a given ZEST
morphological class; in particular, the color contours enclose $\sim
99$\% of the COSMOS galaxies with the specified ZEST morphological
class. To avoid crowding we show in the left panels the location of
the $T=1$ early-type galaxies (red contours), in the central panels
the $T=2$ disk galaxies (blue contours, with decreasing shade of blue
from the bulge-dominated $T=2.0$ galaxies to the $T=2.3$ bulgeless
disks), and in the right panels the $T=3$ irregular galaxies (green
contours).

It is clear from Figure~\ref{fig:projection} that there is a high
level of contamination by different galaxy types in all regions of
these two-dimensional planes. For example, on the $C-M_{20}$ plane,
$T=1$ early-type galaxies form a tight sequence, well visible in the
left panel of Figure~\ref{fig:projection}; however, such a sequence
includes (not only the majority of $T=2.0$ bulge--dominated disk
galaxies, but also) a substantial fraction of lower bulge--to--disk
ratio (i.e., $T=2.2, 2.1$) galaxies.  For example, if the galaxies
that lie above the solid black line on the $C-M_{20}$ plane were
classified as ``early-types", ZEST would return the following galaxy
population mixture in the selected region: $38\%$ of desired $T=1$
early-type galaxies, $23\%$ of $T=2.0$ galaxies, and $26\%$ and $10\%$
of $T=2.1$ and of $T=2.2$ galaxies, respectively. Even summing up
together the T=1 and 2.0 galaxies (which might be desired for some
science applications), still the selected sample would be affected by
a contamination of order $40\%$ contributed by lower ``bulge-to-disk
ratio'' galaxies.  Similar levels of contamination are found when the
other two-dimensional planes of Figure~\ref{fig:projection} or a
simple cut in Sersic index $n$ are used to morphologically-classify
galaxies, as it has been done in most of the previous literature.

\section{A first application of ZEST: The evolution since $z\sim1$ of the LFs 
of morphological early-type, disk and irregular COSMOS galaxies }
\label{sec:LF}

We derive the rest--frame $B-$band LFs of ZEST--classified COSMOS
galaxies brighter than $I=24$ (see Appendix~\ref{app:sample} for
details on the COSMOS sample).  Given the large number statistics of
(the fraction of survey area that we are considering for) COSMOS, we
can compute the LF $\Phi(M,z,T)$ of each morphological class in four
different redshift intervals.

We use the ZEBRA Maximum Likelihood photometric redshifts of Feldmann
et al.\ (2006) to derive the COSMOS LFs.  These photometric redshifts
have an accuracy of $\Delta z/ (1+z)\sim 0.03$ in comparison with the
zCOSMOS spectroscopic redshifts of $I_{AB} \le 22.5$ galaxies
\citep{lilly2006}; the accuracy of the ZEBRA redshifts degrades to
$\Delta z/ (1+z)\sim0.06$ down to our magnitude limit of $I_{AB}=24$
(see Appendix~\ref{app:photoz} for details).  The application of ZEBRA
to our sample shows some dependence of the resulting photometric
redshifts on whether small (of order $\sim 0.05$ magnitudes or
smaller) systematic offsets, which are detected by the code, are
applied to the photometric calibration of the COSMOS Subaru data.
This has however no substantial impact on our resulting LFs, which we
present below as computed both with and without corrections for these
photometric offsets.

The final $I_{AB}\le 24$ COSMOS sample that we study below consists of
30760 galaxies classified with ZEST as $T=1$ (2497 objects), $T=2$
(26873 objects), and $T=3$ (1390 galaxies), which have photometric
redshifts in the range $0.2 < z \le 1.0$.

\subsection{Luminosity Functions: Definitions}
\label{sec:LFdefinitions}
We estimated the rest--frame-$B$ galaxy LF in different redshift bins,
using the $1/V_{max}$ estimator \citep{schmidt1968,felten1976}.

According to the original $1/V_{\rm max}$ formalism, the number of
galaxies per unit comoving volume in the range of absolute magnitudes
${\rm d}M$, at redshift $z$, and morphological class $T$ can be
written as:

\begin{equation}
\int{\Phi(M,z,T)}{\rm d}M =\sum{\frac{1}{V_{\rm max,i}}},
\label{eq:lf}
\end{equation}

\noindent
where the sum is over all galaxies in the specific range of redshift,
absolute magnitude, and morphological class.  $V_{\rm max,i}$ is the
maximum comoving volume within which the galaxy $i$ could still be
detected according to the apparent magnitude limits of the survey,
which, in our case, is given by $16\le I\le 24$. The $V_{\rm max,i}$
is computed for each galaxy according to:

\begin{equation}
V_{\rm max, i}={\rm \Omega}\int^{{\rm min}(z_U,z_{24})}_{{\rm max}(z_L,z_{16})} \frac{{\rm d} V}{{\rm d}z}{\rm d}z,
\label{eq:vmax}
\end{equation}

\noindent
where $z_U$ and $z_L$ are the upper and lower redshift of the
considered redshift bin, and $z_{24}$ and $z_{16}$ are the redshifts
at which a galaxy of a given rest-frame $B-$band absolute magnitude
and a given SED would have $I-$band apparent magnitude of 24 and 16,
respectively.  The values of $z_{24}$ and $z_{16}$ were computed, for
each galaxy, taking into account the $k-$correction resulting from the
best fit SED provided by the ZEBRA fit.  No evolutionary correction
was applied.  $\Omega$ is the effective area of the survey,
corresponding to $0.74$ square degrees. The Poissonian errors from
galaxy counts on $\Phi(M,z,T)$ are given by
$\sigma_{\Phi}=\sqrt{\sum{1/V_{\rm max,i}^2}}$.

Since we are working with a magnitude-selected rather than a
volume-limited sample, the highest and lowest $B-$band magnitudes at
each redshift depend on the galaxy SED. This is evident in
Figure~\ref{fig:kcorr} of Appendix~\ref{app:rfb}, where we show the
color K-correction (i.e., difference between the observed $I-$ and the
$B-$band rest--frame magnitude), as a function of redshift, for
templates of different photometric type (color-coded from red for
early type galaxies up to blue for starbursting galaxies). The effect
vanishes at redshift $z=0.8$, where the central wavelength of the
F814W filter matches exactly the rest--frame $B$-band.  Since
correcting for this bias needs an a priori assumption on the color
distribution of the non--detected galaxies, we limit the computation
of the LF in each redshift bin to the luminosity range for which we
are complete, regardless of galaxy colors. The used magnitude ranges
are listed in Table~\ref{tbl:2}.

\subsection{Corrections for missing sources}  
\label{sec:statcorrection}

About 2.4$\%$ of the ACS-detected galaxies do not have a match in the
ground-based catalog; therefore, no information on their spectral
energy distributions - and thus photometric redshifts - can be
derived.  Furthermore, about 2\% of the ACS-detected sources are in
close pairs, and, particularly, are at a distance from each other
smaller than 0\farcs6; therefore, these sources are detected as a
single object in the ground-based images (see
Appendix~\ref{app:sample} for details). The photometric redshifts
estimated for these objects are thus unreliable, even for those pairs
which are really physically associated and thus at the same distance,
since the ground-based images mix the light from both galaxies, which
could have very different stellar population. Therefore, these
galaxies in close pairs also need to be excluded from the analysis of
the COSMOS LFs. 

In Figure~\ref{fig:missed} we show the $I-$band magnitude distribution
of the ACS--galaxies with no detection in the ground-based catalog
(top panel), and the magnitude distribution of the ACS--galaxies
for which the same groundbased identification was associated with
multiple ACS sources (bottom panel). Both distributions are
normalized to the number of galaxies in the total sample in each
magnitude bin. The fraction of missed objects stays relatively
constant down to a $I-$band of $\sim 22$, but then increases toward
fainter magnitudes. Although the fraction of missed objects is small,
its dependence on the observed $I-$band magnitude may introduce biases
in the computation of the LF.

To assess the potential impact of such biases, we therefore computed
two versions of the LFs, respectively assuming that:

\begin{itemize}

\item[--] All galaxies with no redshift are outside the $0.2<z\le 1.0$
  range.  We refer to these LFs as to the {\it uncorrected LFs}.

\item[--] The missing galaxies have the same redshift distribution
  than the galaxies with available redshift. In this case we can
  calculate the statistical weight ($\chi_i$) needed to correct the
  $V_{max}$ value of galaxies with known redshift.  We calculate
  $\chi_i$ as a function of the observed $I-$band magnitude, following
  the approach used by \citet{willmer2006}. We consider magnitude bins
  of 0.5 magnitudes, and for each magnitude bin we compute: $(i)$ The
  number of galaxies with photometric redshift in the considered
  redshift range ($N_z$, with $0.2<z \le 1.0$); $(ii)$ The number of
  galaxies with $z>z_{h}$ ($N_{zh}$, with $z_{h}= 1.0$); and $(iii)$
  The number of galaxies with $z>z_{l}$ ($N_{zl}$, with $z_{l}= 0.2$).
  Under the assumption that objects without a redshift estimate have
  the same redshift distribution of the entire sample, the probability
  for a galaxy of magnitude $I_i$ of being in the redshift range
  $0.2<z \le 1.0$ is given by the number of galaxies with good
  redshift estimates in that range, divided by the sum of the number
  of objects with good redshift plus the number of objects with
  redshift both lower and higher than the considered limits, i.e.,
  $P_i=N_z/(N_z+N_{zh}+N_{zl})$.  Finally, the weight $\chi_i$ of each
  galaxy with estimated redshift is the sum of the $P_i$ of all
  galaxies in the relevant magnitude bin, divided by the number of
  galaxies in that bin with $0.2<z \le 1.0$ (i.e., $\sum P_i/N_{z}$).
  The $V_{\rm max,i}$ that is associated with each galaxy, weighted to
  account for objects with no available redshift, is simply
  $\chi_{i}/V_{max,i}$, and the Poissonian errors on the LF, in each
  magnitude bin, are given by $\sigma_i^2=\sum \chi_i/V_{{\rm
      max},i}^2$. In the following we will refer to the so-obtained
  LFs as to the {\it corrected LFs}.

\end{itemize}

In Figure~\ref{fig:LF_ML_nocorrection} we show the LFs of the total
sample of $I_{AB}\le 24$ COSMOS galaxies, as well as the LFs for the
different ZEST types.  Specifically, we show the comparison between
the {\it uncorrected} (dotted line) and the {\it corrected} (stars)
LFs.  The figure shows, from top to bottom, the LF integrated over all
morphological types, and the LFs for the $T=$1 (early-type), $T=$2
(disk) and $T=$3 (irregular) galaxies, respectively. Each column shows
the LFs calculated in different redshift bin: $0.2<z\leq 0.4$ (first
column), $0.4<z\leq 0.6$ (second column), $0.6<z\leq 0.8$ (third
column), and $0.8<z\leq 1.0$ (last column). The plotted error bars
show the Poisson errors only.  Uncertainties due to sampling variance
are not included in the figures; these are discussed in
Section~\ref{sec:lowzLF}.  In each redshift bin, the LFs are shown
down to the absolute $B-$band magnitude at which the sample is
complete, regardless of galaxy colors, so that both red and blue
galaxies are sampled in an unbiased way at each redshift, down to the
faintest magnitude bin.

\subsection{Impact  on the LFs  of the  uncertainties in the photometric redshifts} 
\label{sec:photzunc}

In order to check to which extent the limited redshift accuracy
affects the derived LFs, we performed a set of simulations using the
COSMOS mock galaxy catalogs of \citet{kitzbichler2006}; these are
designed to reproduce in detail the COSMOS survey.  From these mock
catalogs, we first extracted all galaxies with observed magnitude
$I_{AB}\le 24$, covering an area on the sky equal to that analyzed in
this paper, and then generated 100 simulations of the galaxy catalog,
perturbing each redshift with an error derived by randomly sampling a
Gaussian with a $\sigma$ equal to the one of the photometric redshift
uncertainty in the relevant redshift bin. We considered $\sigma_z /
(1+z)\sim 0.03$ for galaxies brighter than $I_{AB}=22.5$, and
$\sigma_z / (1+z)\sim 0.06$ for fainter objects (see
Appendix~\ref{app:photoz}). The LFs of the original catalog and of the
100 re-simulated samples were then calculated using the same procedure
described in Section~\ref{sec:LFdefinitions}.  The results of these
tests are presented in Figure~\ref{fig:photozsim}, where, in each
panel, the solid line with Poissonian error bars represents the LF
computed using the ``true" redshifts, and the open diamonds represent
the median of the 100 realizations. The shaded grey area associated
with the simulated volume densities represent the 16$^{th}$ and
84$^{th}$ percentiles of the simulated distributions within each
magnitude bin.  Figure~\ref{fig:photozsim} clearly shows that the
dominant effect of the redshift uncertainty are the systematic trend
of $(a)$ populating the bright end of the LF, where few or no galaxies
are present in the original sample, and $(b)$ slightly underestimating
the density of galaxies around the knee of the LF (i.e., around
$M_{*}$).  However, even at the bright-end of the LF, where the effect
is the strongest, the real and the simulated LFs are well within
$2\sigma_{\Phi}$, implying that the uncertainty in the photometric
redshifts does not affect our conclusions on the evolution of the LFs.
The largest source of uncertainty, especially in the lowest redshift
bin, is due the small volume sampled by the bin, where the
contribution from large scale structure variation are significant.
This explains the relative large difference in the LFs calculated from
two independent mock catalogs (shown with dashed lines in
Figure~\ref{fig:photozsim}).

In Figure~\ref{fig:LF_ML_nocorrection} we show, as stars, the {\it
  corrected} LFs derived adopting the ZEBRA photo-$z$ calculated after
correcting the photometric catalogue for the small systematic offset,
and, shaded in grey, the uncertainty on the LFs that arises when using
the ZEBRA photo-$z$ estimates obtained without correcting the
photometric catalogue (see Appendix~\ref{app:photoz}). Although the
LFs estimated with the two sets of photo-$z$'s differ somewhat (see
Table~\ref{tbl:2} for quantitative differences in the corresponding
Schechter fits), the differences do not substantially impact our
conclusions. Furthermore, Figure~\ref{fig:LF_ML_nocorrection}
illustrates that the LFs derived with the two approaches described in
Section~\ref{sec:statcorrection} are in very good agreement,
demonstrating that the results are independent of the (small) fraction
of objects excluded for not having a photometric redshift estimate.
In the following we therefore focus our discussion entirely on the
{\it corrected} LFs (stars in Figure~\ref{fig:LF_ML_nocorrection}).

\subsection{Results: The evolution of the LFs for the different morphological classes}
\label{sec:lowzLF}

The first column of Figure~\ref{fig:LF_ML_nocorrection} shows that, in
the lowest redshift bin, the shape of the LF varies significantly for
different morphologically--selected galaxy samples. In particular, the
LF of the global galaxy sample and the LF of $T=2$ disk galaxies keep
increasing toward faint magnitudes, and have a rather similar
behavior.  On the other hand, the LFs of the $T=1$ early--type galaxy
population, and of the $T=3$ irregular galaxies appear, within the
errors, to almost flatten for magnitudes $M_B>-19.5$.

We note that our COSMOS LFs computed in the first redshift bin
($z=0.2-0.4$) are rather susceptible to the effect of large scale
structures, as the size of the studied COSMOS field at $z=0.3$ is only
$\sim$20 comoving Mpc.  To quantify this effect we compare the LF for
the global galaxy population with the LFs derived in the same redshift
bin by other surveys.  A fit with a Schechter function to the COSMOS
LF in this redshift bin gives the following best fit parameters:
$\Phi_{*}=0.005\pm 0.001$Mpc$^{-3}$, $M_{B,*}=-21.0\pm 0.2$ and
$\alpha=-1.26 \pm 0.15$.  The faint--end slope $\alpha$ is in good
agreement with the value recently published by \citet[][$\alpha_{\rm
  Willmer}=-1.3$]{willmer2006} on the basis of the DEEP2 spectroscopic
survey \citep{deep}.  The DEEP2 LF is based on spectroscopic
redshifts; therefore, the consistency of the DEEP2 LFs with the one
that we have presented highlights the reliability of our photometric
redshift estimates.  In the $0.2<z<0.4$ bin, the comparison between
our value of $\Phi_{*}$ and those derived from other surveys
demonstrates that the average density in the COSMOS field is
systematically higher than in other studies at similar redshifts. For
example, \citet{willmer2006} finds
$\phi_{*}=26.39^{+1.81}_{-1.62}\times 10^{-4}$ Mpc$^{-3}$, and
\citet{wolf2003} finds $\phi_{*}=18.27\pm 12.98 \times 10^{-4}$
Mpc$^{-3}$ in this redshift regime, for the COMBO17
survey\footnote{The COMBO17 survey covers a total area of 0.78
  deg$^2$, in three fields.  The measured $\Phi_{*}$ in the different
  fields ranges from $13.03\times 10^{-4}$ Mpc$^{-3}$ to $21.17\times
  10^{-4}$ Mpc$^{-3}$ \citep{wolf2003}}.  The overdensity in the
COSMOS field at $0.2<z<0.4$ is confirmed by the spectroscopic (Lilly
et al.\ 2006) and photometric (Feldmann et al.\ 2006) redshift
distributions.  Similarly, a large overdensity at $z=0.75$ is present
in the COSMOS field (Scoville et al.\ 2006), which is also detected by
XMM \citep[][this volume]{hasinger2006}.

Figure~\ref{fig:LF_ML_nocorrection} shows that the LF of the global
galaxy population does not change significantly as a function of
redshift.  Qualitatively, the behavior of the global LF can be
described by a brightening of the global galaxy population from
redshift $z=0.3$ to $z=0.9$, with the volume density of the entire
galaxy population staying basically constant from redshift $z=0.5$ to
$z=0.9$. The Type~2 disk galaxy population follows closely the
behavior of the global sample as a function of redshift, and can be
described with a rather constant number density, and a brightening of
the stellar populations.  

In Figure~\ref{fig:LF_DISK} we show the {\it corrected} LFs of $T=2$
disk galaxies splitted by their {\it bulgeness parameter}. As in
Figure~\ref{fig:LF_ML_nocorrection}, different columns show the
results for different redshift bins from $0.2<z\le0.4$ to $0.8<z\le1$.
The LFs of different types of disk galaxies are plotted in different
rows, from the bulgeless $T=2.3$ disks in the top row, to the
bulge-dominated $T=2.0$ disk galaxies in the bottom row.  In the
lowest redshift bin, as the contribution of the bulge component
increases (i.e., moving from top to bottom in
Figure~\ref{fig:LF_DISK}), the faint end of the LF becomes flatter,
and the LF becomes more similar to that of the Type~1 galaxies.  At
low redshifts, the bulge-dominated disk galaxies contribute most of
the light at magnitudes brighter than $M_B=-21.5$, while their
contribution is strongly reduced at magnitudes fainter than
$M_B=-19.0$.  The LF of $T=2.0$ bulge-dominated galaxies is very
similar in both shape and normalization to the LF of the $T=1$
early-type galaxies.

\subsection{Anchoring the COSMOS evolution to the local universe: Comparison with a
complete SDSS-based sample}
\label{sec:LFsdsscomp}
The presence of the $z<0.4$ overdensity in the COSMOS field hampers
the interpretation of the evolution of the LFs over the studied
redshift range; the small volume sampled by the COSMOS survey at
redshifts $z<0.2$ is furthermore inadequate to constrain the LFs at
such late epochs.  To correctly compare the COSMOS data with the $z=0$
Universe, we used the SDSS-based volume--limited sample of $z=0$
galaxies (appropriately ``redshifted" to $z=0.7$, hereafter indicated
as SDSS$_{z=0.7}$) constructed and extensively discussed by Kampczyk
et al.\ (2006); we briefly summarize its properties in
Appendix~\ref{app:sdss}.

We analyzed the SDSS$_{z=0.7}$-simulated images following the
identical procedure that we adopted for the COSMOS galaxies: We
classified the SDSS galaxies according to their position in the
3--dimensional $PC_i$ space, and then computed the LFs of the
morphologically-selected SDSS$_{z=0.7}$ sample, as described in
Section~\ref{sec:LFdefinitions}.  This SDSS$_{z=0.7}$ sample provides
the properties that local galaxies would show, at redshift $z=0.7$,
assuming no evolution between $z=0.7$ and $z=0$. We therefore stress
that applying the identical classification and LF-derivation
procedures to both our COSMOS and our SDSS-based samples guarantees
that any systematics present in the former is also present in the
latter; thus, the comparison between the two samples allows us to
discover evolutionary trends that are unaffected by such systematics.

The comparison of the COSMOS $0.6<z \le 0.8$ LFs (stars) with the
SDSS$_{z=0.7}$ samples (dash--dotted curves) is shown in
Figure~\ref{fig:compLF_SDSS}. The comparison shows that
pure-luminosity evolution could explain the differences between the
global local SDSS galaxy population and the global COSMOS population
at $z=0.7$.  To quantify this statement, we compute the brightening of
the global local galaxy population that is required to overlap the
global SDSS$_{z=0.7}$ LF with the global COSMOS LF at redshift
$z=0.7$, which is of order $0.95\pm 0.1$ magnitudes.  This amount of
brightening is applied to the SDSS$_{z=0.7}$ LFs in the various plots
of Figure~\ref{fig:compLF_SDSS} (dashed curves).  The agreement
between the brightened SDSS$_{z=0.7}$ LF derived for the global galaxy
population and the global COSMOS $z=0.7$ LF is remarkable.  Thus, the
evolution of the global galaxy population since $z=0.7$ can be
interpreted as due to only a brightening of the stellar populations,
with no evolution in the average volume density of galaxies over this
time period. This result is consistent with recent findings on the
evolution of the global LF derived by \citet{willmer2006} and also
\citet{faber2005} on the basis of the DEEP2 spectroscopic survey.

The Schechter fits to the total, Type~1, Type~2 and Type~3 COSMOS LFs
in the $0.6 < z \le0.8$ redshift bins are reported in
Table~\ref{tbl:2} and, as solid black lines, also in
Figure~\ref{fig:LF_ML_nocorrection}.  In particular, the Schechter
fits to the $z=0.7$ bin are also plotted in the lower redshift bins,
after applying the amount of luminosity evolution that is required to
overlap the $0.6< z\le 0.8$ COSMOS LFs with the local SDSS LF (which
corresponds to a $B-$band evolution proportional to $\sim 1.3\times
z$, as in Figure~\ref{fig:compLF_SDSS}).  Despite the noticeable
effect in the lowest redshift bin of the known structure in the
central part of the COSMOS field, the shapes of the COSMOS LFs at
$z\sim0.7$ well describe, for all the different morphological
populations, the LFs at lower and higher redshifts.

The further inspection of Figure~\ref{fig:compLF_SDSS} shows however,
qualitatively, some interesting evolutionary trends in the individual
morphologically--selected galaxy samples. In particular:

\begin{itemize}

\item[--] Not surprisingly (since they dominate the global galaxy
  population), the evolution of the LF of the Type~2 disk galaxy
  population is very similar to that observed in for the global
  population: the LF of disk galaxies is consistent with a pure
  luminosity evolution of 0.95 magnitudes up to z = 0.7, with a
  constant average volume density.

\item[--] Within the errors, the bright-end ($M_{B}<-21.5$) of the LF
  of morphologically-classified Type~3 irregular galaxies at $z=0.7$
  is also consistent with that of the local SDSS irregular galaxies
  (brightened by a similar amount as above).  At fainter luminosities,
  $M_B>-20.5$ we detect however an excess of a factor of order $\sim
  3$ in the number density of faint irregular galaxies at $z\sim0.7$
  relative to the local universe.

\item[--] At bright magnitudes $M_{B}<-21.5$, there is also a good
  agreement between the LF of $z=0.7$ COSMOS Type~1 early-type
  galaxies and the LF of similarly-classified SDSS galaxies brightened
  by 0.95 magnitudes.  This evolution is similar to that expected
  theoretically from the passive evolution of a coeval stellar
  population formed well before $z \sim 0.7$, and also to the
  evolution that is observed from studies of the surface brightnesses
  of early-type galaxies (Rigler \& Lilly 1994, Schade et al.\ 1999,
  McIntosh et al.\ 2005, Holden et al. 2005).  This result supports a
  scenario where most of the massive early-type galaxies are in place
  at z = 0.7, with the caveat that we cannot be sure that all of the
  $z \sim 0.7$ population are actually the progenitors of early--type
  galaxies (e.g., not all of the T=1 galaxies may be consistent with
  evolving onto the Kormendy relation at $z=0$,Ferreras et al.  2005;
  we investigate this issue further in Scarlata et al.  2006b).  At
  fainter magnitudes, $M_B>-20.5$, however, there is a drop of a
  factor $\sim 2$ between $z=0$ and $z=0.7$ in the number density of
  morphologically-classified early-type galaxies.  This result is
  potentially in contrast with a significant contribution of
  elliptical-elliptical mergers to build the most massive early-type
  galaxies at relatively recent epochs, and favors a down-sizing
  scenario for early-type galaxy formation, in which the most massive
  ellipticals form and dynamically-relax at earlier epochs than the
  lower-mass ellipticals. This is also suggested by some studies of
  ellipticals in the local universe (e.g., Carollo et al.\ 1993;
  Thomas \& Davies 2006) and at high redshifts (e.g., Treu et al.
  2005, Daddi et al. 2005).
 
\end{itemize}

The ZEST misclassification at the faintest magnitudes discussed in
Section~\ref{sec:ZESTerr} is not responsible for the observed deficit
of Type~1 early-type galaxies and increase in the number density of
Type~3 galaxies, since: (1.)  at most $\sim 30\%$ of Type~1 galaxies
could be misclassified as Type~2 or Type~3 galaxies; this fraction is
too small to explain (a) the missing factor of two of Type~1 galaxies
at redshift $z=0.7$; and (b) the factor of 3 excess in Type~3 galaxies
observed in the same redshift bin.  (2.)  The SDSS$_{z=0.7}$ galaxy
sample accurately simulates the COSMOS ACS images.  Therefore, the
same fraction of COSMOS and SDSS$_{z=0.7}$ galaxies would be
misclassified by ZEST due to low S/N.  Thus, the relative comparison
of the two samples is not affected by misclassification due to low
S/N.

We postpone to future publications more sophisticated analyses of the
LFs of COSMOS galaxy samples that are based on the
structural/morphological classification obtained with ZEST.

\section{Summary}
\label{sec:discussion}

The Zurich Estimator of Structural Types (ZEST) uses the Sersic index
$n$ of the fits to the galaxy surface brightness distributions, and
five basic non-parametric diagnostics to quantify the properties of
galaxy structure. The novelty of ZEST is to use a Principal Component
Analysis to retain the full information provided by the entire set of
diagnostics while reducing to three the dimensionality of the
structural parameter space (of axes $PC_1$, $PC_2$, and $PC_3$). The
ZEST scheme morphologically classifies galaxies in three main types
(early-types, disk galaxies and irregular galaxies), assigns a
bulgeness parameter related to the B/D ratio (quantized in four bins)
to disk galaxies, and ranks galaxies according to the elongation,
irregularity and clumpyness of their light distribution. Most
important to stress is that, in contrast with other approaches to
morphological classifications that utilize only two or three of the
basic structural parameters that enter the ZEST PCA, the ZEST
$PC_1$-$PC_2$-$PC_3$ classification scheme substantially reduces the
contamination to each morphological class from different galaxy types.
We have used a sample of $\sim 56000$ $I_{AB}\le 24$ galaxies in the
central 0.74 deg$^2$ of the COSMOS field to calibrate the ZEST
classification grid, and, so far, ascertained the robustness of the
ZEST classification down to the faintest magnitudes of our sample;
work to stabilize the results down to even fainter magnitudes is in
progress.

As a first application, we use our ZEST-classified sample of COSMOS
galaxies to study the evolution since $z=1$ of the Luminosity Function
of the early-type, disk (of different bulgeness parameter) and
irregular galaxies. Our analysis shows that the average volume density
of disk galaxies remains constant through the studied redshift range,
although the stellar populations of these systems are brightened at
earlier epochs.  In contrast, only the bright, $M_B<-21.5$ end of both
the irregular and the early-type galaxies remains roughly consistent
with the brightened LF of local galaxies. At faint magnitudes, both
morphologically-classified irregular and early-type galaxies show
substantial evolutionary effects from $z=0$ to $z=0.7$ and above. In
particular, there is an excess of a factor of order $\sim 3$ of faint
morphologically-classified irregular galaxies, and a deficit of a
factor of about 2 of faint ($M_B> -20.5$) morphologically--classified
early--type galaxies at $z\sim0.7$ relative to the local universe.

The ZEST classification of our sample of 56000 COSMOS galaxies has
already been used to study the number density evolution of disks of
different sizes (Sargent et al.\ 2006), the LF-evolution of plausible
progenitors of elliptical galaxies (Scarlata et al.\ 2006b), and the
evolution of the merger rate out to $z\sim1$ (Kampczyk et al.\ 2006).
The ZEST approach will be used to automatically classify the whole
2deg$^2$ of COSMOS down to $I_{AB}=24$ (Carollo et al.\ 2006).

\clearpage

\appendix

\section{Details on the COSMOS galaxies}
\label{app:cosmos}

\subsection{The sample}
\label{app:sample}
A detailed discussion of the ACS data processing is given in
\citet{kokomero2006}; the description of the generation of the ACS
catalog that we use in this paper is given in \citet{leauthaud2006}.
Briefly, the ACS catalog was produced by applying SExtractor
\citep[version 2.4.3][]{sextractor} to the reduced ACS images. Edge
regions in each image were removed with no loss of data, since the
overlap between neighboring ACS tiles was larger than the excluded
edge regions. Two SExtractor runs were performed in order to avoid
both blending of multiple independent sources and de--blending of
large galaxies, i.e., a ``cold'' run with a configuration optimized
for the detection of large, bright objects, and a ``hot'' run with a
configuration optimized for faint, small objects. The two resulting
samples were then merged together to produce a final catalog by
retaining all the ``cold'' detections plus the ``hot'' detections
which fell outside the SExtractor segmentation map of any galaxy
detected in the ``cold" run.

We visually inspected the catalog to remove any residual
over--deblending of large galaxies and also false detections.
Specifically, we cross-correlated the ACS catalog with the COSMOS CFHT
$I$-band catalog down to $I=24.5$ using the celestial coordinates of
the objects ($5\sigma$ magnitude limit of $I_{\rm CFHT}=24.9$; Capack
et al. (2006))\footnote{For each ACS source we searched for
  CFHT-detected galaxies within a circle of 0\farcs6 radius. If more
  than one CFHT source was found within this radius, the closest CFHT
  galaxy was assigned to the ACS source as its ground-based
  counterpart.}, and flagged for visual inspection the ACS-detected
objects that: ($i$) Had no match in the available ground-based COSMOS
catalog; ($ii$) Were detected in the ground-based catalog, but had a
$>0.5 mag$ difference between the ground-based and HST $I$ magnitude
estimates; and ($iii$) Were linked with a CFHT source that had already
been associated with another ACS source.  About 4.5\% of the entries
in the original ACS-based catalog were removed after the visual
inspection.

Stars were then removed from the ACS-based catalog using the
SExtractor CLASS\_STAR parameter, and particularly removing sources
with CLASS\_STAR$>0.6$. In Figure~\ref{fig:star} we show the
SExtractor estimate of the radius $R_{tot}$ including 100\% of the
light of an object, as a function of the SExtractor ACS $I-$band
magnitude.  Objects with CLASS\_STAR$>0.6$ are identified as filled
circles.  These form a tight, well identifiable sequence at all
magnitudes. The final, cleaned, ACS-based catalog that we classify in
this paper includes $\sim 55651$ galaxies down to $I_{AB}=24$.

\subsection{Data analysis}
\label{app:dataanal}
The steps of data analysis that we performed on the 56000 COSMOS
galaxies of our sample are summarized below.

\begin{itemize}

\item[--] For each ACS--detected galaxy, we removed in the ACS images the
  contamination from nearby objects that could affect the computation
  of the non--parametric coefficients.  To create a ``cleaned" version
  of each galaxy image, a stamp centered around the ACS--detected
  galaxy was extracted from the original ACS tile, with size equal to
  $3\times R_{\rm tot}$. SExtractor was used to identify the
  individual sources in the stamp, and the pixels belonging to
  contaminant galaxies were replaced by background noise that matched
  the properties of the noise in the original ACS image.

\item[--] The Petrosian radius \cite[][$R_{P}$]{petrosian1976} of each
  galaxy was calculated from the surface brightness profile measured
  from its ``cleaned" stamp\footnote{$R_{P}$ is defined as the radius
    at which the ratio $\eta=\mu(R)/\langle \mu(<R)\rangle$ between
    the surface brightness at that radius and the surface brightness
    averaged within that radius is equal to a given number. For a
    surface brightness distribution described by a de~Vaucouleurs or
    an exponential profile, a value of $\eta=0.2$ is reached at
    $R_P\sim 1.8 R_{1/2}$ and $R_P\sim 2.2 R_{1/2}$, respectively
    (with $R_1/2$ the half--light or ``effective" radius of the
    galaxy, \cite{graham2005}).}.  The surface brightness profiles
  were derived using elliptical annuli centered on the galaxies, in
  order to minimize the contribution by background pixels. The
  position angle and ellipticity of the annuli were kept fixed at all
  radii, to the values measured from the image moments
  \citep{sextractor}.  The annuli were uniformly spaced in $\log(R)$
  to increase the $S/N$ in the external region of galaxies. Each
  galaxy profile was sampled by at least 10 points.  We calculated
  $R_P$ for all galaxies with an Half--Width--Half--Maximum (HWHM, as
  estimated by SEXtractor) larger than 1.5 pixels, so as to ensure
  that enough pixels were available for the measurement of the
  parameters. This cut excludes $\sim 0.2$\% of the entire sample of
  COSMOS galaxies under consideration.  In Figure~\ref{fig:prhlr} we
  show, for the sample of $I_{AB}\le 22.5$ galaxies analyzed by
  \citet[][this volume]{sargent2006}, the comparison between the
  Petrosian radius derived in our analysis and the half--light radius
  derived by Sargent and collaborators from GIM2D fits
  \citep{simard1998} to the galaxy surface brightness distributions.
  The {\it solid} and {\it dashed} lines represent the relation
  between $R_{P}$ and $R_{1/2}$ expected for a Sersic profile with
  $n=1$ and $n=2$, respectively.  The very good correlation between
  the two measured quantities indicates that our Petrosian radii
  reliably reproduce the galaxy sizes, and offer the advantage of
  expanding the COSMOS galaxy sample with computed robust size
  measurements down to $I_{AB}=24$.
  
\item The pixels that are associated with a galaxy were defined as all
  the pixels within elliptical apertures of semi-major axis equal to
  the Petrosian radius of the galaxy.  With this definition, the same
  fraction of light is considered, for any given shape of the surface
  brightness profile, at different redshifts, therefore allowing to
  consistently compare measurements of galaxy populations at different
  epochs.

\end{itemize}

In Figure~\ref{fig:examples} we illustrate with a few examples the
procedure outlined so far. In particular, we show four COSMOS galaxies
with increasing magnitudes, from $I_{AB}=18.8$ (top row) to
$I_{AB}=22.3$ (bottom row).  For each galaxy we show the original
stamp (first column), the cleaned stamp (second column), the
segmentation map (third column), and, in the last column, the surface
brightness profile (top panel) and the ratio $\eta=\mu(R)/\langle
\mu(<R) \rangle$ (bottom panel) as a function of radius $R$ in
arcseconds (solid circles).  The solid horizontal line in the $\eta-R$
plots represents $\eta=0.2$.  For the galaxy in the bottom row, $\eta$
decreases to a minimum value that is however larger than $\eta=0.2$,
and then increses again at larger radii.  This is due to the
contamination from a nearby source, still visible even in the cleaned
stamp, which adds flux in the outermost isophotes (as easily seen in
the $\mu$ radial profile).  To obtain reliable Petrosian radii, we
corrected these cases by fitting a linear relation to the $\mu -R$
profile in the outer radial bins (dashed line), and then subtracting a
constant value to the surface brightness profile.  The open circles in
the $\eta -R$ plot of the bottom-row galaxy in
Figure~\ref{fig:examples} show the $\eta$ profile obtained after the
application of this procedure, and demonstrate that we recover robust
values of $R_P$ even in these potentially troublesome cases.

\subsection{Photometric redshifts for COSMOS galaxies}
\label{app:photoz}

Ancillary ground-based photometry in several passbands is available
for the COSMOS field from different facilities (see summary in
Scoville et al.\ 2006, this volume). This allows the estimate of
accurate photometric redshifts for the COSMOS galaxies. Due to the
differences in adopted catalogues, and the intrinsic uncertainties in
photometric redshift estimates, the COSMOS collaboration has decided
to adopt several approaches to determine photometric redshifts for
COSMOS galaxies.  Mobasher et al. (2006, this volume) present a
comparison among several estimates for the COSMOS photo$-z$, obtained
by different groups within the COSMOS team.

In our work, we adopt the photometric redshifts for the COSMOS
galaxies of our ACS-selected $I_{AB}\le 24$ galaxy sample by Feldmann
et al.\ (2006). These are derived with the Zurich Extragalactic
Bayesian Redshift Analyzer (ZEBRA).  ZEBRA produces two separate
estimates for the photometric redshifts of individual galaxies: A
Maximum Likelihood estimate and a 2-D Bayesian estimate. In this paper
we use the Maximum Likelihood photometric redshifts. These were
derived using: $\bullet$ Subaru photometry in six broad band filters
$B$, $V$, $G$, $r'$, $i'$, and $z'$ ($5\sigma$ magnitude limit of
$\sim$27 for point sources in all bands; see Taniguchi et al.~2006,
this volume, for details); $\bullet$ $u^{*}$ photometry from the
MegaCam at the CFHT ($5\sigma$ magnitude limits for point sources of
$u^*=27.4$); and $\bullet$ $K_s$ ($2.2\mu m$) photometry from the
National Optical Astronomy Observatory (NOAO) wide--field IR imager
Flamingos (mounted on the Kitt Peak 4m telescope) and the Infrared
Side Port Imager (ISPI) at the Cerro Tololo Inter--American
Observatory (CTIO) Blanco 4m telescope.

The sources in the ground-based catalog were detected from the
original (i.e, non--matched and non--homogenized PSFs) Subaru $i'$
images, since these had, on average, the best seeing condition of
$\sim$0\farcs6 FWHM, allowing for an optimal deblending of close
galaxy pairs.  Photometric fluxes were instead estimated on
PSF-matched images.  All magnitudes were calculated within circular
apertures of 3\farcs0 diameter.  The final groundbased catalog of
Capak et al.\ (2006) includes $>10^6$ galaxies down to $i'=26$ over
the entire COSMOS field.

In deriving the LFs for the COSMOS galaxies, we checked that, in
addition to being insensitive to slightly different photo-$z$
estimates derived by applying or ignoring small shifts in the
calibration of the groundbased photometry (see Section~\ref{sec:LF}),
the results remain unchanged also when using the full ZEBRA Bayesian
probability distribution for the photometric redshift estimates.
Furthermore, the LFs do not vary when computing them $(a)$ after
excluding from the sample all galaxies with ZEBRA $\chi^2$ values
above the 95th percentile of the $\chi^2$ distribution, and $(b)$ by
including all galaxies in the sample, independent of the $\chi^2$
value of their ZEBRA fits.

The ZEBRA photo-z estimates of the COSMOS galaxies are less accurate
for $I_{AB}>22.5$ sources not only due to their larger photometric
errors, but also due to the lack of photo-z calibration at such faint
magnitudes.  The ongoing spectroscopic follow up of COSMOS on the ESO
VLT, zCOSMOS, will secure spectroscopic redshifts for $>35'000$ COSMOS
galaxies \citep{lilly2006}. So far, however, only about 1500 COSMOS
galaxies have VLT spectroscopic redshifts, most of which in the
$0.2<z<1.1$ redshift range, at magnitudes brighter than I$_{AB}=22.5$.
In order to estimate the accuracy of the ZEBRA photometric redshifts
at magnitudes fainter than $I_{AB}=22.5$, we simulated a faint version
of the spectroscopic redshift catalog by dimming the $I_{AB}<22.5$
galaxies with available spectroscopic redshift down to our magnitude
limit of $I_{AB}=24$. We then recomputed the ZEBRA redshifts for these
artificially-fainted sources.  The accuracy of the ZEBRA photo-z's
down to $I_{AB}=24$ is $\sigma/(1+z)\sim 0.06$.

\subsection{Rest-frame quantities}
\label{app:rfb}

Total magnitudes for the COSMOS galaxies in our sample were available
in the ACS $I-$band. In order to compute the rest--frame absolute
$B-$band total magnitude ($M_B$), we computed for each COSMOS galaxy
the color $B -I_{z}$, where $B$ is the observed magnitude at $\lambda
_{B,{\rm obs}}=(1+z)\times\lambda _B$, and $I_{z}$ indicates the
observed $I-$band magnitude for a galaxy at redshift $z$.  The central
wavelength of the ACS filter corresponds to the central wavelength of
the $B-$band at redshift $z\sim 0.8$, therefore $B-I_{z=0.8}=0$.
Defining the color $k-$correction as $K(z)=B -I_{z}$, the rest--frame
absolute magnitude $M_B$ can be espressed as:

\begin{equation}
M_B=I_{z} -5.0\log{(d_L(z)/10{\rm pc})}-25.0 + K(z) +2.5\log{(1+z)},
\end{equation}

\noindent
where $d_L$ is the luminosity distance at redshift $z$. We computed
the color $B -I_{z}$ using the best fit spectral energy distributions
derived for each galaxy as part of the ZEBRA photometric redshift
calculation.

In Figure~\ref{fig:kcorr} we show $B -I_{z}$ as a function of redshift
for our six basic spectral templates going from an early--type galaxy
template ($T=1$, solid line), to the starburst galaxy template ($T=6$,
long-dash dot line).  As expected, at redshift $z\sim 0.8$ the
$k-$correction is zero for all photometric types.

In the calculation of the LFs, $V_{\rm max}$ was computed for each
galaxy taking into account its ZEBRA photometric type. Specifically,
given the galaxy rest--frame $M_B$ absolute magnitude, we derived the
observed expected $I-$band magnitude as a function of redshift using
the galaxy best--fit template derived from the ZEBRA photometric
redshift. The $I-$band observed magnitude as a function of redshift is
required to find $z_{16}$ and $z_{24}$, which define the redshift
range within which the galaxy could still be seen in the survey, and
therefore $V_{\rm max}$.

\section{Uncertainties and systematics in the morphological measurements}
\label{app:simulations}

To quantify the errors on the measured parameters for the COSMOS
galaxies, and to assess systematic trends as a function of galaxy
brightness, we performed extensive simulations based on real bright
COSMOS galaxies. In particular, we selected 35 bright isolated COSMOS
galaxies with magnitude between $I=17.5$ and $I=18$ and, for each of
them, we created $40$ S/N-degraded versions of their images. This was
done by dividing the original frame by a factor $f$ such that the
final magnitude of the object was $2.5 \times {\rm log}(f)$ fainter
than the original one. The factor $f$ was chosen so that each galaxy
uniformly sampled the magnitude range from $18$ to $24$. The scaled
images were then added back to an ACS tile, in randomly-selected
positions.  Before adding the rescaled image, the original ACS tile
was multiplied by a factor $k$, to preserve the noise properties in
the final image. Specifically, with $\sigma_{\rm orig}$ the background
noise in the image containing one of the 35 bright galaxies, the
background noise after the rescaling is $\sigma_{\rm orig}/f$. Adding
the rescaled image to the ACS original image increases the noise in
the final image to $\sigma_{\rm final}^2=\sigma_{\rm orig}^2 +
\sigma_{\rm orig}^2/f^2$.  The factor $k$ is therefore computed by
requiring that $\sigma_{\rm final}^2=\sigma_{\rm orig}^2=\sigma_{\rm
  orig}^2/k^2 + \sigma_{\rm orig}^2/f^2$.

In Figure~\ref{fig:sim} we show the results of this set of
simulations. In particular we show the fractional change of each
parameter as a function of the magnitude of the S/N-degraded galaxy.
Different lines connect the variation of the parameters for each
galaxy as a function of magnitude.

For all parameters (except $\epsilon$) the scatter remains low (at a
level of 10--20\%) and no strong systematic effects are observed down
to $I\sim 22$. At magnitudes $I>22$ the scatter increases and all
parameters are systematically underestimated up to a maximum of $\sim
20$\% in the faintest magnitude bin. The behavior of $\epsilon$ shows
a different trend, with the scatter starting to increase already at
bright magnitudes (at a level of $\sim 10$\%) and systematic effects
start to be important at $I\sim 23$.  The asymmetry $A$ is
systematically underestimated at the 10\% level for $I-$band
magnitudes of order $21$, and degrades down to a 20\% underestimate
for $I_{AB}=24$.

\subsection{Effects of the Point-Spread-Function}

For very compact objects, the structural parameters computed for the
COSMOS galaxies from the HST-ACS images can be affected by the
instrumental PSF (FWHM $\sim 0.1''$).  In order to determine the
limiting size/magnitude at which the effects of the PSF become
important in our classification algorithm, we performed a set of
simulations based on synthetic galaxy images.  Specifically, we
constructed artificial images of about 2500 galaxies with magnitudes
in the range $21\le I \le 24$, surface brightness profiles described
either by a deVaucouleur (i.e., $n=4$) or an exponential ($n=1$)
profile, and with half-light radii $R_{1/2}$ in the range $0.05'' \le
R_{1/2} \le 1.0''$.  These artificial images were convolved with the
Tinytim (Krist \& Hook 1997) HST PSF, and with a PSF constructed from
stars present in the COSMOS ACS data.  Similar results are obtained
with either of these PSFs; in the following, we discuss the results
obtained with the Tinytim PSF.

We analyzed and classified, in a similar way as it was done for the
COSMOS data, both the original and the PSF-convolved artificial
galaxies.  Comparing the results obtained on the original and on the
PSF-convolved images we find that galaxies with the exponential
($n=1$) surface brightness profile are always classified as Type~2
disk galaxies, and no change of class is observed over the entire
range of half-light radii and magnitudes explored. For $n=4$ galaxies,
we find that all galaxies are classified as Type~1 (early-type
galaxies) when the analysis is performed on the not PSF-convolved
images. When the convolved images are considered, a fraction of about
15\% af all galaxies with half-light radius larger than 0\farcs17
changes class from Type~1 to Type~2; 85\% of these systems have
$I_{AB}> 23$. For galaxies with $n=4$ profile, all objects with
measured $R_{1/2}<$ 0\farcs17 change class from Type~1 to Type~2.

The fraction of objects in our COSMOS sample with measured
$R_{1/2}<$0\farcs17 is about $4$\% down to our magnitude limit of
$I_{AB}=24$. For (the steepest-profile galaxies in) this small
fraction of the sample, the ZEST morphological classification is
therefore likely affected by convolution with the ACS-HST PSF.

\section{The Kampczyk's SDSS-based $z=0$ comparison sample}
\label{app:sdss}

The volume of Universe sampled by the Cycle~12 ACS COSMOS data within
the $0<z\le 0.1$ redshift range is only $\sim 5.5\times 10^3$Mpc$^3$,
and includes $<50$ galaxies. Therefore, cosmic variance and small
number statistics in this low-redshift bin are inadequate to provide a
meaningful $z\sim0$ calibration for studies of galaxy evolution.

To study the LF evolution down to $z=0$ of the ZEST-classified
morphological galaxies, we therefore used the SSDS-based dataset of
artificial galaxy images constructed and presented by Kampczyk et al.\
(2006). These simulate how the local Universe would be observed in the
COSMOS survey at a redshift $z=0.7$.  Specifically, the artificial
dataset is a volume--limited sample of 1813, $M_B<-18.5$,
$0.015<z<0.025$ galaxies extracted from the SDSS--DR4 galaxy
catalog\footnote{The SDSS survey provides homogenous color information
  and spectroscopic redshfits for more than $10^6$ galaxies with
  magnitude $r<17.77$ (Strauss et al.~2002).  We expect the Kampczyk
  et al.\ SDSS-based sample to be unbiased towards environmental
  effects and fully representative of the local galaxy population down
  to the $M_{B}=-18.5$ magnitude limit, since: $(a)$ The
  photometric$+$spectroscopic SDSS data are mostly complete down to
  $r\sim 17.8$ (Strauss et al.~2002), i.e., well below the considered
  absolute magnitude cut.  Specifically, the surface brightness cut
  applied for the selection of the SDSS spectroscopic targets ($\mu<
  23$ mag arcsec$^{-2}$) and mechanical constraints of the
  spectrograph (which exclude galaxies in pairs closer than $55''$)
  remove less than 1\% and $\sim$6\% of targets brighter than
  $r=17.77$, respectively. Most importantly, the fraction of excluded
  galaxies does not depend on the galaxy magnitude; and $(b)$ It is
  extracted from a sample of $\sim 700,000$ SDSS galaxies over a
  volume of $2.5\times 10^5$ Mpc$^3$, which encompasses the full range
  of local environmental densities, from voids to galaxy clusters.}.
The absolute $B$-band agnitude $M_B$ was derived from the SDSS $g$
magnitude, by applying a $k-$correction computed on the basis of the
observed $g-r$ color and redshift of each galaxy.

The SDSS $g$-band at $z=0$ almost perfectly matches the ACS F814W
($I$) band at $z=0.7$.  To simulate how the local galaxies (observed
in the SDSS $g$-band) would appear, in the absence of evolution, in
the F814W COSMOS ACS images at a redshift $z=0.7$, it is thus only
necessary to take into account the different pixel scales and PSFs,
and cosmological surface brightness dimming.  The
SDSS$_{z=0.7}$-simulated images do not include either surface
brightness or size evolution; they were randomly added to the COSMOS
ACS images to reproduce the ACS-COSMOS noise properties and to
simulate galaxy-overlapping due to projection effects.

The simulated SDSS$_{z=0.7}$ galaxies were analyzed with the same
procedure used for the real COSMOS data. First, we run SExtractor to
detect the sources and to measure their $I-$band magnitudes, position
angles and ellipticities.  The same SExtractor configuration
parameters that were used to generate the COSMOS--catalog
\citep{leauthaud2006} were adopted for the extraction of the
SDSS$_{z=0.7}$ objects.  About 6\% of the SDSS$_{z=0.7}$ galaxies were
not recovered by SExtractor; 0.1\% were instead detected, but with a
magnitude fainter than our $I_{AB}=24$ limit, and were therefore
excluded from the SDSS$_{z=0.7}$ sample.  For each simulated galaxy,
we then measured the structural parameters described in
Section~\ref{sec:parameters} and applied the ZEST morphological
classification described in Section~\ref{sec:pcaglobal}.

The rest--frame $B-$band LF for the constructed SDSS$_{z=0.7}$ sample
is shown in Figure~\ref{fig:sdssLF} ({\it solid histogram}).  Being
based on a volume--limited, random selection of the local galaxy
population, this LF will have the same shape as the global $z=0$
$B-$band LF.  We therefore used, to normalize the volume density, the
total number of SDSS galaxies with $M_B<-18.5$, derived integrating
the $B-$band LF recovered from the $g-$band LF of Blanton et al.
(2003). To convert $g$-band to $B$-band magnitudes, we considered two
extreme SEDs: a young (0.1~Gyr) single burst stellar population with
solar metallicity, and an old (12~Gyr) single burst stellar population
with the same metallicity.  The corresponding transformations are:
$B=g+0.02$, and $B=g+0.36$.  In Figure~\ref{fig:sdssLF} we show the
LFs calculated using the normalization derived adopting the ``old" SED
(dotted line), the ``young" SED (dashed line), and the average of the
two (solid histogram). In discussing the comparison of the COSMOS and
SDSS$_{z=0.7}$ LFs, we adopt the SDSS $B-$band LF derived with the
average SED correction.

\clearpage

\begin{deluxetable}{lccccc}
\tabletypesize{\scriptsize}
\tablecaption{Results of the ZEST PC analysis\label{tbl:1}}
\tablewidth{0pt}
\tablehead{
\colhead{} & 
\colhead{$PC_1$} & 
\colhead{$PC_2$} & 
\colhead{$PC_3$} & 
\colhead{$PC_4$} & 
\colhead{$PC_5$} \\
\colhead{(1)} & 
\colhead{(2)} & 
\colhead{(3)} & 
\colhead{(4)} & 
\colhead{(5)} & 
\colhead{(6)}}
\startdata
Eigenvalue&2.46& 1.19& 0.92& 0.25& 0.17\\
Proportion&0.49& 0.24& 0.18& 0.05& 0.03\\
Cumulative&0.49& 0.73& 0.92& 0.97& 1.00\\
&&&&&\\
\hline
\hline
Variable  & $PC_1$&  $PC_2$ & $PC_3$ & $PC_4$ & $PC_5$\\
 \hline
 Concentration (=$x_1$) &$-$0.54  &   0.35 &0.18 &$-$0.34&$-$0.66\\  
 $M_{20}$     (=$x_2$)  & 0.60    &$-$0.04 &0.03 & 0.39  &$-$0.70\\
 Gini      (=$x_3$)     &$-$0.56  &$-$0.20 &0.14 & 0.79  &$-$0.02\\
 Ellipticity  (=$x_4$)  & 0.20    &   0.57 &0.74 &  0.16 &   0.26\\
 Asymmetry   (=$x_5$)   & 0.02    &$-$0.71 &0.64 &$-$0.29&$-$0.08\\
\enddata
\tablecomments{Columns 2--6 refer to the five principal components in ZEST.
  The first row gives the eigenvalue (i.e., variance) of the data
  along the direction of the corresponding PC.  The second and third
  rows show the fraction of variance and the cumulative fraction of
  each PC, respectively.  The bottom part of the table lists the
  weights assigned to each input variable, in the linear combination
  that gives the direction of the principal component (i.e.,
  $PC_i$=$\alpha$x$_1$+$\beta$x$_2$+$\gamma$x$_3$+$\delta$x$_4$+$\phi$x$_5$,
  with the coefficients $\alpha$,...,$\phi$ given by the numbers
  listed, per each PC, in the bottom part of the table).}
\end{deluxetable}

\begin{deluxetable}{lccc}
\tabletypesize{\scriptsize}
\tablecaption{The ZEST classification scheme\label{tbl:zest}}
\tablewidth{0pt}
\tablehead{
\colhead{} & 
\colhead{ {\bf Type 1} }&
\colhead{ {\bf Type 2}} & 
\colhead{ {\bf Type 3}} \\
\colhead{} & 
\colhead{ {\bf = EARLY TYPES} }&
\colhead{ {\bf = DISK GALAXIES}} & 
\colhead{ {\bf = IRREGULAR GALAXIES}} \\
\colhead{} & 
\colhead{ {\bf (no visible disk)} }&
\colhead{ } & 
\colhead{ } \\
}
\startdata
&&&\\
 {\bf BULGENESS} & -&  {\bf from .0 (massive bulge) } & -\\
   &  &  {\bf  to .3 (bulgeless disk)} & \\
 &&& \\
 \hline
 &&& \\
 {\bf ELONGATION} & {\bf from 0 (face on) } &  {\bf from 0 (face on) }  & - \\
   & {\bf  to 3 (edge on)} &  {\bf  to 3 (edge on)}  & \\
 &&& \\
  \hline
 &&& \\
 {\bf IRREGULARITY } & {\bf from 0 (regular) } &  {\bf from 0 (regular) }  & - \\
   & {\bf  to 2 (highly irregular)} &  {\bf  to 2 (highly irregular)}  & \\
 &&& \\
  \hline
 &&& \\
{ \bf CLUMPYNESS } & {\bf from 0 (smooth) } &  {\bf from 0 (smooth) }  &  {\bf from 0 (smooth) } \\
   & {\bf  to 2 (very clumpy)} &  {\bf  to 2 (very clumpy)}  &  {\bf  to 2 (very clumpy)}  \\
 &&& \\
 \hline
 \hline
 &&& \\
 {\bf SIZE$^*$ } & {\bf R$_P$ } &  {\bf R$_P$ }  & {\bf R$_P$} \\
&&& \\
\hline
 \enddata
 \tablecomments{Summary of the ZEST classification scheme. Type 1
   (early-type galaxies) are spheroids with no visible disk (including
   face-on S0 galaxies, for which the identification of the disk
   component is difficult). Type 2 are disk galaxies, and Type 3 are
   irregular galaxies.  A clumpiness parameter is assigned to each
   (ZEST unit cube and thus) galaxy in the sample.  Early-type and
   disk galaxies are assigned an elongation parameter in four steps
   from 0 (face on) to 3 (edge on), and an irregularity parameter
   (from 0 for regular to 2 for highly irregular galaxies).  Type 2
   disk galaxies are split in four bins of bulgeness parameter, namely
   2.0, 2.1, 2.2 and 2.3 from bulge-dominated to bulgeless disks,
   respectively.  Relatively inclined S0 galaxies occupy cubes
   classified as T=2.0. \\ $^*$The last row indicates that a measure
   of galaxy size (the Petrosian radius, \citep{petrosian1976}) is
   available for all ZEST--classified galaxies, as a byproduct of our
   analysis.}
  \end{deluxetable}

\begin{deluxetable}{crccccccccc}
\tabletypesize{\scriptsize}
\tablecaption{COSMOS ZEST classification grid\label{tbl:class}}
\tablewidth{0pt}
\tablehead{
\colhead{$$} & 
\colhead{$PC_1=-6$} & 
\colhead{$-5$} & 
\colhead{$-4$} & 
\colhead{$-3$} & 
\colhead{$-2$} &
\colhead{$-1$} & 
\colhead{$0$} & 
\colhead{$1$} & 
\colhead{$2$} & 
\colhead{$3$} \\
\colhead{$PC_2=$} & 
\colhead{$$} & 
\colhead{$$} & 
\colhead{$$} & 
\colhead{$$} & 
\colhead{$$} &
\colhead{$$} & 
\colhead{$$} & 
\colhead{$$} & 
\colhead{$$} & 
\colhead{$$} }
\startdata
\multicolumn{11}{c}{$PC_3=-2$}\\
\hline
$-4$&   &   &   &   &   &   &   &   &   &   \\
$-3$&   &   &   &   &   &   &   &   &   &   \\
$-2$&   &   &   &   &   &   &   &   &   &   \\
$-1$&   &   &   &   &   &2.2&2.2&2.3&2.3&2.3\\
 $0$&   &   &   &   &2.1&2.2&2.2&2.3&2.3&2.3\\
 $1$&   &   &   &   &2.2&2.2&2.2&2.2&2.2&2.2\\
 $2$&   &   &   &   &   &   &   &   &   &   \\
 $3$&   &   &   &   &   &   &   &   &   &   \\
\hline
\multicolumn{11}{c}{$PC_3=-1$}\\
\hline
$-4$&   &   &   &   &   &   &   &   &   &   \\
$-3$&   &   &   &   &   &   &   &   &   &   \\
$-2$&   &   &   &   &  1&2.2&2.2&2.3&  3& 3 \\
$-1$&   &   &1  &1  &2.1&2.1&2.2&2.3&2.3&2.3\\
 $0$ &  &   &1  &2.0&2.1&2.1&2.2&2.3&2.3&2.3\\
 $1$ &  &   &2.0&2.0&2.1&2.1&2.2&2.3&2.3&2.3\\
 $2$ &  &   &   &   &   &   &   &   &   &   \\
 $3$ &  &   &   &   &   &   &   &   &   &   \\
\hline
\multicolumn{11}{c}{$PC_3=0$}\\
\hline
$-4$&   &   &   &   &   &   &   &   &   &   \\
$-3$&   &   &   &   &2.1&  3&  3&  3&  3&  3\\
$-2$&   &   &1  &  1&2.0&2.1&  3&  3&  3&  3\\
$-1$&   &  1&1  &  1&2.1&2.1&2.2&2.3&2.3&  3\\
 $0$&  1&  1&1  &2.0&2.1&2.1&2.2&2.3&2.3&2.3\\
 $1$&   &  1&1  &2.0&2.1&2.1&2.2&2.3&2.3&2.3\\
 $2$&   &   &2.0&2.1&2.1&2.1&2.2&2.2&2.3&2.3\\
 $3$&   &   &   &   &   &   &   &   &   &   \\
\hline
\multicolumn{11}{c}{$PC_3=1$}\\
\hline
$-4$&   &   &   &   &   &   &   &   &   & \\
$-3$&   &   &   &   &   &  3&  3&  3&  3&3\\
$-2$&   &   &  1&2.0&2.0&  3&  3&  3&  3&3\\
$-1$&   &  1&  1&2.0&2.0&2.1&2.2&  3&  3&3\\
$0$&   1&  1&2.0&2.0&2.1&2.1&2.2&2.3&2.3&3\\
$1$&   1&  1&2.0&2.0&2.1&2.2&2.2&2.3&2.3&2.3\\
$2$&   &    &2.0&2.1&2.1&2.2&2.2&2.3&2.3&2.3\\
$3$&   &    &   &   &   &2.2&2.2&2.2&2.3&   \\
\hline
\multicolumn{11}{c}{$PC_3=2$}\\
\hline
$-4$&   &   &   &   &   &  3&  3&  3&  3&  3\\
$-3$&   &   &   &   &  1&  3&  3&  3&  3&  3\\
$-2$&   &   &   &  1&  1&  3&  3&  3&  3&  3\\
$-1$&   &   &  1&2.0&2.0&  3&  3&  3&  3&  3\\
$0$ &   &2.0&2.0&2.0&2.1&2.1&2.2&2.3&  3&  3\\
$1$ &2.0&2.0&2.0&2.0&2.1&2.2&2.2&2.3&2.3&2.3\\
$2$ &   &2.0&2.0&2.0&2.1&2.2&2.3&2.3&2.3&2.3\\
$3$ &   &   &   &2.1&2.1&2.2&2.3&2.3&2.3&   \\
\multicolumn{11}{c}{$PC_3=3$}\\
\hline
$-4$&   &   &   &   &   &   &  3&  3&  3&\\
$-3$&   &   &   &   &   &   &  3&  3&  3&\\
$-2$&   &   &   &2.0&2.0&  3&  3&  3&  3&\\
$-1$&   &   &   &2.0&2.1&2.1&  3&  3&  3&\\
$0$ &   &   &2.0&2.0&2.1&2.1&2.2&2.3&2.3&\\
$1$ &   &   &2.0&2.0&2.1&2.1&2.2&2.3&2.3&\\
$2$ &   &   &2.1&2.1&2.1&2.2&2.2&2.3&2.3&\\
$3$ &   &   &   &   &   &   &   &   &   &\\
\enddata
\tablecomments{COSMOS-calibrated ZEST grid in each plane of constant
  value of PC3, going from $PC_3=-2$ first panel of the Table, to
  $PC_3=3$, last panel. In each panel, we give for each $PC_1,$ $PC_2$
  the ZEST classification as Type 1,2, and 3. For Type 2 disk
  galaxies, we also present the ``bulgeness'' parameter associated
  with the cube.}
  \end{deluxetable}

\begin{deluxetable}{cccccc}
\tabletypesize{\scriptsize}
\tablecaption{Schechter Function best--fit parameters.\label{tbl:2}}
\tablewidth{0pt}
\tablehead{
\colhead{$z$(Range)} & 
\colhead{$-M_{B}$(Range)} & 
\colhead{N$_{\rm gal}$} & 
\colhead{$\Phi^*$} &
\colhead{$M_B^*$} & 
\colhead{$\alpha$} \\
\colhead{} & 
\colhead{} & 
\colhead{} & 
\colhead{(Mpc$^-3$ Mag$^{-1}$)} &
\colhead{(mag)} & 
\colhead{} \\
\colhead{(1)} & 
\colhead{(2)} & 
\colhead{(3)} & 
\colhead{(4)} & 
\colhead{(5)} & 
\colhead{(6)} \\
} 
\startdata
\multicolumn{6}{c}{All galaxies}\\
0.2--0.4 &17.0--23.5 &  6682  &0.0050 $\pm$0.0006  &$-$21.03 $\pm$0.25  & $-$1.26 $\pm$  0.15  \\
0.6--0.8 &18.5--23.5 & 10092  & 0.0049$\pm$0.0003  &$-$21.24 $\pm$0.12  & $-$1.22 $\pm$  0.10 \\
	 &	     &  8146  & 0.0058$\pm$0.0003  &$-$21.09 $\pm$0.12  & $-$0.95 $\pm$  0.10  \\
\multicolumn{6}{c}{$T=1$ galaxies}\\
0.6--0.8 &18.5--23.5 &   740  & 0.00095$\pm$0.0001  &$-$21.07 $\pm$0.20  & $-$0.26 $\pm$  0.12 \\
	 &           &   761  & 0.00095$\pm$0.0001  &$-$21.12 $\pm$0.20  & $-$0.18 $\pm$  0.12 \\
\multicolumn{6}{c}{$T=2$ galaxies}\\
0.6--0.8 &18.5--23.5 &  8820  & 0.0047$\pm$0.0003  &$-$21.04 $\pm$0.12  & $-$1.25$\pm$  0.10 \\
 	 & 	     &  7016  & 0.0053$\pm$0.0003  &$-$20.89 $\pm$0.12  & $-$0.96$\pm$  0.10 \\
\multicolumn{6}{c}{$T=3$ galaxies}\\                                                              
0.6--0.8 &18.5--23.5 &   513  & 0.0007$\pm$0.0001 &$-$20.54 $\pm$ 0.4  &  $-$0.47$\pm$  0.20 \\
	 &   	     &   357  & 0.0005$\pm$0.0001 &$-$20.27 $\pm$ 0.4  &     0.07$\pm$  0.20 \\
\hline
\hline
\multicolumn{6}{c}{$T=2.3$ galaxies}\\                                                              
0.6--0.8 &18.5--23.5 &  3542  & 0.0036$\pm$0.0002 & $-$20.38 $\pm$0.20  & $-$1.04$\pm$  0.10 \\
	 &	     &  2574  & 0.0031$\pm$0.0002 &$-$20.31 $\pm$0.20  & $-$0.73$\pm$  0.10 \\
\multicolumn{6}{c}{$T=2.2$ galaxies}\\   
0.6--0.8 &18.5--23.5 &  3575  & 0.0019$\pm$0.0002 &$-$20.86 $\pm$0.25  & $-$1.35$\pm$  0.10 \\
         &           &  2863  & 0.0024$\pm$0.0002 &$-$20.66 $\pm$0.25  & $-$1.02$\pm$  0.10 \\
\multicolumn{6}{c}{$T=2.1$ galaxies}\\                                                              
0.6--0.8 &18.5--23.5 &  1240  & 0.0008$\pm$0.0001 &$-$21.20 $\pm$0.35  & $-$1.05$\pm$  0.10 \\
         &           &  1110  & 0.0009$\pm$0.0001 &$-$21.01 $\pm$0.35  & $-$0.81$\pm$  0.10 \\
\multicolumn{6}{c}{$T=2.0$ galaxies}\\                                                              
0.6--0.8 &18.5--23.5 &   463  & 0.0007$\pm$0.0002 &$-$20.98 $\pm$0.25  &$-$0.1$\pm$  0.10 \\
         &           &   466  & 0.0006$\pm$0.0002 &$-$20.81 $\pm$0.25  &  0.28$\pm$  0.10 \\
\enddata
\tablecomments{The columns are: (1) Redshift range; (2) $B-$band
  absolute magnitude range; (3) Total number of galaxies considered in
  the fit; (4) Schechter function normalization and error; (5)
  $M_{B}^*$ and error; (6) Faint-end slope of the Schechter function.
  In the $z=0.7$ bin, two values are provided for each parameter,
  reflecting the two different estimates for the photometric
  redshifts, derived with (upper values) and without (lower values)
  corrections to the photometric catalogues, respectively (see text).}
  \end{deluxetable}

%
%
%
\clearpage
\begin{figure*}[ht]
\epsscale{1.0}
\plotone{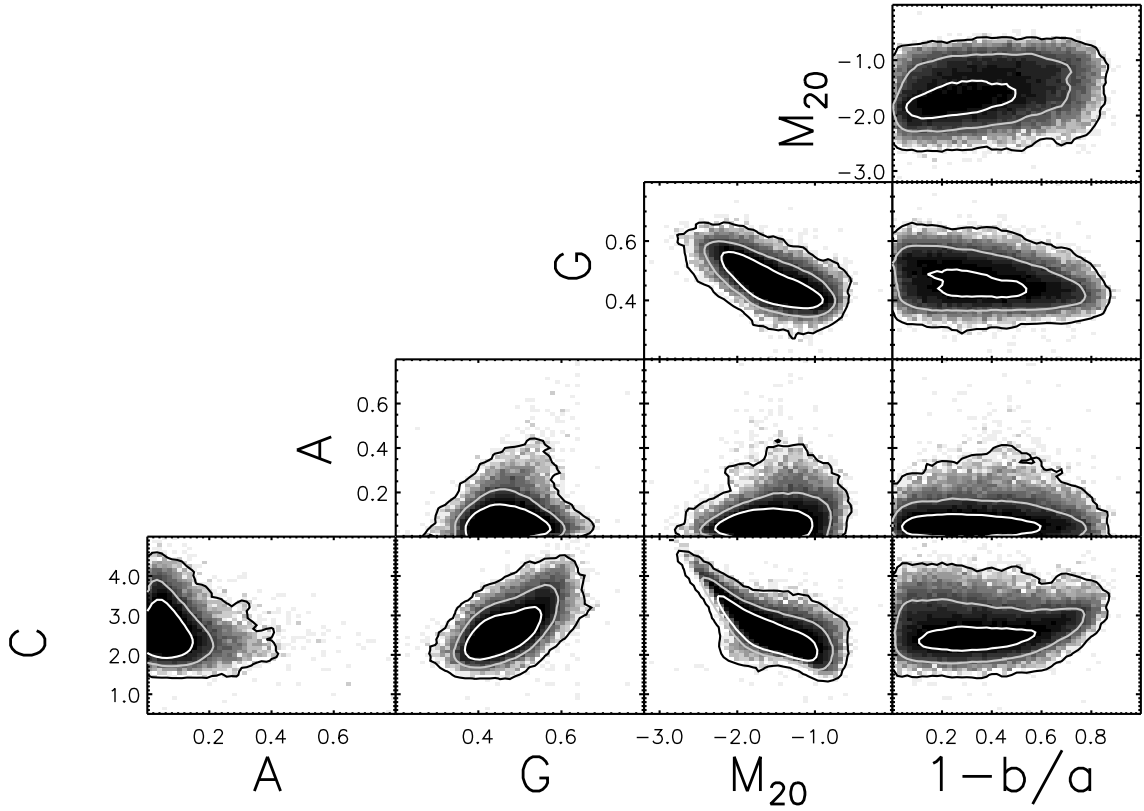}
\caption{Relations between the non-parametric diagnostics ($M_{20}$,
  $G$, $A$, $C$, and $\epsilon =1-b/a$).  Contours enclose $\sim$30\%
  (white contour), 80\% (grey contour), and 98\% (black contour) of
  the galaxies.  The main correlations among some of the parameters,
  such as $M_{20}$, $C$, and $G$, are clearly visible in these
  diagrams.}
\label{fig:allpar}
\end{figure*}

\clearpage
\begin{figure}[ht]
\epsscale{0.8}
\plotone{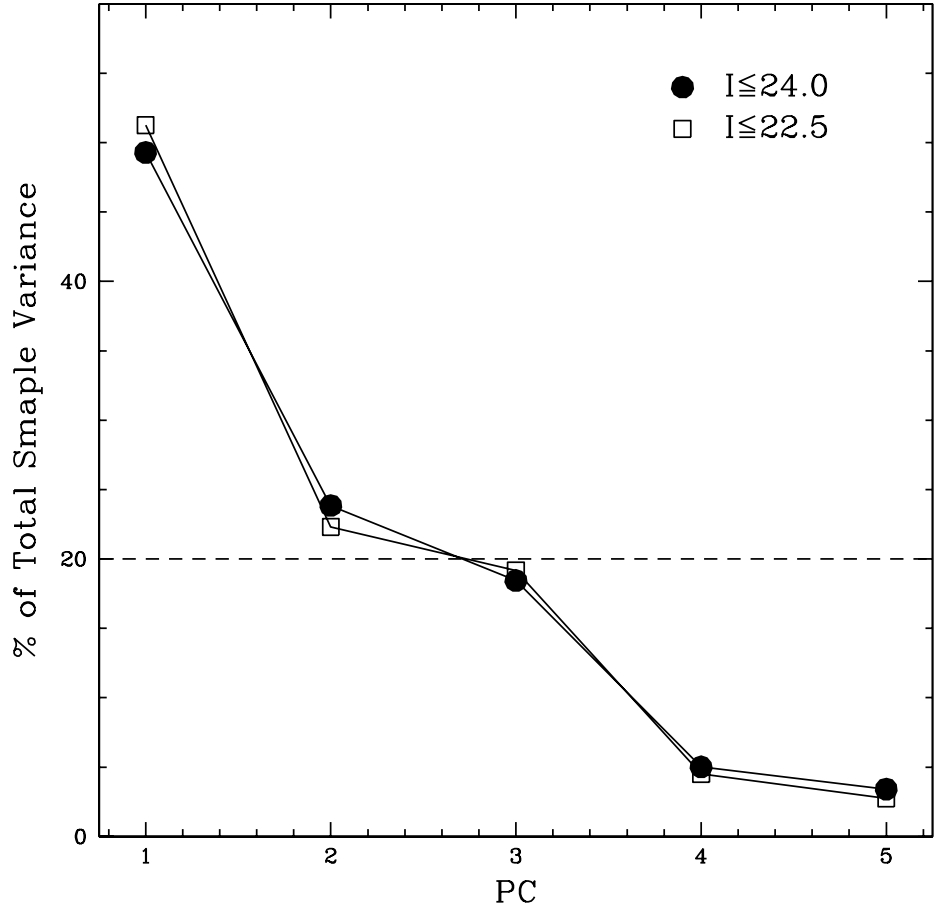}
\caption{Fraction of the total variance explained by each principal
  component as a function of the corresponding principal components
  for all galaxies down to $I=24$ in the ACS--catalog (solid circles).
  Open squares refer to the same analysis performed only on those
  objects with magnitude brighter than $I=22.5$. The horizontal line
  indicates 20\% of the total variance, i.e., the value for the five
  eigenvalues for a sample of $100$\% uncorrelated variables. }
\label{fig:pca}
\end{figure}

  \clearpage

 \begin{figure}[ht]
\epsscale{0.8} 
\plotone{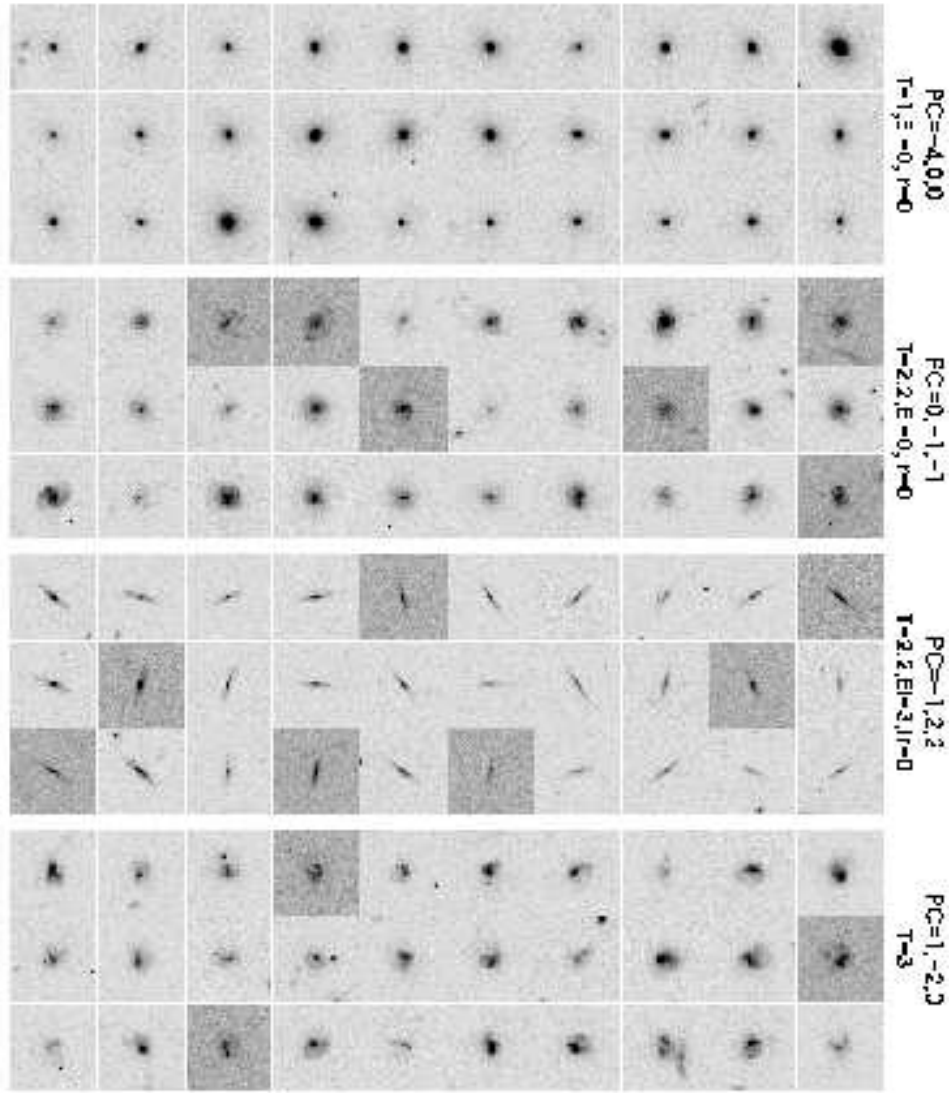}
\caption{Illustration of the power of the ZEST grid to separate COSMOS
  galaxies with different structural properties in the
  three-dimensional $PC_1$, $PC_2$, $PC_3$ space. Shown are four
  separate unit cubes of $PC_1$-$PC_2$-$PC_3$, centered around the
  values reported in the labels. In every unit cube, the few galaxies
  are representative of the population of objects of that specific
  bin. }
\label{fig:examp}
\end{figure}

\clearpage
\begin{figure}
\epsscale{0.8} 
\figurenum{4}
\centerline{\includegraphics[width=15cm,angle=-90]{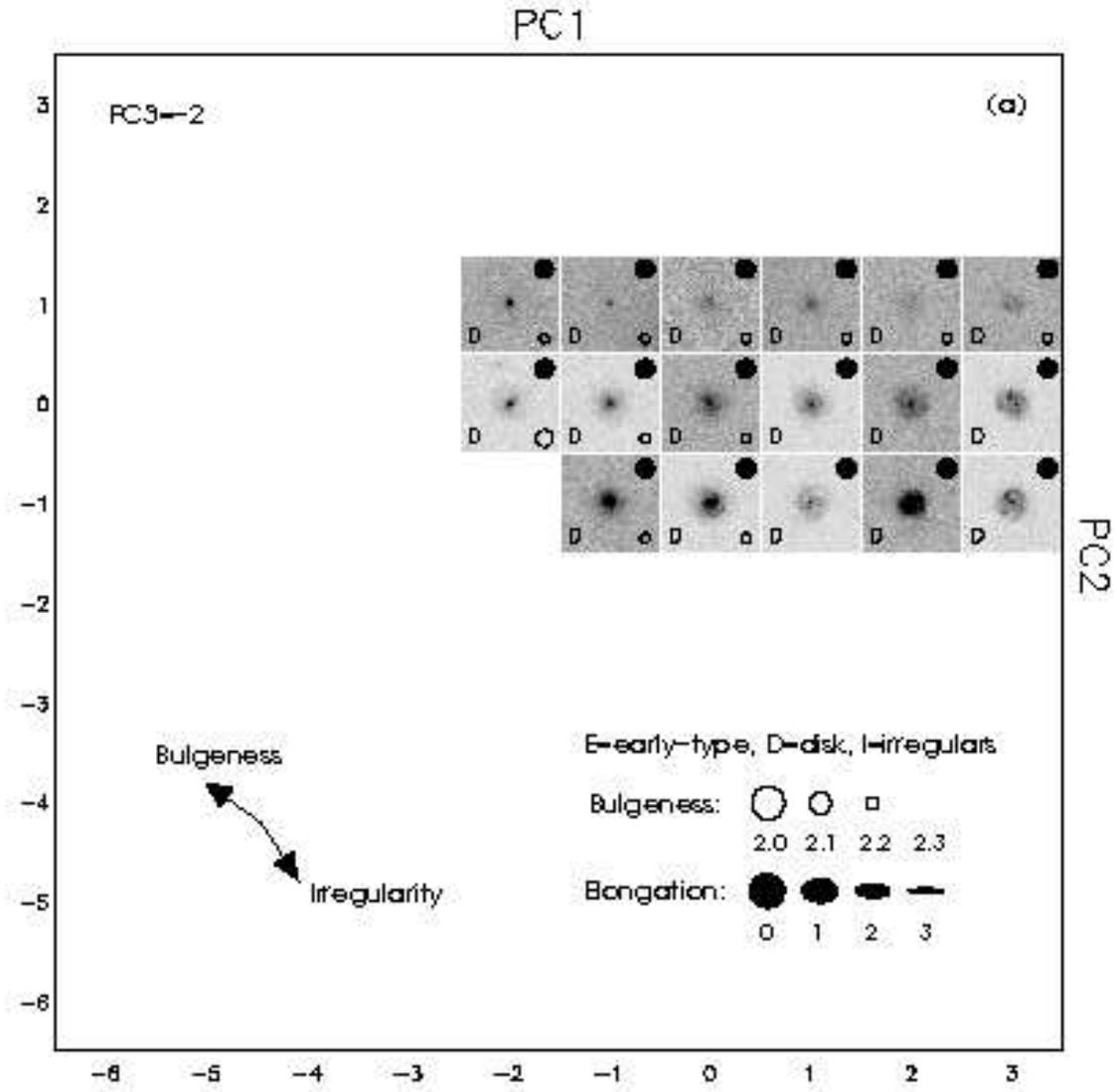}}
\caption{The stamp of a representative galaxy is shown for each cube
  of PC, arranged in planes of constant $PC_3$ in six different panels
  ($a$ to $f$). In particular the panels show sequentially the
  $PC_3=-2,-1,0,1,2,3$ planes. Galaxy structural properties -and thus
  galaxy types- change smoothly through the PC$_1$-PC$_2$-PC$_3$
  space.}
\label{fig:examples6a}
\end{figure}

\clearpage
\centerline{\includegraphics[width=15cm,angle=-90]{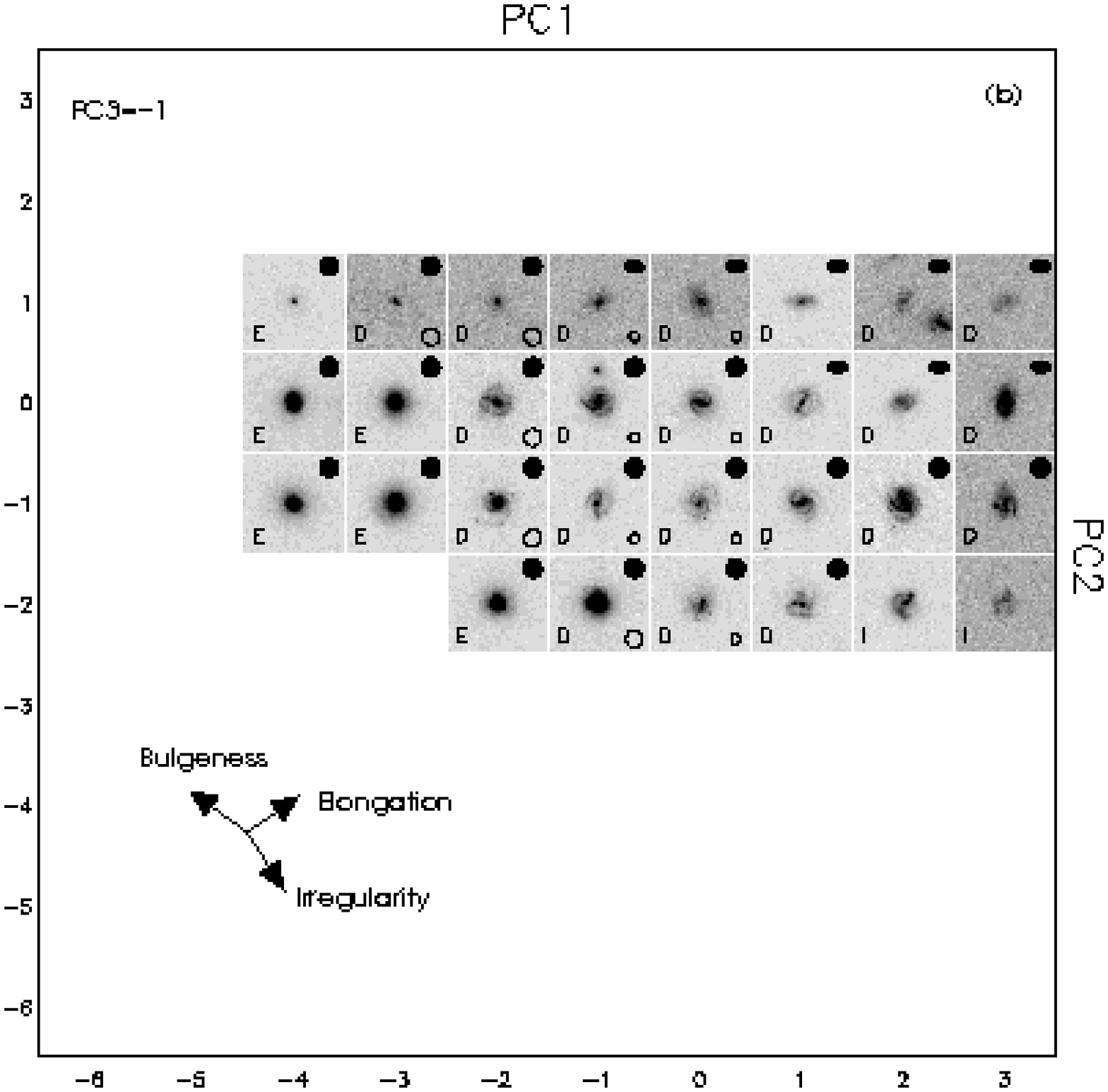}}
\centerline{Fig. 4 --- Continued.}
\clearpage
\centerline{\includegraphics[width=15cm,angle=-90]{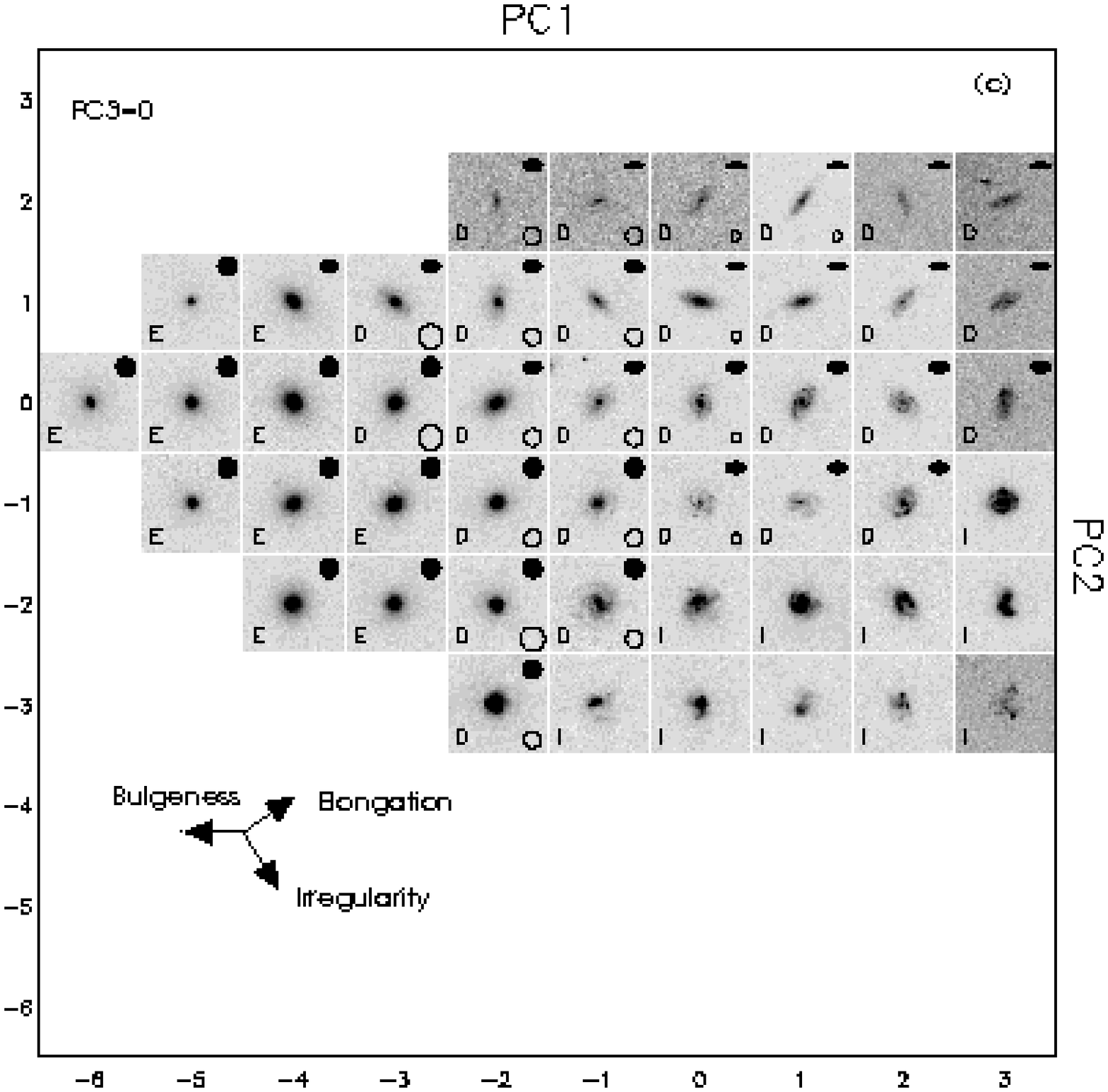}}
\centerline{Fig. 4 --- Continued.}
\clearpage
\centerline{\includegraphics[width=15cm,angle=-90]{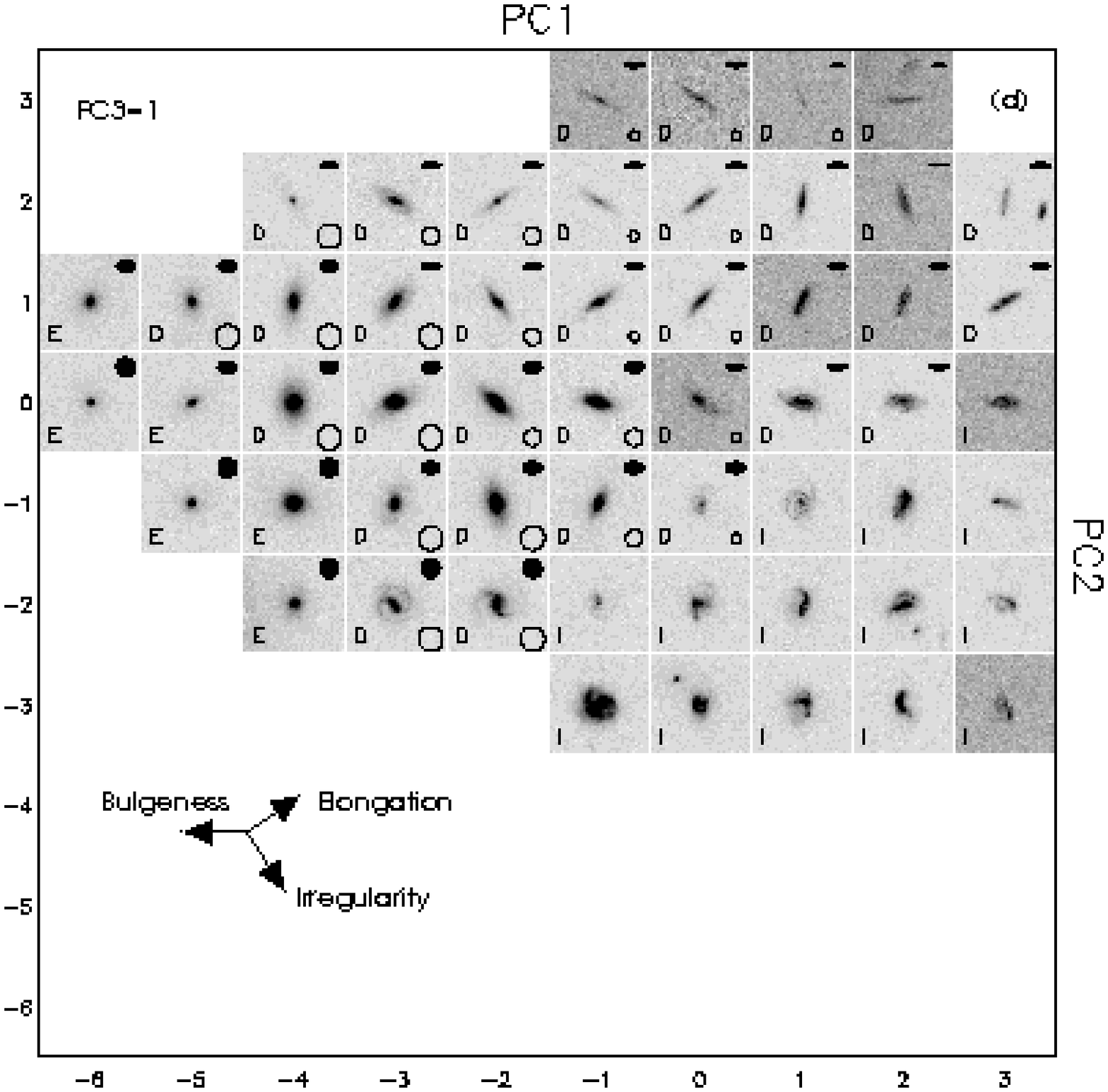}}
\centerline{Fig. 4 --- Continued.}
\clearpage
\centerline{\includegraphics[width=15cm,angle=-90]{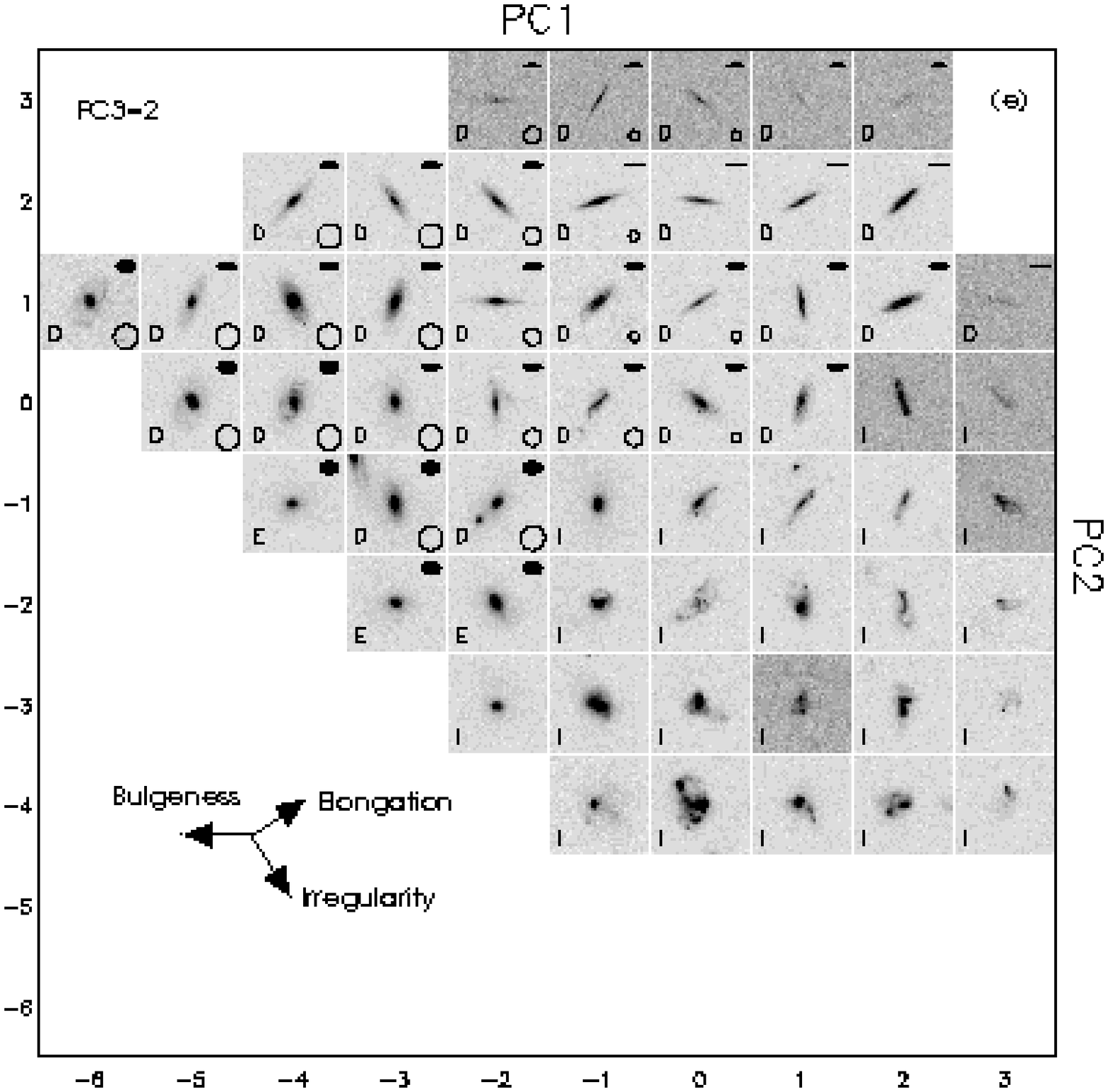}}
\centerline{Fig. 4 --- Continued.}
\clearpage
\centerline{\includegraphics[width=15cm,angle=-90]{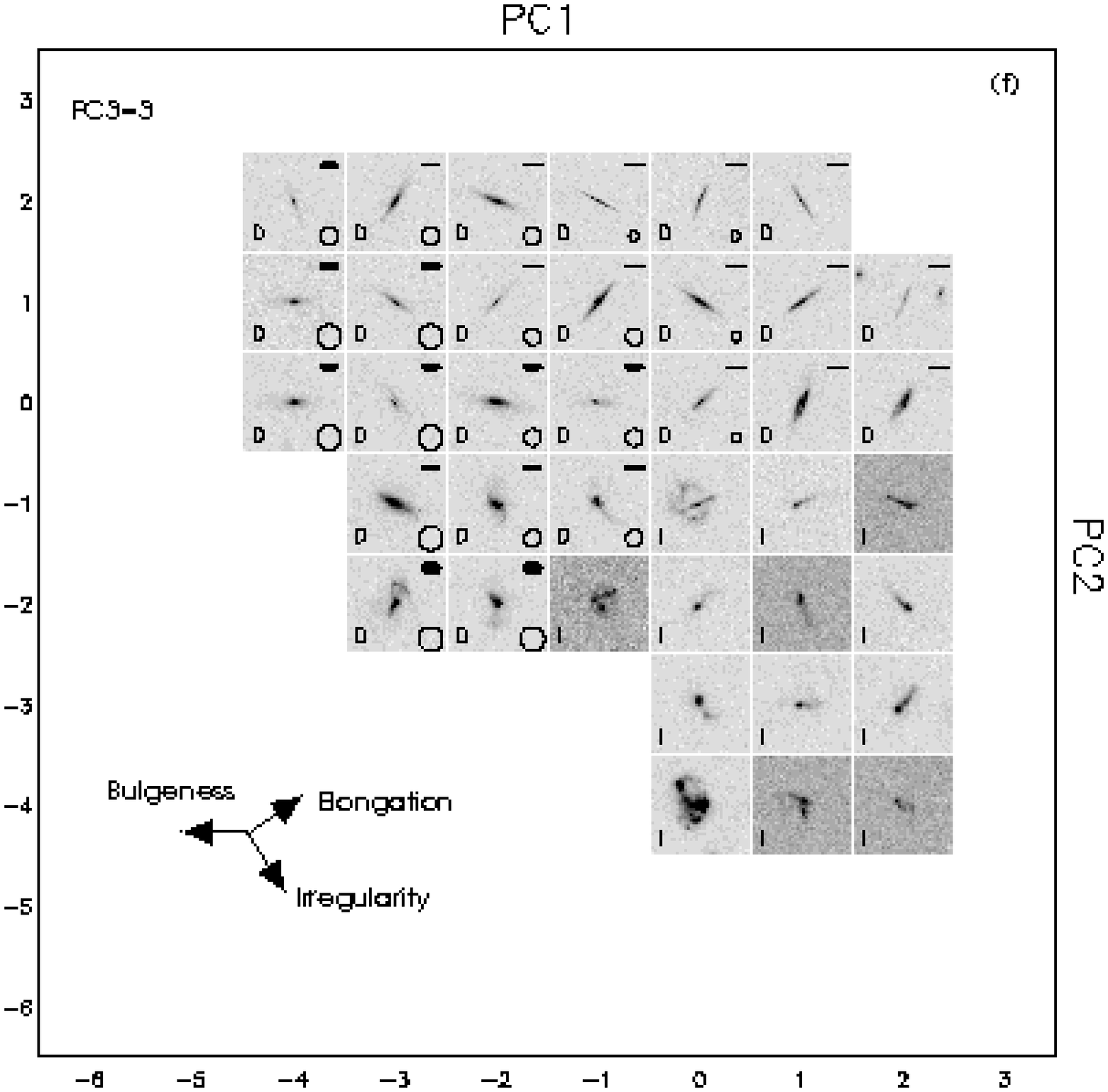}}
\centerline{Fig. 4 --- Continued.}

\clearpage
\begin{figure*}
\epsscale{0.8} 
\figurenum{5}
\includegraphics[width=18cm,angle=0]{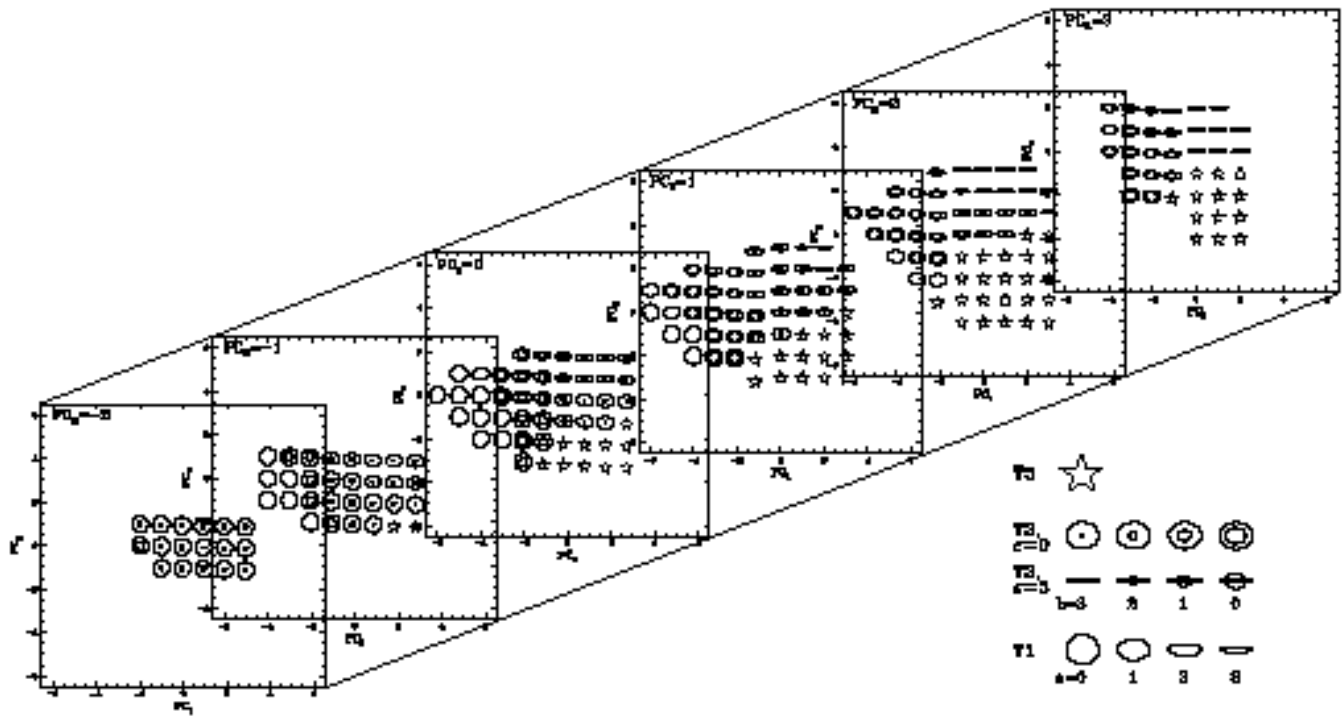}
\caption{Schematic view of the ZEST morphological classification.
  Shown are slices of the PC space at constant value of $PC_3$. Each
  slice shows $PC_1$ vs $PC_2$; the value of $PC_3$ increases from
  bottom-left to upper-right. In each unit cube the symbol indicates
  the ZEST classification associated with that cube (according to the
  legend shown in the bottom-right corner of the Figure). In
  particular, stars represent $T=3$ galaxies, and single ellipses
  represent $T=1$ galaxies. The double ellipses indicate $T=2$
  galaxies, for which the size of the inner ellipse increases for
  increasing bulgeness parameter.}
\label{fig:class}
\end{figure*}

\clearpage
\begin{figure*}
\epsscale{1.0} 
\figurenum{6}
\plotone{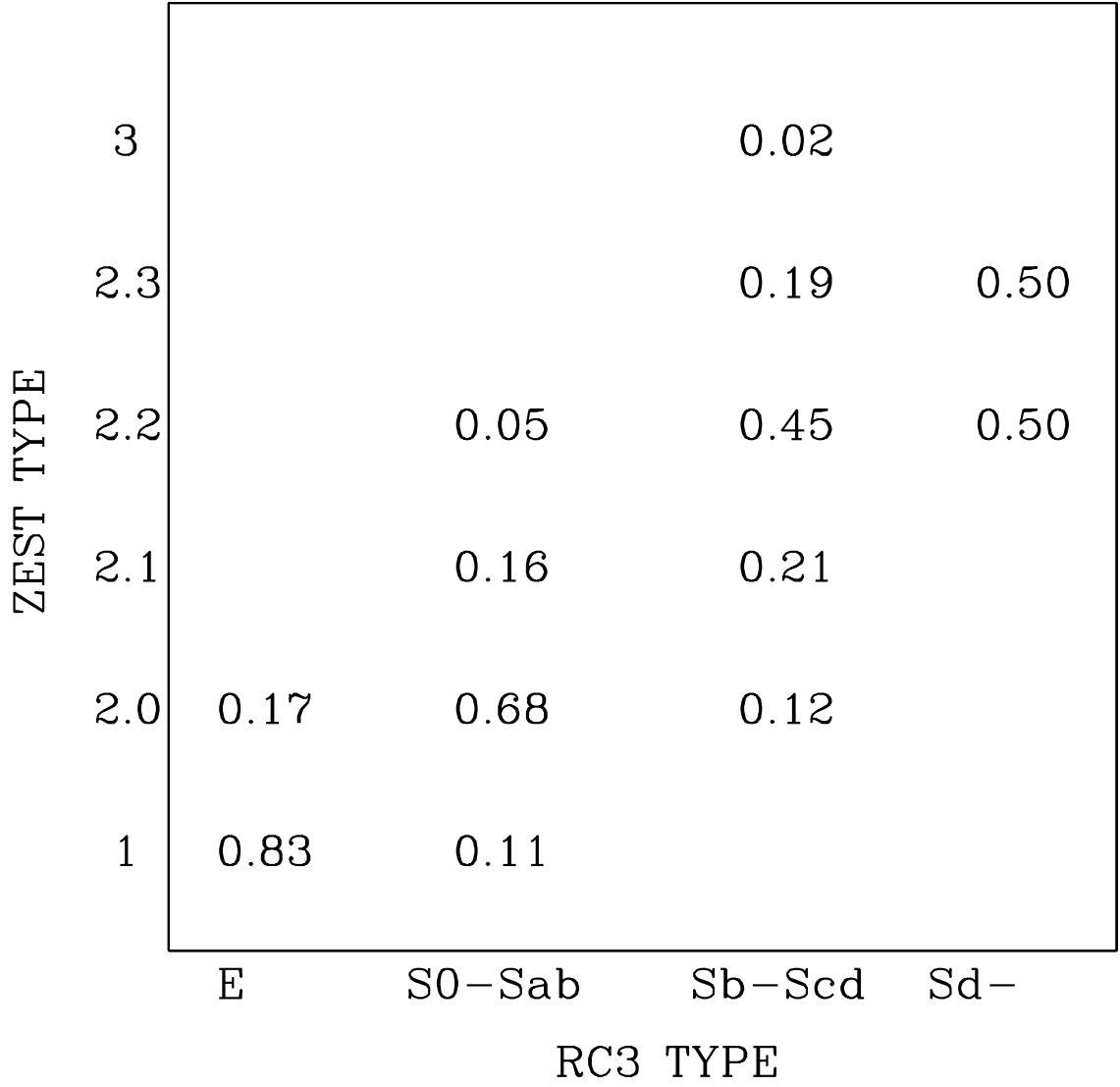}
\caption{The fraction of objects with a given RC3 classification (E,
  S0-Sab, Sb-Scd, and Sd and later) that have respectively T=1,
  2.0,2.1,2.2,2.3, and T=3 ZEST morphological Type.}
\label{fig:visautonew}
\end{figure*}

\clearpage
\begin{figure*}
\epsscale{1.0} 
\figurenum{7}
\plotone{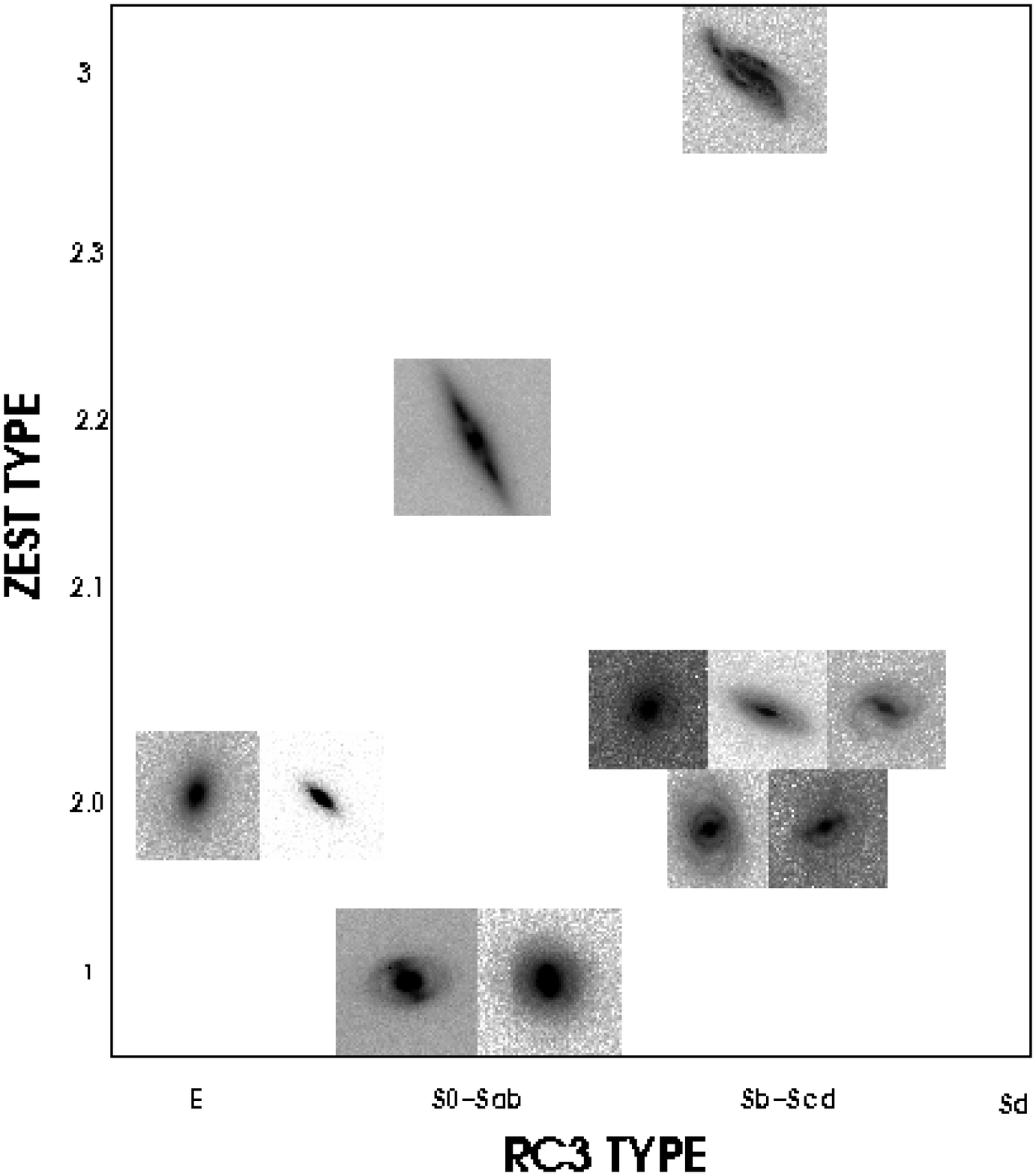}
\caption{Images of the $z=0$ Frei et al.\ galaxies with relatively
  discrepant ZEST and RC3 classifications. }
\label{fig:visautonew2}
\end{figure*}

\clearpage
\begin{figure}
\epsscale{0.8} 
\figurenum{8}
\plotone{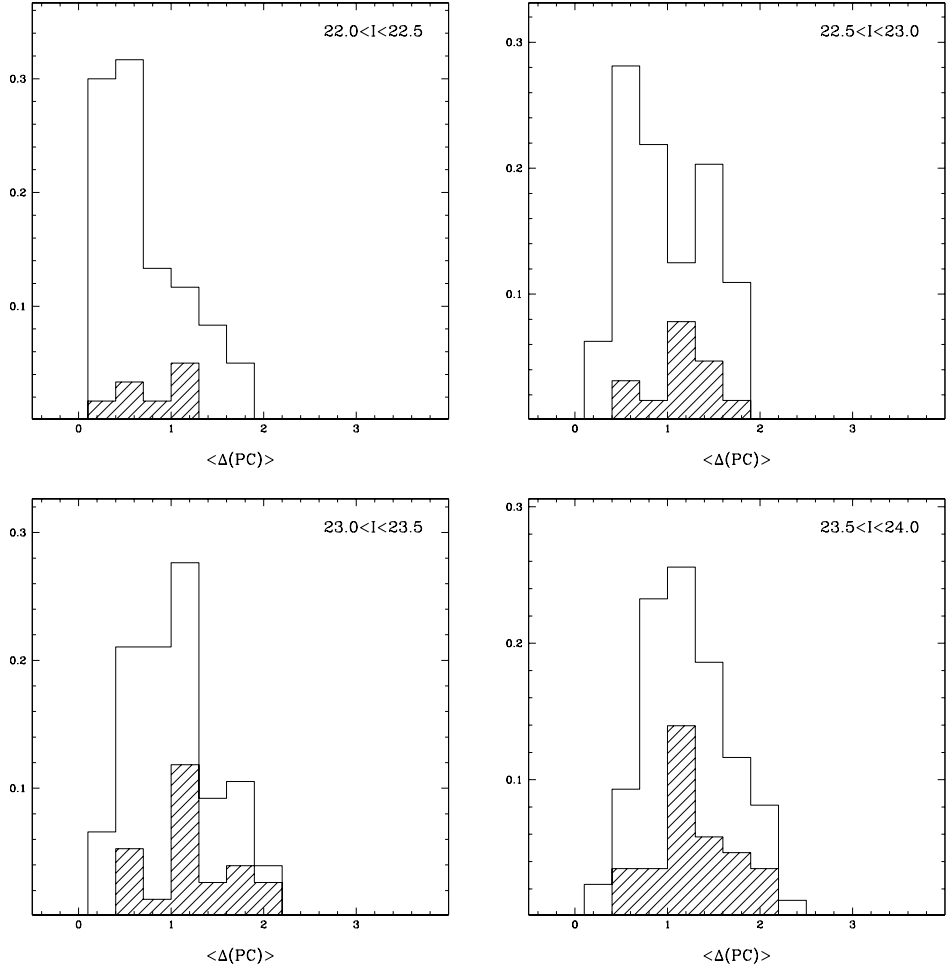}
\caption{The solid histograms show the distribution of $\langle
  \Delta(PC)\rangle=\sum_{i} |PC_i-PC_{i,0}|)/3$, with $i=1,2,3$,
  $PC_{i,0}$ the initial values of the PCs, and $PC_{i}$ the measured
  PCs after signal-to-noise image degradation of a set of $17.5<I<18$
  COSMOS galaxies. We show four different magnitude bins for the S/N
  degraded galaxies, from $I=22$ (top-left panel) down to $I=24$
  (bottom-right panel. The hatched histograms show the $\langle
  \Delta(PC) \rangle$ distribution for the galaxies with $\Delta T \ne
  0$.}
\label{fig:simclass}
\end{figure}

\clearpage
\begin{figure}
\epsscale{0.8} 
\figurenum{9}
\begin{center}
\epsscale{0.8}
\plotone{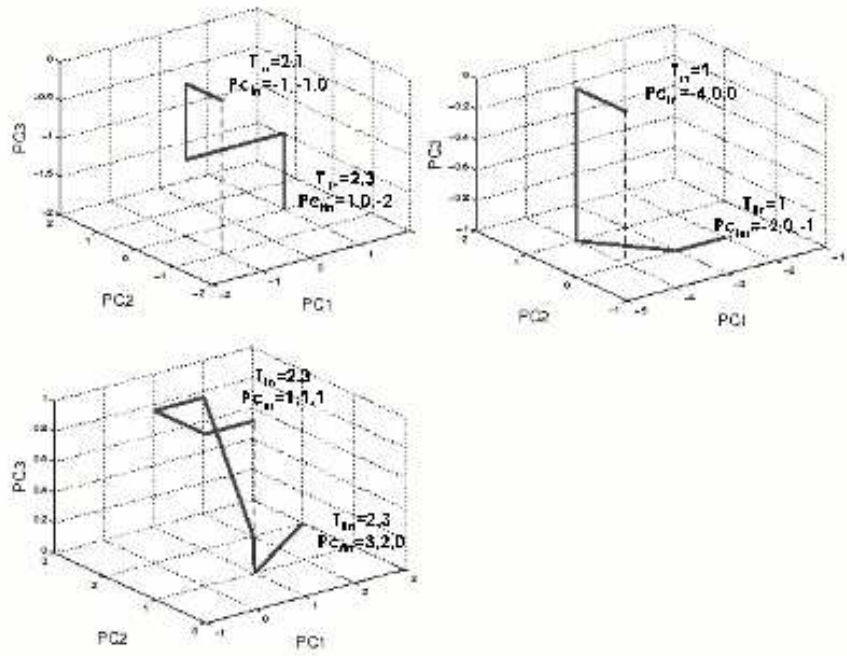}
\end{center}
\caption{Three examples of how galaxies move in the PC space as the
  signal-to-noise of the galaxy image decreases. The start- and
  end-coordinates in PC space are indicated (PCin, and PCfin,
  respectively). }
\label{fig:tracks}
\end{figure}

\clearpage

\begin{figure*}
\epsscale{0.6} 
\figurenum{10}
\includegraphics{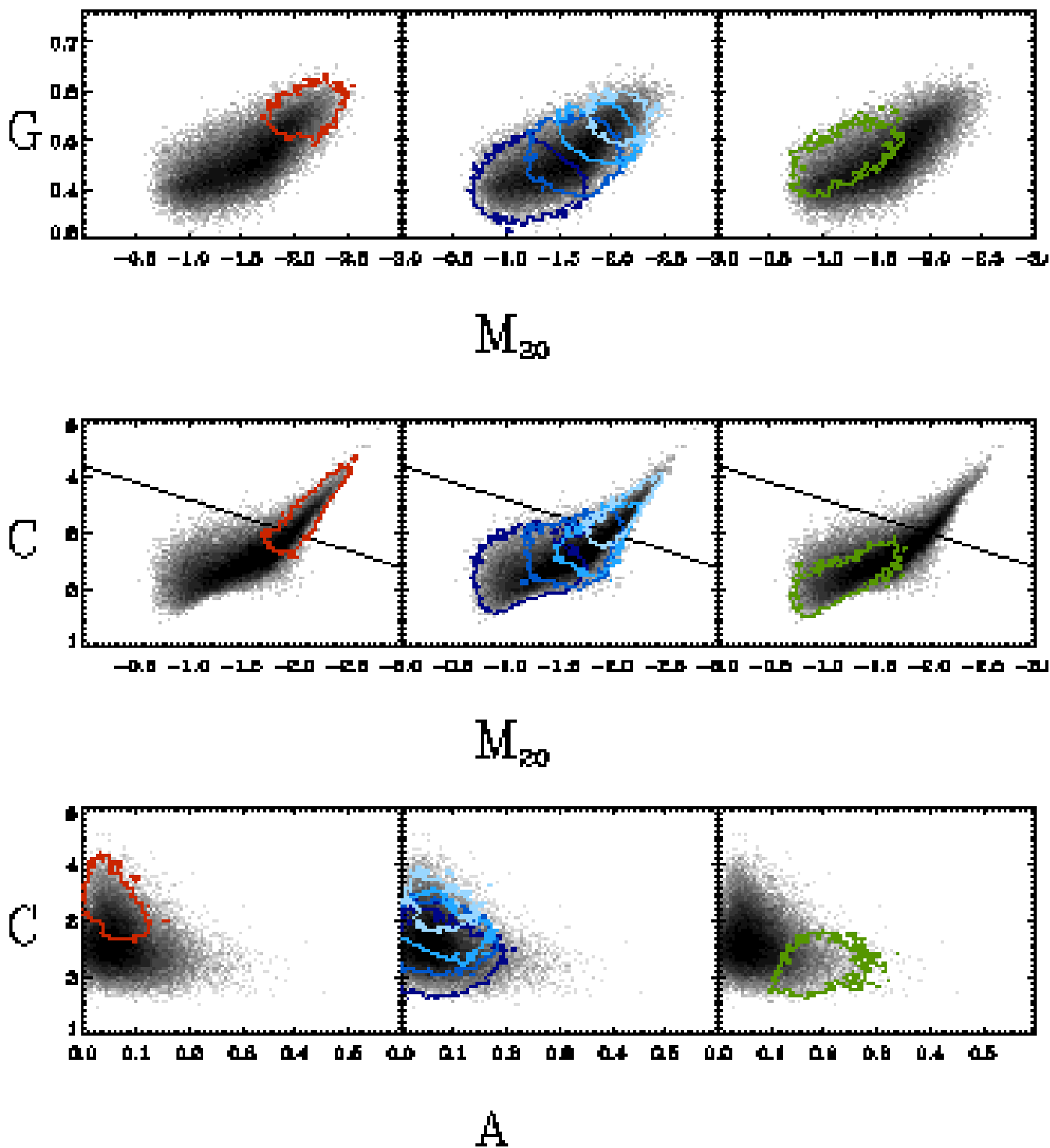}
\caption{Distribution of our $\sim$56000 $I_{AB} \le 24$ COSMOS
  galaxies in the $G-M_{20}$ plane (top panels), $C-M_{20}$ plane
  (middle panels), and $C-A$ plane (bottom panels).  In each column we
  highlight the position of a different ZEST morphological class by
  drawing the contours enclosing 99\% of the objects in that ZEST
  class.  In the first, second and third columns we show respectively
  $T=1, 2$ and $3$ galaxies (red, blue and green contours). The blue
  color for $T=2$ galaxies ranges from dark blue for bulge-dominated
  $T=2.0$ galaxies to light blue for $T=2.3$ bulgeless disks. }
\label{fig:projection}
\end{figure*}

\clearpage
\begin{figure}[ht]
\figurenum{11}
\epsscale{0.8}
\plotone{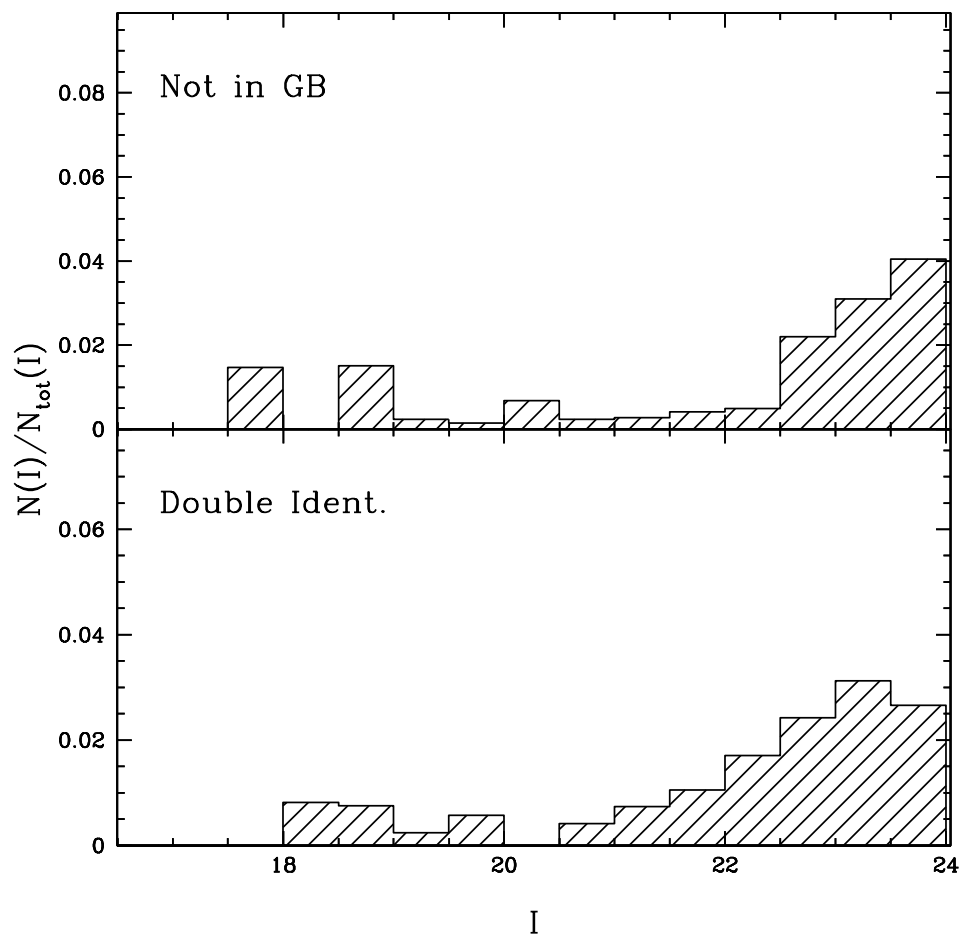}
\caption{ $I-$band magnitude distribution for the ACS--galaxies with
  no detection in the ground--based catalog (top panel), and for
  the ACS--galaxies whose ground--based match was associated with more
  than one ACS-detected galaxy. Both distributions are normalized to
  the magnitude distribution of all galaxies in the sample. The
  fraction of missed objects stays relatively constant down to
  $I_{AB}\sim 22$ and then increases toward fainter magnitudes.}
\label{fig:missed}
\end{figure}

\clearpage
\begin{figure*}[ht]
\figurenum{12}
\epsscale{0.8} 
\begin{center}
\includegraphics{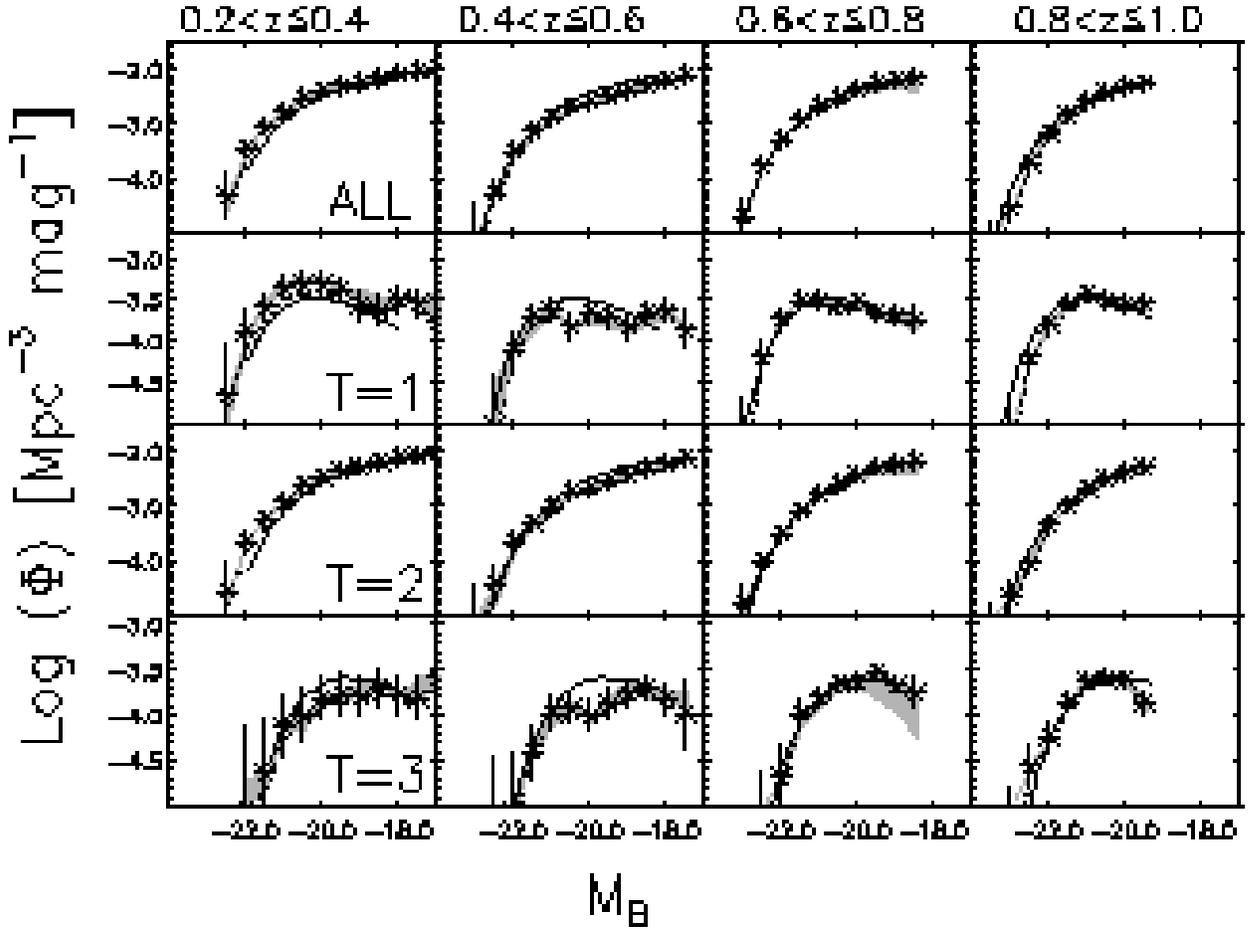}
\end{center}
\caption{The $1/V_{\rm max}$ {\it corrected} (stars) and {\it
    uncorrected} (dotted line) LFs derived using the ZEBRA Maximum
  Likelihood photometric redshifts for the global sample (top row),
  for $T=$1 early-type galaxies (second row), for $T=$2 disk galaxies
  (third row), and for $T=$3 irregular galaxies (bottom row).  In each
  row, we show the LFs in four redshift bins: $0.2<z\leq 0.4$ (first
  column), $0.4<z\leq 0.6$ (second column), $0.6<z\leq 0.8$ (third
  column), and $0.8<z\leq 1.0$ (last column).  Error bars in each
  luminosity bin take into account Poissonian errors only. In each
  redshift bin LFs are plotted down to the $B-$band at which the
  sample is complete, regardless of the color of the galaxy.  Solid
  curves show the best--fit Schechter function to the COSMOS LFs in
  the redshift interval $0.6 < z \le 0.8$, (de-)brightened in each
  redshfit range by $\sim 1.3\times z$ magnitudes (see text).  The
  stars show the LFs obtained using the ZEBRA photo--$z$'s derived
  after correcting the photometric catalogue for detected systematic
  calibration errors; grey shaded regions show the uncertainty in the
  LFs due to the use of photo-$z$'s derived without applying any
  correction to the photometric catalogue.}
\label{fig:LF_ML_nocorrection}
\end{figure*}

\clearpage

\begin{figure}
\epsscale{0.8}
\figurenum{13}
\plotone{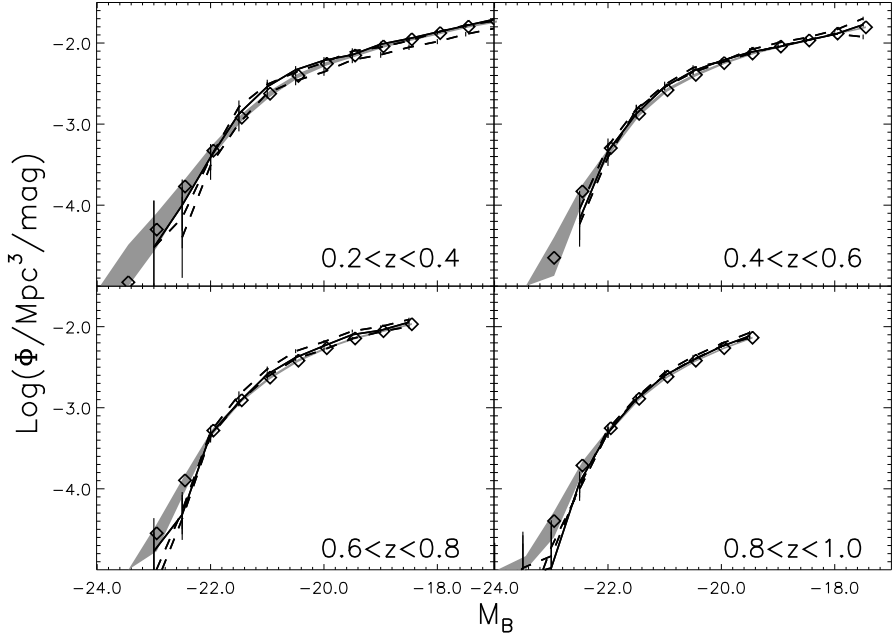}
\caption{Results of the simulations performed to assess the impact of
  the photometric redshift errors in the calculation of the LF, in
  four redshift bins, as indicated in each panel. Open diamonds
  represent the median volume density computed from the 100 mock
  realizations; the associated grey area indicates the 16$^{th}$-to
  $84^{th}$ percentiles of the distribution in each magnitude bin.
  Points are shifted of 0.05 magnitude from the center of the
  magnitude bin for presentation purposes. The solid line with
  errorbars represents the LF of the original mock catalog used to
  generate the 100 mock datasets. For comparison, we also show the LFs
  derived using two independent mock catalogs (dashed lines).  These
  LFs show that cosmic variance in the lowest redshift bin dominates
  the uncertainty budget in the calculation of the LF errors.}
\label{fig:photozsim}
\end{figure}

\clearpage
\begin{figure*}[ht]
\epsscale{0.8} 
\figurenum{14}
\begin{center}
\plotone{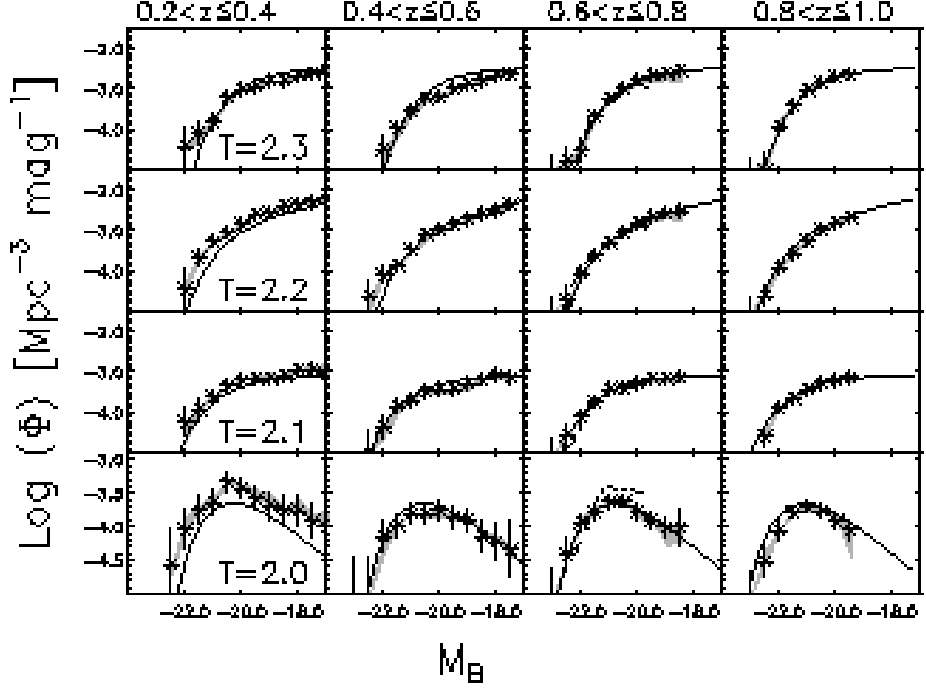}
\end{center}
\caption{{\it Corrected} $1/V_{\rm max}$ LF computed for for $T=2$
  galaxies, split by their bulgeness parameter, from $T=2.3$ bulgeless
  disks (top row), to $T=2.0$ bulge-dominated galaxies (bottom row).
  Dot--dashed curves in the redshift bin $0.6<z\le0.8$ show the LFs
  derived for the SDSS$_{z=0.7}$ comparison sample.  The solid curves
  show the best--fit Schechter function to the COSMOS LFs in the
  redshift interval $0.6 \le z \le 0.8$, brightened in each redshfit
  range by $\sim 1.3\times z$ magnitudes (see text). Grey shaded
  regions are as in Figure~\ref{fig:LF_ML_nocorrection}.}
\label{fig:LF_DISK}
\end{figure*}

\clearpage
\begin{figure*}[ht]
\epsscale{0.8} 
\figurenum{15}
\begin{center}
\includegraphics{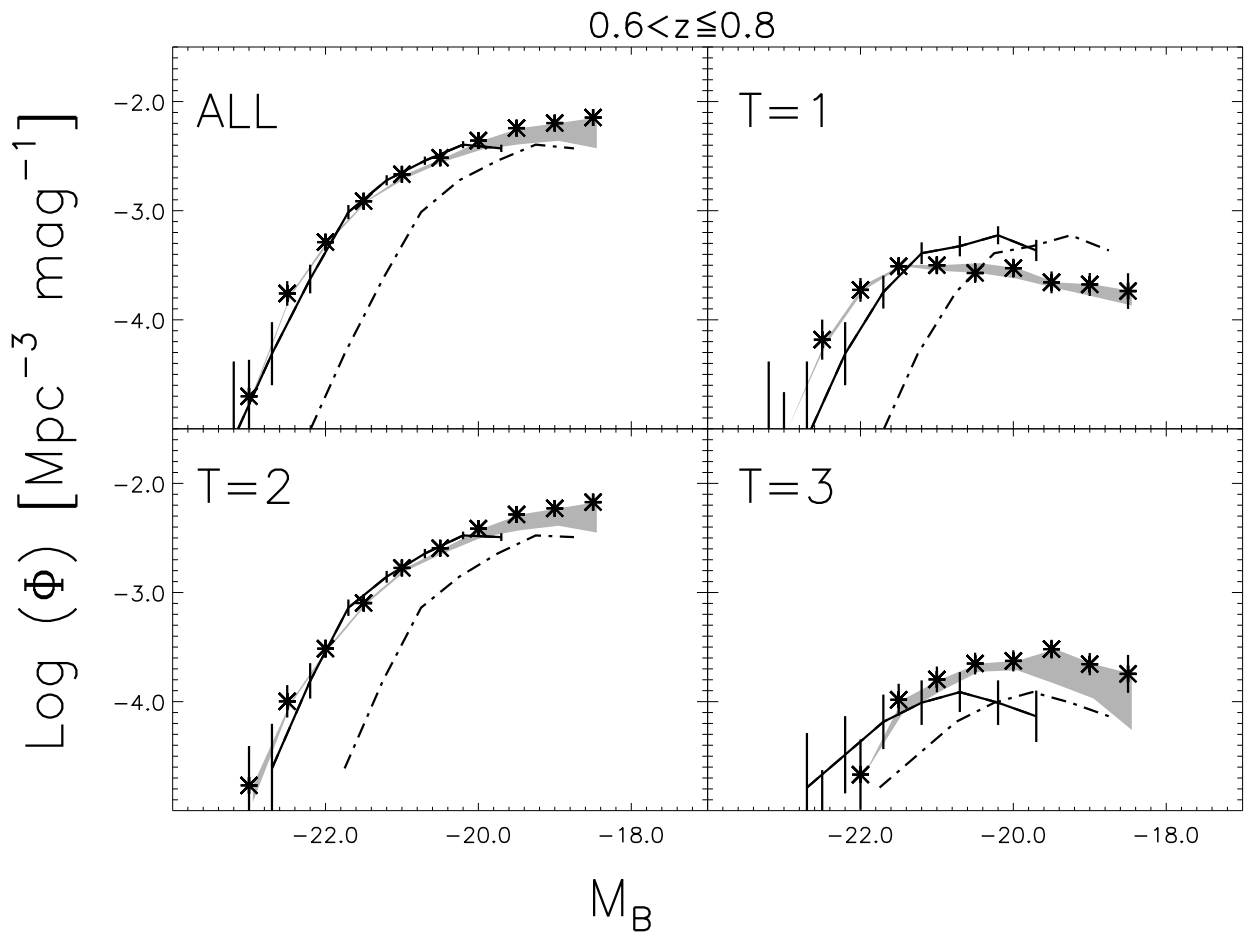}
\end{center}
\caption{Comparison between the {\it corrected} COSMOS LFs (stars) and
  the (redshifted) SDSS$_{z=0.7}$ LFs, representing the no--evolution
  predictions at redshift $z=0.7$ (dash--dotted line).  The solid
  curves shows the SDSS$_{z=0.7}$ LFs brightened by 0.95 magnitudes
  (see text for details). Grey shaded regions are as in
  Figure~\ref{fig:LF_ML_nocorrection}.}
\label{fig:compLF_SDSS}
\end{figure*}

%
%

\clearpage

\begin{figure}[ht]
\epsscale{0.8}
\figurenum{16}
\plotone{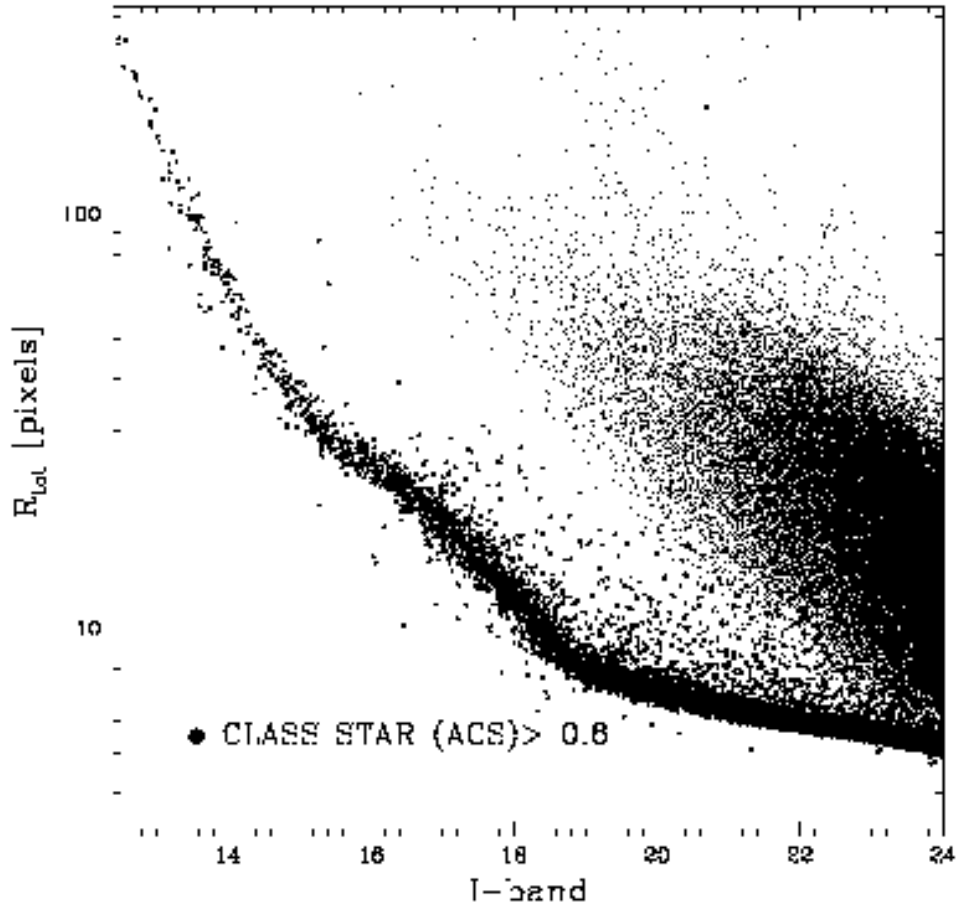}
\caption{The total radius $R_{tot}$ versus $I-$band magnitude for all
  COSMOS galaxies with $I_{AB} \le 24$. All objects with
  CLASS\_STAR$>0.6$ are identified as solid circles. Down to
  $I_{AB}=24$, stars form a tight sequence in this plane, which can be
  identified using the SExtractor CLASS\_STAR parameter.}
\label{fig:star}
\end{figure}

\clearpage

\begin{figure}[ht]
\epsscale{0.8}
\figurenum{17}
\plotone{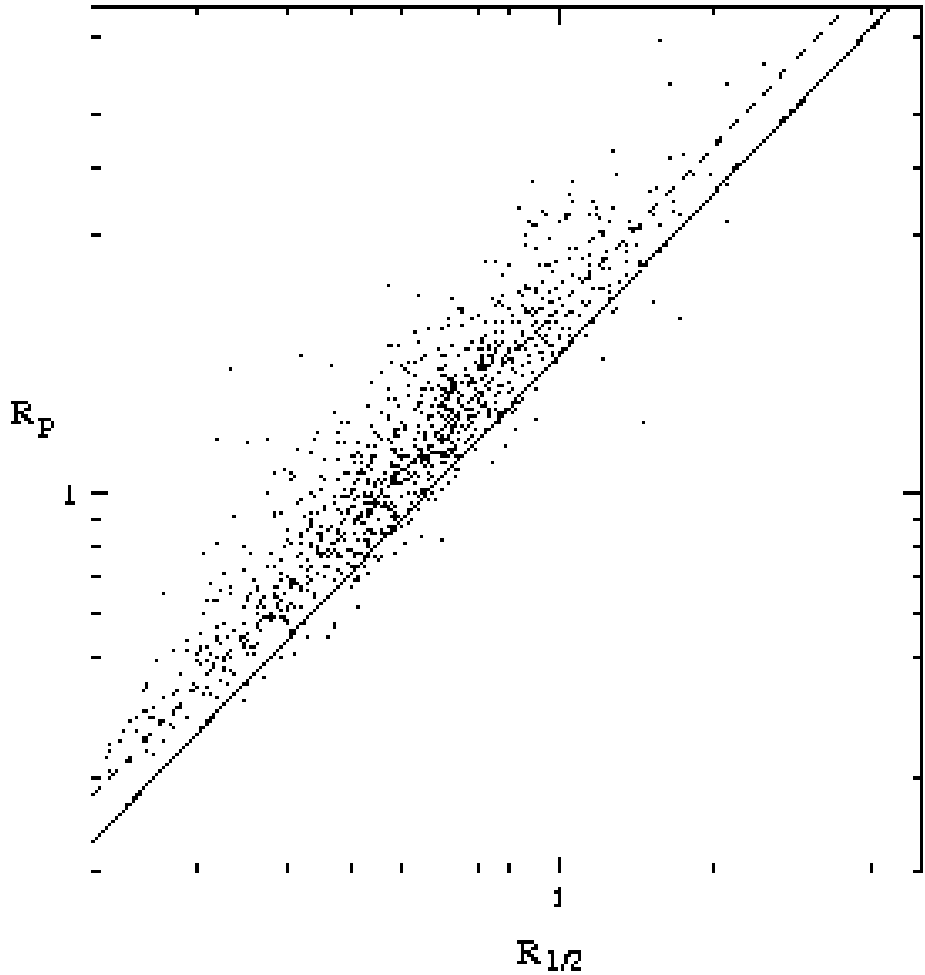}
\caption{Comparison between the Petrosian radii of $I_{AB}\le 24$
  COSMOS galaxies, as derived in our analysis, and the half--light
  radii derived by Sargent et al. (2006) for the $I_{AB}\le 22.5$
  subsample (using GIM2D fits to the galaxy surface brightness
  distribution). The solid and dashed lines represent the relation
  between $R_{P}$ and $R_{1/2}$ expected for a Sersic profile with
  $n=1$ and $n=2$, respectively.}
\label{fig:prhlr}
\end{figure}

\clearpage

\begin{figure*}[ht]
\epsscale{0.8}
\figurenum{18}
\includegraphics{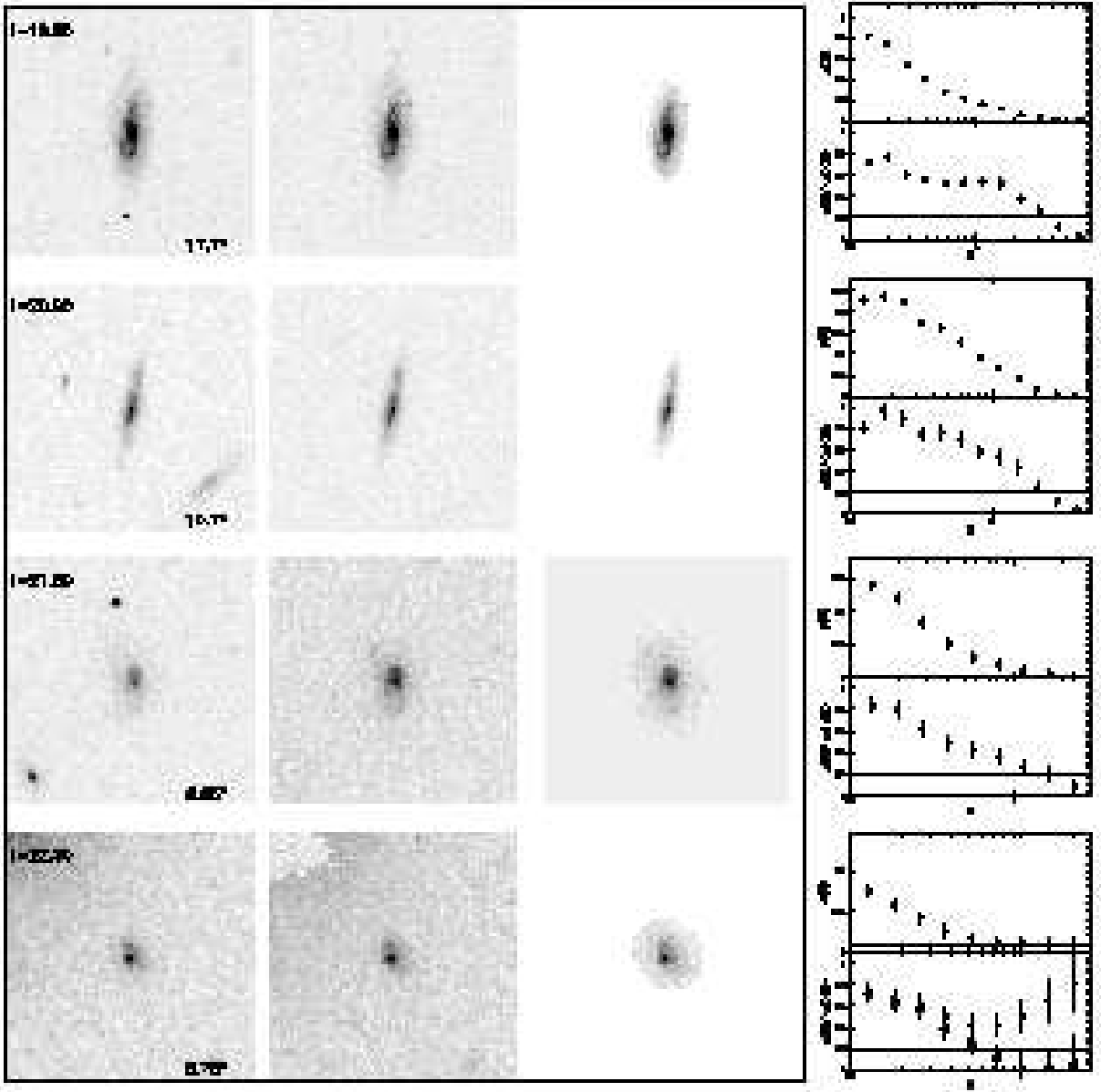}
\caption{Examples of the steps performed to compute the Petrosian
  radius $R_{P}$. In particular, we show four galaxies with increasing
  magnitudes, from $I=18.80$ (top row) to $I=22.3$ (bottom row).  For
  each galaxy we show the original stamp (first column), the cleaned
  stamp (second column), and the segmentation map (third column). In
  the last column we show the surface brightness profile (top panel)
  and the ratio $\eta=\mu(R)/\langle \mu(<R)\rangle$ (bottom panel) as
  a function of $R$ in arcseconds (solid circles).  The solid
  horizontal line in the $\eta-R$ plots shows $\eta=0.2$.  In the last
  example, $\eta$ decreases to a minimum value larger than $\eta=0.2$
  and then increases again.  This behavior is due to contamination to
  the external isophotes from light coming from the wings of the
  nearby source, still visible in the cleaned stamp, despite the
  attemp to remove it before proceeding with the analysis.  This light
  causes the surface brightness profile to flatten at large radii, as
  seen in the $\mu-R$ diagram. To correct for this effect, we subtract
  the residual light contribution from the nearby source.  The
  correction is derived by fitting a linear relation (shown as dashed
  line in the $\mu -R$plot) to the external part of the light profile
  in the $\mu -R$ plot, and then subtracting a constant value to the
  surface brightness profile.  Open circles in the $\eta -R$ plots
  show the $\eta$ profiles obtained after this correction is applied.}
\label{fig:examples}
\end{figure*}

\clearpage

\begin{figure}[ht]
  \epsscale{0.8} 
\figurenum{19}
\plotone{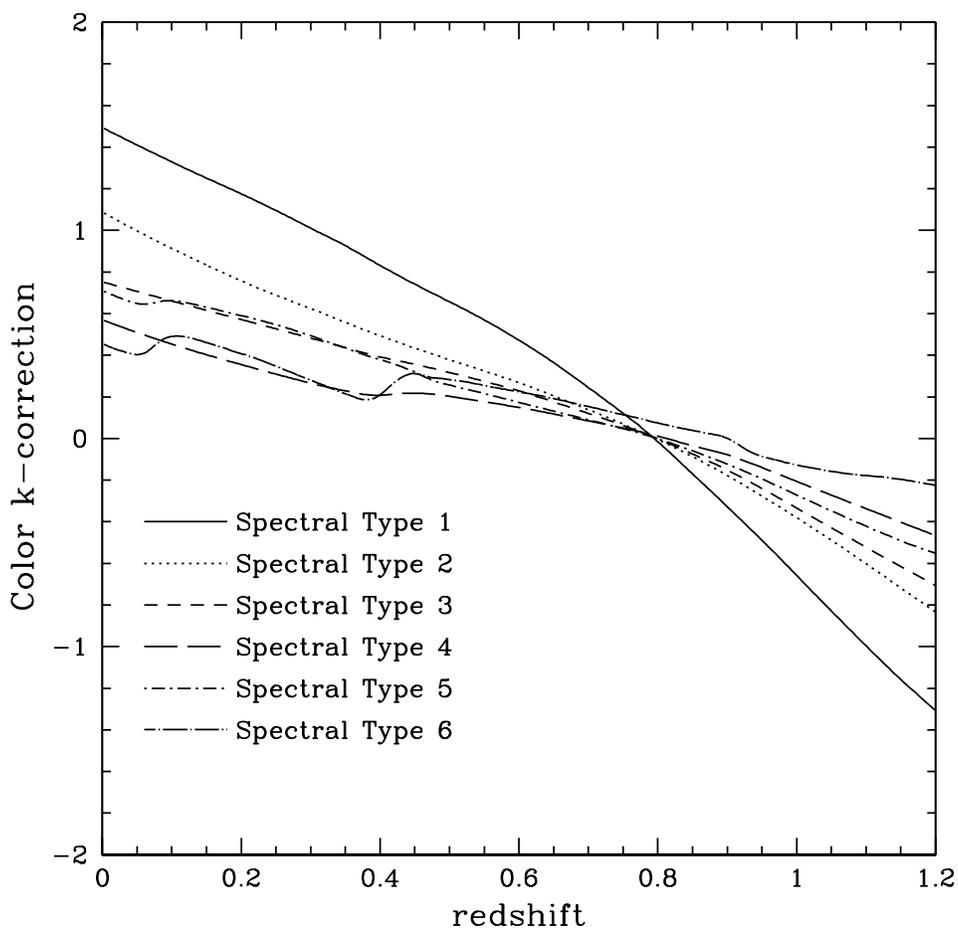}
\caption{Color $k-$correction (defined as $k(z)=B-I_z$, see text for
  details) as a function of redshift for different spectral templates,
  going from an early--type galaxy template ($T=1$, solid line), to
  the starburst galaxy template ($T=6$, long-dash dot line).  At
  redshift $z\sim 0.8$ the color $k-$correction is virtually zero for
  all photometric types, since at this redshift the rest--frame
  $B-$band coincides with the $I-$band filter from which the $B-$band
  absolute magnitudes are derived.}
\label{fig:kcorr}
\end{figure}

\clearpage

\begin{figure}[ht]
\epsscale{0.8} 
\figurenum{20}
\plotone{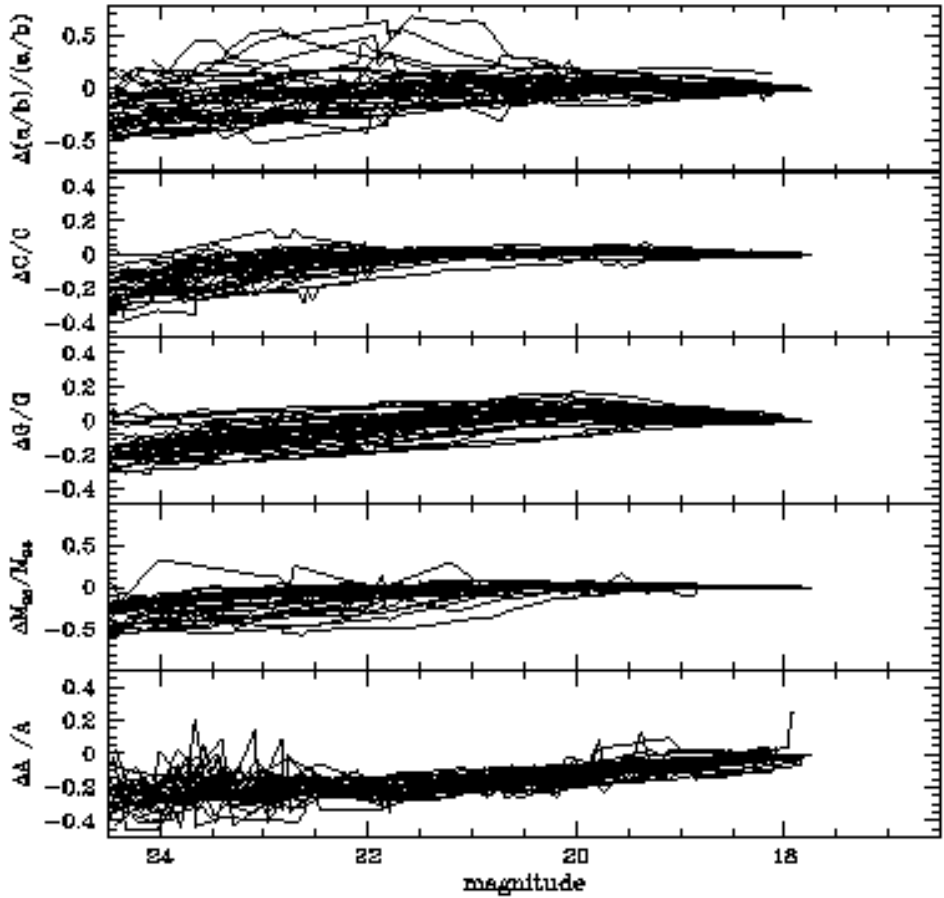}
\caption{Results of the simulations performed by degrading the S/N of
  real galaxies (see text for details). Lines in each panel show the
  fractional variation of the parameter as a function of magnitude
  (each line shows a different galaxy).  From top to bottom we show
  ellipticity, concentration, Gini, $M_{20}$, and asymmetry.
  Systematic effects are $< 20$\% over the whole range of observed
  magnitudes.  }
\label{fig:sim}
\end{figure}

\clearpage

\clearpage

\begin{figure}[ht]
\epsscale{0.8} 
\figurenum{22}
\plotone{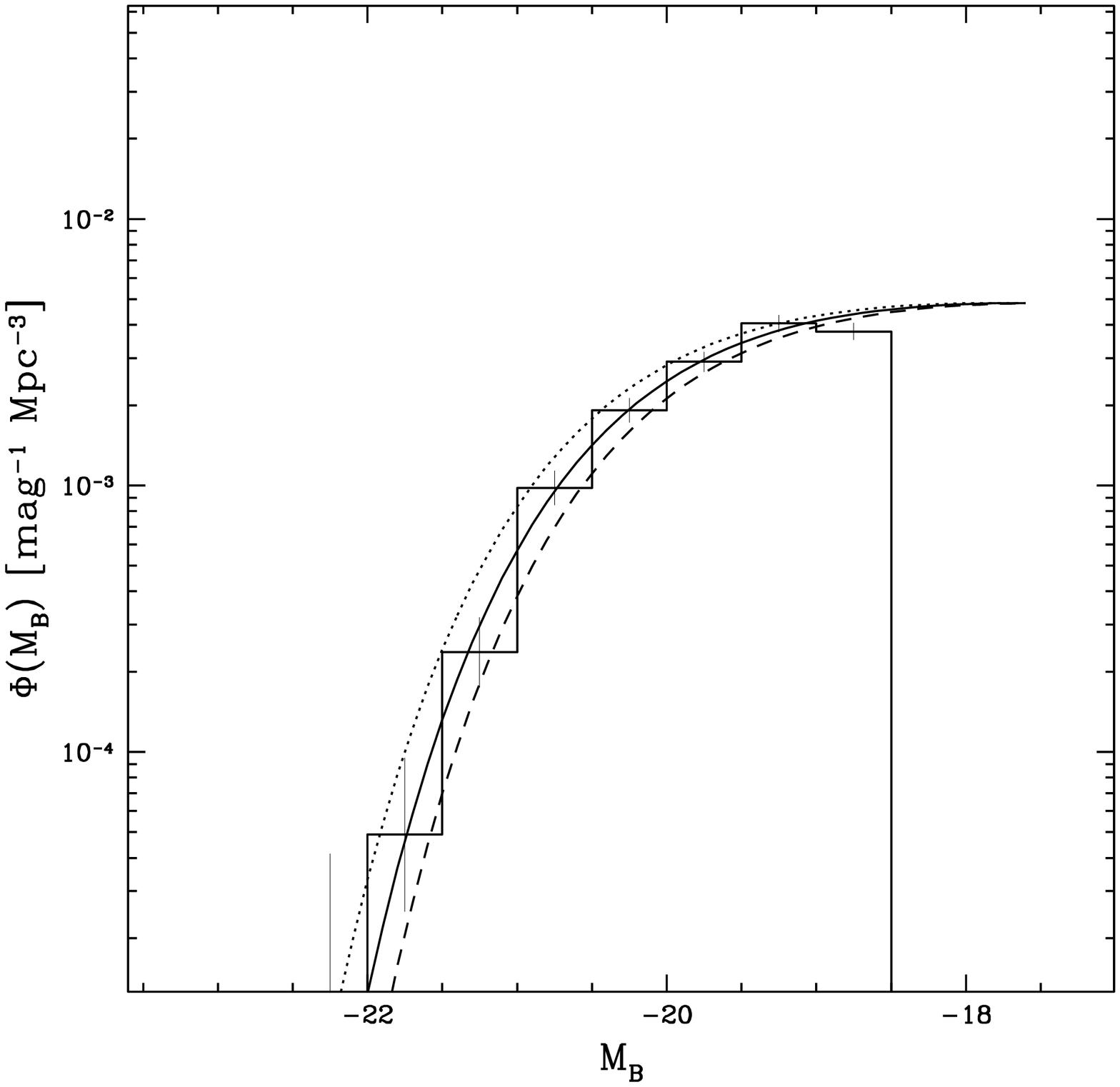}
\caption{The rest--frame $B-$band LF for the SDSS galaxy sample of
  Kampczyk et al., which we use to quantify the evolutionary trends
  since $z\sim 1$ present in our COSMOS sample (solid histogram). This
  is normalized using the \citet{blanton2003} $g-$band LF, converted
  to the $B$-band magnitude using the average transformation (solid
  line) between a 12Gyr-old single burst SED with a solar metallicity
  (dotted line) and a single-burst 0.1Gyr-old SED with a similar
  metallicity (dashed line). }
\label{fig:sdssLF}
\end{figure}

\clearpage
 \acknowledgments
 
 The HST COSMOS Treasury program was supported through NASA grant
 HST-GO-09822.  We gratefully acknowledge the contributions of the
 entire COSMOS collaboration.  More information on the COSMOS survey
 is available \\ at {\bf \url{http://www.astro.caltech.edu/~cosmos}}.
 We thank the NASA IPAC/IRSA staff for providing online archive and
 server capabilities for the COSMOS datasets. CS acknowledges support
 from the Swiss National Science Fondation.

{\it Facilities:} \facility{HST (ACS)}


\clearpage  
\begin{thebibliography}{}
\bibitem[Abraham et al.(1996)]{abraham1996} Abraham, R.~G., van den
  Bergh, S., Glazebrook, K., Ellis, R.~S., Santiago, B.~X., Surma, P.,
  \& Griffiths, R.~E.\ 1996, \apjs, 107, 1

\bibitem[Abraham et al.(2003)]{abraham2003} Abraham, R.~G., van den
  Bergh, S., \& Nair, P.\ 2003, \apj, 588, 218  

\bibitem[Bershady et al.(2000)]{bershady2000} Bershady, M.~A., 
Jangren, A., \& Conselice, C.~J.\ 2000, \aj, 119, 2645 

\bibitem[Bertin \& Arnouts(1996)]{sextractor} Bertin, E., \& 
Arnouts, S.\ 1996, \aaps, 117, 393 
  
\bibitem[Blanton et al.~(2003)]{blanton2003} Blanton, M.~R., et al.\ 
  2003, \apj, 594, 186
 
\bibitem[Bruzual \& Charlot (2003)]{bc03} Bruzual, G. \& Charlot, S.,
  2003, \mnras, 344, 1000
 
\bibitem[Capak et al.~(2006)]{capack2006} Capack, P. et al.~ 2006,
  \apjs, this volume
  
\bibitem[Carollo et al.~(1993)]{carollo1993} Carollo, C.~M., Danziger,
  I.~J., \& Buson, L.\ 1993, \mnras, 265, 553
 
\bibitem[Carollo et al.~(2006)]{carollo2006} Carollo, P. et al.~ 2006,
  in preparation
  
\bibitem[Cassata et al.(2005)]{cassata2005} Cassata, P., et al.\ 2005,
  \mnras, 357, 903

\bibitem[Conselice et al.(2000)]{conselice2000} Conselice, C.~J.,
  Bershady, M.~A., \& Jangren, A.\ 2000, \apj, 529, 886

\bibitem[Daddi et al.(2005)]{2005ApJ...631L..13D} Daddi, E., et al.\
  2005, \apjl, 631, L13

\bibitem[Dahari(1985)]{1985ApJS...57..643D} Dahari, O.\ 1985, \apjs,
  57, 643

\bibitem[Davis et al.(2003)]{deep} Davis, M., et al.\ 2003, \procspie,
  4834, 161

\bibitem[de Vaucouleurs et al.(1991)]{devaucouleurs1991} de
  Vaucouleurs, G., de Vaucouleurs, A., Corwin, H., Buta, R. J.,
  Paturel, G., \& Fouque, P. 1991, {\it Third Reference Catalogue of
    Bright Galaxies}, New York: Springer--Verlag

\bibitem[Emsellem et al.(2004)]{emsellem2004} Emsellem, E., et al.\
  2004, \mnras, 352, 721

\bibitem[Faber et al.~(2005)]{faber2005} Faber, S.M., et al.\ 2005,
  \apj, submitted, astroph/0506044
  
\bibitem[Feldmann et al.~(2006)]{feldmann2006} Feldmann, R. et al.,
  2006 \mnras, accepted, astroph/0609044

\bibitem[Felten (1976)]{felten1976} Felten, J.~E.\ 1976, \apj, 207,
  700
 
\bibitem[Ferreras et al.~(2005)]{ferreras2005} Ferreras, I., et al.,
  2005, \apj, in press

\bibitem[Frei et al.(1996)]{frei1996} Frei, Z., Guhathakurta, P.,
  Gunn, J.~E., \& Tyson, J.~A.\ 1996, \aj, 111, 174
 
\bibitem[Glasser~ (1962)]{glasser1962} Glasser, G.J. 1962, J. Amer.
  Stat. Assoc. 57, 648, 654
  
\bibitem[Graham \& Driver (2005)]{graham2005} Graham, A.~W., \&
  Driver, S.~P.\ 2005, Publications of the Astronomical Society of
  Australia, 22, 118
  
\bibitem[Hasinger et al.~(2006)]{hasinger2006} Hasinger, G., et al.\
  2006, \apjs, this volume

\bibitem[Holden et al.(2005)]{2005ApJ...626..809H} Holden, B.~P., et
  al.\ 2005, \apj, 626, 809

\bibitem[Jolliffe(1972)]{jolliffe1972} Jolliffe, 1972 Applied Statistics, 21, 160.
    
\bibitem[Kaiser (1960)]{kaiser1960} Kaiser, H. F. 1960. Educ. Psychol.
  Meas., 20, 141
  
\bibitem[Kampczyk et al.~ (2006)]{kampczyk2006} Kampczyk et al.~ 2006,
  \apjs, this volume (astro-ph/0611187)

\bibitem[Kitzbichler et al.~ (2006)]{kitzbichler2006} Kitzbichler et
  al.~ 2006, \apjs, submitted

\bibitem[Koekemoer et al.~ (2006)]{kokomero2006} Koekemoer, A. et al.~
  2006, \apjs, this volume

\bibitem[Krist \& Hook(1997)]{1997hstc.work..192K} Krist, J.~E., \&
  Hook, R.~N.\ 1997, The 1997 HST Calibration Workshop with a New
  Generation of Instruments, p.~192, 192

\bibitem[Leauthaud et al.~ (2006)]{leauthaud2006} Leauthaud, A. et
  al.~ 2006, \apjs, this volume

\bibitem[Lilly et al.(2006)]{lilly2006} Lilly, S. et al.~ 2006, \apjs,
  this volume

\bibitem[Lotz, Primack \& Madau (2004)]{lotz2004} Lotz, J.~M.,
  Primack, J. \& Madau, P. 2004, \aj, 128, 163
 
\bibitem[Lotz et al.(2006)]{lotz2006} Lotz, J.~M., Madau, P., 
Giavalisco, M., Primack, J., \& Ferguson, H.~C.\ 2006, \apj, 636, 592 
  
\bibitem[McIntosh et al.(2005)]{2005ApJ...632..191M} McIntosh, D.~H.,
  et al.\ 2005, \apj, 632, 191


\bibitem[Michard \& Marchal~(1994)]{michard1994} Michard, R.; \&
  Marchal, J.\ 1994, \aj, 105, 481-501

\bibitem[Mizuno \& Oikawa(1996)]{mizuno1996} Mizuno, T., \& Oikawa,
  K.-I.\ 1996, \pasj, 48, 591
    
\bibitem[Oke (1974)]{oke1974} Oke, J.~B.\ 1974, \apjs, 27, 21

\bibitem[Petrosian (1976)]{petrosian1976} Petrosian, V.\ 1976, \apjl,
  209, L1
     
\bibitem[Rigler \& Lilly(1994)]{1994ApJ...427L..79R} Rigler, M.~A., \&
  Lilly, S.~J.\ 1994, \apjl, 427, L79
 
\bibitem[Sargent et al.~(2006)]{sargent2006} Sargent M., et al.\ 2006,
  \apjs, this volume (astroph/0609042)
  
\bibitem[Schechter(1976)]{schechter1976} Schechter, P.\ 1976, \apj,
  203, 297
 
\bibitem[Schmidt (1968)]{schmidt1968} Schmidt, M.\ 1968, \apj, 151,
  393

\bibitem[Scarlata et al.~(2006b)]{scarlata2006b} Scarlata, C., et al.\
  2006, \apjs, this volume

\bibitem[Scorza \& Bender(1995)]{scorza1995} Scorza, C., \& Bender,
  R.\ 1995, \aap, 293, 20

\bibitem[Scoville et al.~(2006)]{scoville2006a} Scoville, N., et al.\
  2006, \apjs, this volume

\bibitem[Scoville et al.~(2006)]{scoville2006b} Scoville, N., et al.\
  2006, \apjs, this volume

\bibitem[Sersic(1968)]{sersic1960} Sersic, J.~L.\ 1968, Cordoba,
  Argentina: Observatorio Astronomico, 1968,

\bibitem[Simard(1998)]{simard1998} Simard, L.\ 1998, ASP
  Conf.~Ser.~145: Astronomical Data Analysis Software and Systems VII,
  145, 108

\bibitem[Simard et al.(2002)]{simard2002} Simard, L., et al.\ 2002,
  \apjs, 142, 1

\bibitem[Thomas \& Davies~(2006)]{thomas2006} Thomas, D., \& 
Davies, R.~L.\ 2006, \mnras, 366, 510 
 
\bibitem[Treu et al.(2005)]{treu2005} Treu, T., et al.\ 2005, \apj,
  633, 174
 
\bibitem[Willmer et al.~(2006)]{willmer2006} Willmer, C.~N.~A., et
  al.\ 2006, \apj, 647, 853
 
  
\bibitem[Wolf et al.(2003)]{wolf2003} Wolf, C., Meisenheimer, K., Rix,
  H.-W., Borch, A., Dye, S., \& Kleinheinrich, M.\ 2003, \aap, 401, 73

\bibitem[York et al.(2000)]{york2000} York, D.~G., et al.\ 2000, \aj,
  120, 1579

 \end{thebibliography}
 \end{document}